%% file: sumof3wellsELS.tex
\documentclass[11pt]{article}
\usepackage{amsmath,amssymb,amsthm}
\usepackage{graphicx}
\numberwithin{equation}{section}
\usepackage{fullpage}
\usepackage{MnSymbol}
\usepackage[numbers,sort&compress]{natbib} 
\usepackage{subfig}
\usepackage[export]{adjustbox}
\usepackage{enumitem}
\include{abbrevs}

\graphicspath{{Figures/}}

\title {Bifurcations of relative periodic orbits in NLS/GP with a triple-well potential}

\author{Roy H. Goodman\\
Department of Mathematical Sciences\\
New Jersey Institute of Technology\\
University Heights \\
Newark, NJ 07102
\thanks{goodman@njit.edu}
}

\date{\today}

\begin{document}

\maketitle

\begin{abstract}
The nonlinear Schr\"odinger/Gross-Pitaevskii (NLS/GP) equation is considered in the presence of three equally-spaced potentials. The problem is reduced to a finite-dimensional Hamiltonian system by a Galerkin truncation. Families of oscillatory orbits are sought in the neighborhoods of the system's nine branches of standing wave solutions.  Normal forms are computed in the neighborhood of these branches' various Hamiltonian Hopf and saddle-node bifurcations, showing how the oscillatory orbits change as a parameter is increased. Numerical experiments show agreement between normal form theory and numerical solutions to the reduced system and NLS/GP near the Hamiltonian Hopf bifurcations and some subtle disagreements near the saddle-node bifurcations due to exponentially small terms in the asymptotics. 
\end{abstract}

\section{Introduction}
This paper considers the nonlinear Schr\"odinger/Gross-Pitaevskii equation (NLS/GP) with a triple-well potential as a model problem for investigating the occurrence and bifurcations of approximate relative periodic orbits (RPOs) in the setting of a Hamiltonian partial differential equation, i.e.\ solutions that are periodic modulo a symmetry of the system. The largest part of the paper is dedicated to understanding how these orbits bifurcate when the system undergoes a Hamiltonian Hopf (HH) bifurcation, but will discuss a number of other families of RPOs in this system that are unrelated to this bifurcation. This paper will analyze, primarily,  a finite-dimensional system of Hamiltonian ODE derived as a Galerkin truncation of NLS/GP with the given potential.

The HH bifurcation is a mechanism that gives rise to oscillatory instabilities in conservative systems including nonlinear dispersive wave equations. It occurs when varying some parameter $\alpha$ in the system across a threshold $\alpha_{\rm HH}$ causes pairs of eigenvalues $\pm i \omega_1$ and $\pm i \omega_2$ on the imaginary axis to collide and split to form so-called Krein quartets of eigenvalues $\lambda = \pm \mu \pm i \omega$. The motion with $\alpha$ on one side of $\alpha_{\rm HH}$ is oscillatory due to the purely imaginary eigenvalues, while on the other side, the real parts lead to exponential growth or decay, accompanied by oscillation due to the imaginary parts.

Our object of study is the cubic nonlinear Schr\"odinger/Gross-Pitaevskii equation,
\begin{equation}
i \partial_t u = -\partial_x^2 u + V(x) u -|u|^2 u
\label{NLS}
\end{equation}
with a particular potential. This equation is ubiquitous in mathematical physics, arising due to the balance between nonlinearity and dispersion in nearly monochromatic wavepackets. In the optics context, the equation models the propagation of a nearly monochromatic electrical field along a waveguide where $t$ represents the longitudinal distance and $V(x)$ describes the geometry of the waveguide in the transverse direction. The sign of the cubic \emph{Kerr} nonlinearity implies that in regions of higher intensity, the refractive index is increased~\cite{Newell:2003}. This system also arises in describing the evolution of a Bose-Einstein condensate (BEC), a state of matter that occurs at extremely low temperatures at which a gas of identical particles obeying Bose statistics lose their individual identities and share a common wavefunction. When the BEC is confined by strong magnetic or optical fields to a quasi-one-dimensional cigar-shaped region, its wavefunction satisfies NLS/GP with the potential $V(x)$ describing the weaker potential along the axis of the cigar~\cite{Chen:2016,ESY,PS:03}.

\subsection{Mathematical Set-Up}
\begin{figure}[htbp] 
   \centering
   \includegraphics[width=4in]{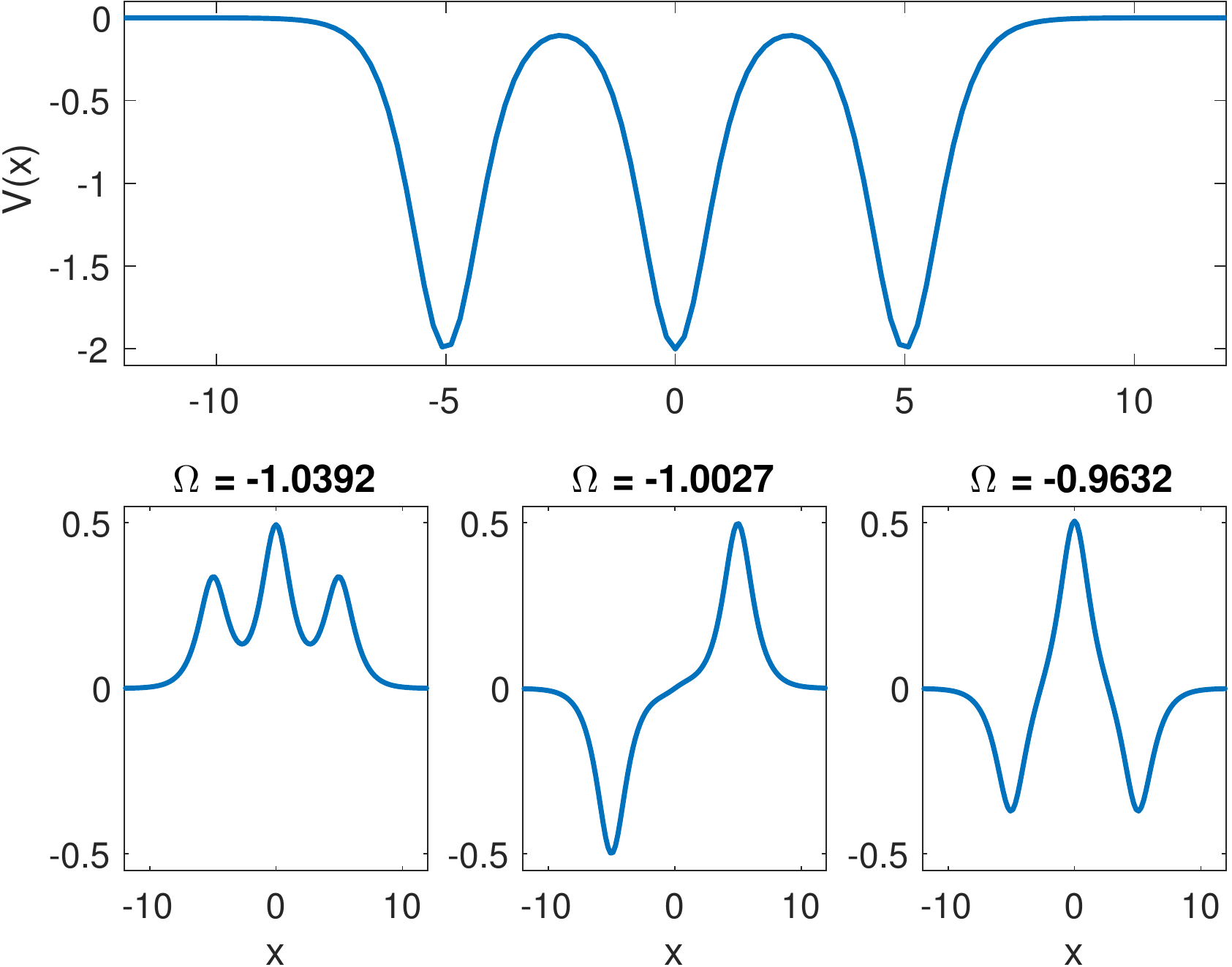} 
   \caption{The potential $V_3(x)$ of equations~\eqref{V3} and~\eqref{V0sech} with $L=5$ and its three eigenfunctions: the ground state $U_1$ which is positive everywhere, the antisymmetric first excited state or dipole mode $U_2$, and the second excited state $U_3$.}
\label{fig:potential}
\end{figure}

Here we focus on a particular potential $V(x)$ given as the sum of three identical, symmetric, strongly localized potentials $V_0(x)$, hereafter the ``triple-well'' potential
\begin{equation}
V_3(x) = V_0(x+L) + V_0(x) + V_0(x-L). 
\label{V3}
\end{equation}
We assume that the single potential $V_0(x)$ supports a single discrete eigenfunction for the linear system 
$$
\left(-\partial_x^2 + V_0(x) \right) U_0(x) = \Omega_0 U_0(x),
$$
and that $V_0(x)$ satisfies even symmetry $V_0(x)=V_0(-x)$. An example to keep in mind is 
\begin{equation}
V_0(x)=-2\sech^2{x}, 
\label{V0sech}
\end{equation}
for which $\Omega_0=-1$ and $U_0(x)=\sech(x)$.
The potential $V_3(x)$ and its three eigenfunctions are displayed in Figure~\ref{fig:potential}. One can show that in the large-$L$ limit the eigenvalues take the form
\begin{equation}
(\W_1,\W_2,\W_3) = (\W_2 -\D + \ep, \W_2, \W_2 + \D + \ep),
\label{W1W2W3}
\end{equation}
where, exponentially as $L\to \infty$,
\begin{equation*}
\W_2 \to \W_0 , \, \D \to 0,\, \ep \to 0, \text{and } \ep \ll \D
%\label{convergence}
\end{equation*}
and $\D$ itself approaches zero exponentially as a function of $L$.
The frequency $\D$ is by definition positive, while  $\ep$ can be found by asymptotics to be positive. This arrangement of eigenvalues predisposes a nonlinear mode associated with the frequency $\W_2$ to HH bifurcations, as we shall see.

\begin{rem}
\label{numeric_values}
In all the numerical calculations, we take $V_0$ as in equation~\eqref{V0sech} and $L=5$, which gives approximate parameters $\D=3.801\times 10^{-2}$ and $\ep = 1.529\times10^{-3}$, so that $\D \approx 25 \ep$, ensuring a large separation of scales.
\end{rem}

System~\eqref{NLS} is Hamiltonian and can be written
\begin{equation*}
i \partial_t u = \frac{\delta \cH}{\delta u^*},
\end{equation*}
where $\cH$ is the conserved energy functional
\begin{equation*}
\cH = \int\left(\abs{u_x}^2 + V(x)\abs{u}^2 -\frac{1}{2} \abs{u}^4 \right) \ dx.
\end{equation*}
The system conserves the squared $L^2$ norm, also known as the optical power or, in the BEC context, the particle number:
\begin{equation}
\cN = \int \abs{u}^2 dx.
\label{cN}
\end{equation}

We are interested in the behavior near \emph{nonlinear bound states}, or \emph{standing waves}, solutions to equation~\eqref{NLS} of the form
$$
u(x,t) = \cU(x) e^{-i \Omega t},
$$
consisting of a real-valued function $\cU(x;\cN) \in H^2(\RR)$ and  a frequency $\Omega(\cN)$ satisfying
\begin{equation}
\label{stationary}
\Omega \cU  = -\cU'' + V(x) \cU - \cU^3.
\end{equation}
These generally exist as one-parameter families where we think of $\cU$ and $\Omega$ as being functions of the optical power~$\cN$. 

In absence of  the cubic term, equation~\eqref{NLS}, the familiar Schr\"odinger equation of quantum mechanics, possesses a set of  discrete eigenfunctions and frequencies $U_j(x)$ and $\w_j$.  By the implicit function theorem, as applied in~\cite{KirKevShl:08}, for each eigenpair, there exists for small $\cN$ a unique \emph{nonlinear normal mode} (NNM),  a nonlinear bound state $(\cU_j(x;\cN), \Omega_j(\cN))$ satisfying
\begin{equation}
\label{PDE_NNM}
\cU_j(x,\cN) \sim \sqrt{\cN} \left(U_j(x) + O(\cN)\right); \;   \int {U_j(x)}^2 dx = 1; \; \Omega_j(\cN) \to \w_j \text{ as } \cN \to 0.
\end{equation}

Potentials of the form~\eqref{V3} have three eigenfunctions, which in the limit of large $L$ approach the form
\begin{equation*}
\begin{split}
U_1(x) &= \frac{1}{2} \left( U_0(x+L) + \sqrt{2} U_0(x) + U_0(x-L) \right) \\
U_2(x) &= \frac{1}{\sqrt{2}}\left( U_0(x+L) -U_0(x-L) \right) \\
U_3(x) &= \frac{1}{2}\left( -U_0(x+L) + \sqrt{2} U_0(x) - U_0(x-L) \right).
\end{split}
%\label{linearModes}
\end{equation*}
Numerically computed eigenfunctions are shown in Figure~\ref{fig:potential}. The modes $U_1$ and $U_2$ are known as the \emph{ground state} and the \emph{dipole mode}, respectively. The modes $U_1$ and $U_3$ are symmetric, while $U_2$ is antisymmetric. In addition $U_2$ and $U_3$ are known, respectively, as the first and second excited states.

\begin{rem}
\label{rem:naming}
We adopt a naming convention from~\cite{KapKevChe:06} for the \emph{nonlinear} standing waves. A mode is labeled by a three-character string composed of {\rm 0}, {\rm+}, and $-$ if, in the limit of large $\cN$, the solution has, respectively, 0, $U_0(\cdot)$, or $-U_0(\cdot)$ in the indicated location. Thus the continuations of $U_2$ and $U_3$ are labeled, respectively, as $\mop$ and $\mpm$. Less obviously, the continuation of $U_1$ is labelled $\opo$ because for large amplitudes, the concentration in the first and third wells decreases to zero, as discussed in Section~\ref{sec:evenFixed}. Interchanging the symbols ``-'' and ``+'' leaves the mode unchanged, while reversing the order of the labels gives a mode's mirror image.
\end{rem}

\subsection{Mathematical/Physical motivation and prior work}
\label{sec:priors}

The study of NLS/GP with multi-well potentials generally falls into two categories. The first is analysis of existence, bifurcation, and stability of stationary solutions to equation~\eqref{stationary}. A phenomenon of particular interest here is \emph{localization} or \emph{self-trapping}, the concentration of energy in one (or at least a small subset) of the wells as the total amplitude is increased. The second is investigation of the time-dependent dynamics. Here, we speak of \emph{tunneling}, as the solution moves between the various wells. This can be periodic or chaotic.

The first question is addressed, for small $\cN$, for the \emph{double}-well potential $V_2(x)$ by Kirr et al.~\cite{KirKevShl:08}. This system has two NNMs, a ground state and a dipole mode. As the amplitude parameter $\cN$ is increased past a critical value, the ground state becomes unstable and a pair of stable asymmetric standing waves emerges in a symmetry-breaking (supercritical pitchfork) bifurcation, with $U(x)$ concentrated in either the left or right well. The behavior in the limit of large $\cN$ is shown in~\cite{Kirr:2011eu,Kirr:2016uy}. Here there exist localized solutions that are centered not at the minima of the potential but at points in between, including at the maxima.

Stationary solutions of the \emph{triple}-well problem are enumerated by Kapitula et al~\cite{KapKevChe:06}. In addition to the three NNM's, they found six additional solutions, which arise in three saddle-node bifurcations as $\cN$ is increased. They show analytically that the symmetry-breaking seen in the double-well potential is not possible with a triple well. They numerically calculated stability of each solution as $\cN$ is increased, showing that the NNM  loses stability and then regains stability in a pair of HH bifurcations. Localization in either of the outer wells occurs due to a saddle-node bifurcation and  localization in the central well occurs continuously, without any bifurcations, as $\cN$ is raised. Our recent study~\cite{Goodman:2011} estimates the critical values of $\cN$ for the HH bifurcations.  Sacchetti~\cite{Sacchetti:2012tx} uses semiclassical methods to prove the correctness of a finite-dimensional truncation for the general $n$-well system, and applies this result to the \emph{quadruple} well, finding that in this case, too, the ground state undergoes a pitchfork bifurcation, which may be sub- or supercritical, depending on the exponent of the nonlinearity. 

The second question, regarding time-dependent dynamics and tunneling, has been addressed in several studies of the double-well problem~\cite{Goodman:2015by,Marzuola:2010,Pelinovsky:2012}. These show the  dynamics of supercritical Hamiltonian pitchfork bifurcations: below the critical amplitude, the dynamics resemble a particle oscillating in a single-well potential, while above this amplitude, the dynamics resemble those of a double-well potential. For other nonlinearities, the pitchfork may be supercritical or subcritical~\cite{Fukuizumi:2011ku,Pelinovsky:2012}. The finite-dimensional models derived in these papers have a two-dimensional phase space. This allows the dynamics to be understood completely by plotting the level sets of the conserved energy. These studies confirm that solutions to the reduced model that are sufficiently bounded away from the separatrix are shadowed over long times by PDE solutions but do not address the question of solutions on or near the separatrix, nor the interesting question of how radiative dissipation in the PDE could cause the solution to cross the separatrix.

The triple well potential $V_3(x)$ leads to significantly more complicated dynamics.  Our previous study~\cite{Goodman:2011} estimates the critical values of $\cN$ for the HH bifurcation and begins a study of the four-dimensional phase space of a finite-dimensional model. In particular, it explores some of the dynamics that arise along with the HH bifurcation, and numerically demonstrates chaotic dynamics. It studies an averaged form of the equations which gives some insight to the geometry underlying the observed behavior. 

These phenomena have also been reported in laboratory studies. Notably, Albiez et al.\ have reported both self-trapping and tunneling in a cigar-shaped BEC in a double-well potential~\cite{Albiez:2005}. Kapitula et al.\ perform experiments in an optically-induced optical waveguide~\cite{KapKevChe:06}. They find symmetry-breaking in an analogue of the double well problem, and verify certain mathematical details of the triple-well problem.

The current study aims to extend previous analyses and demonstrate other structures present in the dynamics. \emph{While it was possible and logical in~\cite{Marzuola:2010} both to describe the relative periodic orbits in a reduced ODE model and to prove their validity in the full double-well potential in the same paper, the dynamics the finite-dimensional reduction of the triple well are sufficiently complicated to require their own paper, and to leave the proof of their validity for later.}

A similar analysis to ours, with overlapping but not identical results, appears in the recent work of Yang~\cite{Yang:2016hj}, who uses the method of multiple scales to derive a normal form for the HH bifurcation directly from the NLS/GP equation, without assuming a triple-well potential. Most interestingly, he finds that in the case of a resonance between the continuous spectrum and the discrete, some coefficients in the normal form equations become complex-valued, which leads to a blow-up behavior similar to one seen by Kevrekidis et al.~\cite{Kevrekidis:2015bq}.

\subsection{The discrete self-trapping equations}
\label{sec:DST}
It should be further noted that mathematically and physically, the system under study has much in common with the discrete self-trapping (DST) equations 
\begin{equation}
i \frac{d}{d t} \PPsi = -D \PPsi - \mathbf{F}(\PPsi); \;
\PPsi \in \ \RR^{n},
\label{DST}
\end{equation}
where $\mathbf{F}(\PPsi) = \left(\abs{\Psi_1}^2 \Psi_1, \abs{\Psi_2}^2 \Psi_2, \ldots, \abs{\Psi_n}^2 \Psi_n \right)^{\rm T}$ and $D$ is the discrete second derivative or Laplacian. This system, also referred to as discrete NLS, was introduced by Eilbeck et al.\ to model, among other things, the oscillations of small molecules~\cite{Eilbeck:1985tu}. Eilbeck and collaborators were the first to enumerate the stationary solutions for small values of $n$ and to calculate their stability numerically~\cite{Carr:1985,Eilbeck:1985tu}. Susanto has recently surveyed some important results for this system~\cite{Susanto:2009km}.

Exact periodic orbits in the dimer case $\PPsi \in \RR^2$ have been known since the 1980's~\cite{Banacky:1988ft,Kenkre:1986fe}. The trimer has been studied extensively as a model of BEC in a triple-well potential~\cite{Liu:2007cl,Zhang:2001jg}. While many groups have noticed that the system can be reduced from three to two degrees of freedom, most of the papers have focused on stationary solutions and numerical computations using tools like Poincar\'e sections. Johansson computes some  RPOs in this system and identifies two HH bifurcations of one stationary solution~\cite{Johansson:2004uj}, noting a qualitative difference between the two, equivalent to the two different normal forms discussed here. Panayotaros has also studied the trimer, using topological methods to investigate the stability of a different stationary solution~\cite{Panayotaros:2012cs}. Basarab has done some work to apply the methods discussed in the present paper to the DST trimer~\cite{Basarab:2016}.

\subsection{Preview and Organization}
In this paper, we study the dynamics of a finite-dimensional ODE system that model the dynamics of solutions to NLS/GP~\eqref{NLS} whose initial conditions are a small-amplitude linear combination of the $U_j$. By exploiting a phase invariance, this ODE system can be reduced to a two-degree-of-freedom Hamiltonian system which, at small amplitudes, is close to the semisimple -1:1 resonance (to be defined below).  The phase space is four-dimensional, so it is not possible to simply draw a phase plane, but by reduction techniques, we are able to determine some important features. In particular, whereas~\cite{KapKevChe:06} enumerated nine families of standing wave solutions, we enumerate additional families of \emph{relative periodic orbits} in which energy moves periodically between the wells and describe how these families of orbits change as the optical power $\cN$ is increased. 

A primary tool used here is \emph{canonical normal forms}, originally developed in the study of celestial mechanics. One of our goals is simply to understand what occurs in nonlinear waves that undergo HH bifurcations. To this end, we apply two separate types of normal form calculations to the finite-dimensional model system. The first of these normal forms, which applies small amplitudes, gives more global information than we expected and describes three families of relative periodic orbits. The second, which is more standard, describes how the topology of relative periodic orbits changes as two HH bifurcation values are crossed. Three saddle-node (SN) bifurcations are also studied, using both normal forms and numerical continuations.

The remainder of the paper is organized as follows. 
Section~\ref{sec:prelim} begins with an overview of the notation used and continues with a brief review of the necessary concepts from Hamiltonian mechanics. 
Section~\ref{sec:modelODE} describes the finite-dimensional system of ODEs used to model the behavior of solutions to NLS and a further reduced system which we will study in greater detail. It contains asymptotic expressions for the nonlinear normal mode solutions in the weakly nonlinear limit and also describes the reduction from three degrees of freedom to two.
In Section~\ref{sec:allBoundStates}, we briefly enumerate all the branches of standing wave solutions and their bifurcations for reference in later sections. This necessarily repeats some of the calculations of~\cite{KapKevChe:06}.
Section~\ref{sec:linear} describes the linearization about the dipole mode $\mop$, which undergoes two separate HH bifurcations, and derives asymptotic expressions for the value of the optical power $\cN$ at which the bifurcation occurs.
Section~\ref{sec:ChowKim} describes the normal form of Chow and Kim~\cite{Chow:1988}, which applies in the neighborhood to the semisimple -1:1 resonance. We apply this normal form in a neighborhood of $\mop$ for optical power which is on the order of the small parameter $\epsilon$ defined in equation~\eqref{W1W2W3}. The normal form describes the topological changes that occur in branches of relative periodic orbits in which energy oscillates periodically among the three eigenmodes.
In a neighborhood of the HH bifurcation, a more standard normal form is applicable. This comes in two flavors that could be called supercritical and subcritical.  In Section~\ref{sec:HH} we examine the dynamics near the mode $\mop$ at its two HH bifurcations and find that they are of opposite type.
Section~\ref{sec:saddlenode} constructs a normal form that applies in the neighborhood of two saddle-node bifurcations. This bifurcation is shown to have two subtypes, similar to what is seen for the HH bifurcation, as well as three different families of small-amplitude periodic orbits.
Section~\ref{sec:Numerics} contains numerical studies that show the normal form equations accurately describe the bifurcations of the branches of relative periodic orbits that occur in the finite-dimensional Hamiltonian reduction. In the neighborhood of the first HH bifurcation, we find long-lived nearly periodic orbits of the PDE that undergo this same bifurcation. It shows similar agreement near the second HH bifurcation. Numerical investigation of the branches in a neighborhood of two saddle-node bifurcations is found to be a bit more subtle than the normal form calculation would indicate, owing to terms beyond all orders in the asymptotics. These numerical results are discussed in relation to the existing literature.
Finally, Section~\ref{sec:conclusion} concludes with some discussion summarizing the findings, putting them in context of other research, and describing possible future directions.

\section{Mathematical Preliminaries and Notation}
\label{sec:prelim}
\subsection{Notation}
\begin{itemize}[itemsep=0.3mm]
\item An overbar, $\bar z$ represents the complex conjugate of $z$.  
\item Boldface quantities, e.g.\ $\pp$, $\qq$, are vectors.
\item Angle brackets $\innerProd{\cdot}{\cdot}$ represent various types of inner products.
\item $\cP_j(\RR^n)$ is the space of homogeneous polynomials of degree $j$, i.e.\ polynomials $ {\mathfrak p}(\yy)$ satisfying ${\mathfrak p}( c \yy) = c^j {\mathfrak p}(\yy)$. Monomials in $\cP_j$ are written in multi-index notation $\yy^{\aalpha} = y_1^{\alpha_1}\ldots y_n^{\alpha_n}$ where $\aalpha \in \ZZ^n_+$ and $\abs{\aalpha} = \alpha_1 + \cdots +\alpha_n =j$. The natural inner product on this space is
\begin{equation}
\innerProd{F}{G}_{\cP_j} = F(\partial_{\yy})G(\yy)\rvert_{\yy = 0}.
\label{dotPJ}
\end{equation}
In particular, when applied to monomials,
$
\innerProd{\yy^{\aalpha}}{\yy^{\bbeta}}_{\cP_j} = \aalpha!\delta_{\aalpha,\bbeta},
$
where the multi-index factorial is $\aalpha! = \prod_{j=1}^N{\alpha_j!}$. Analogous definitions hold for $\CC^n$.
\item $I_n$, or just $I$, is the $n\times n$ identity matrix. $J_{2n}$, or just $J$, is the $2n\times 2n$ matrix $\left( \begin{smallmatrix}
0 & I_{n} \\ -I_{n} & 0 
\end{smallmatrix} \right)$.
\item The canonical Poisson bracket with respect to $(\pp,\qq)$ coordinates is
\begin{equation*}
\PB{F}{G} =
\sum_{j=1}^n  \left(
\frac{\partial F}{\partial q_j} \frac{\partial G}{\partial p_j} - 
\frac{\partial F}{\partial p_j} \frac{\partial G}{\partial q_j}
\right). 
\end{equation*}
\item The adjoint operator of a quadratic Hamiltonian $H_0(\pp,\qq) \in \cP_2(\RR^{2n})$ is defined as
\begin{equation*}
\ad{H_0} F = \PB{F}{H_0}.
\end{equation*}
\item Noting that $\ad{H_0}: \cP_j \to \cP_j$, define $\ad{H_0}^{(j)}$ to be the restriction of $\ad{H_0}$ to $\cP_j$.
\item The kernel and range of a linear operator $L$ are denoted by $\ker{L}$ and $\ran{L}$, respectively.
\end{itemize}

\subsection{Notions from Hamiltonian Mechanics}
Familiarity with the basics of Hamiltonian mechanics will be assumed, but we highlight here the main concepts  applied in the paper. Most of the ideas used in this paper can be learned from~\cite{Meyer:2010}. 

\subsubsection{The basics}
An $n$-degree-of-freedom Hamiltonian system consists of a set of $2n$ ordinary differential equations 
\begin{equation}
\label{Ham}
\dot{q}_j = \partial_{p_j} H(\qq,\pp,t); \;
\dot{p}_j = -\partial_{q_j} H(\qq,\pp,t),
\end{equation}
where $\pp,\qq\in \RR^n$ or $\yy = (\qq,\pp)$ lies in a $2n$-dimensional manifold, and $H(\qq,\pp,t)$ is a $C^2$ function $H:\RR^{2n}\times \RR \to \RR $. The variables $\qq$ and $\pp$ known respectively as the position and momentum vectors. In $\yy$ coordinates, $\dot \yy = J \nabla H(\yy;t)$. 

\begin{figure}[htbp] 
   \centering
   \includegraphics[width=.7\textwidth]{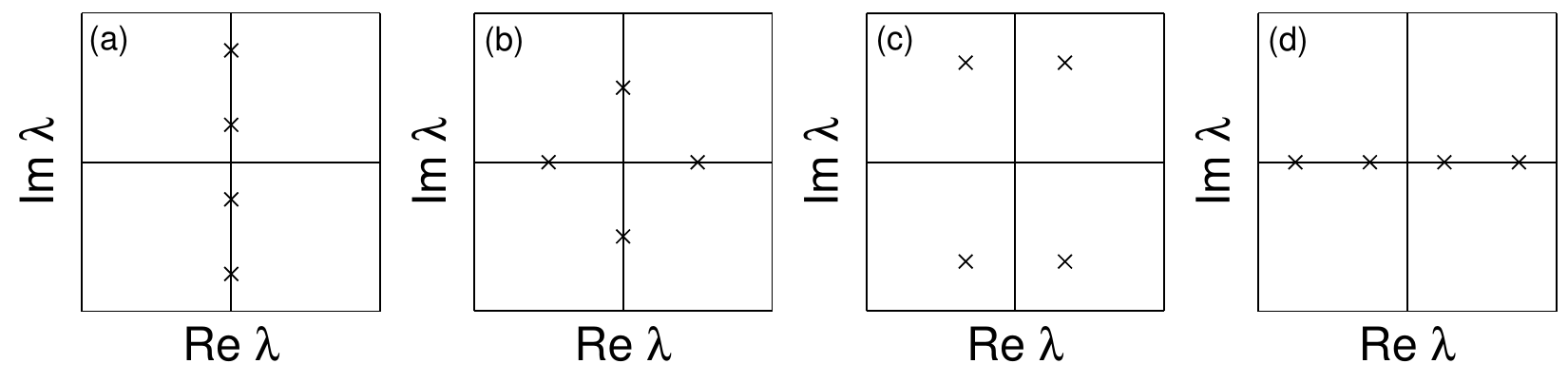}
   \caption{The four possible non-zero arrangements of eigenvalues in a two-degree-of-freedom Hamiltonian system.}
\label{fig:eigsH}
\end{figure}

If $H(\yy) \in \cP_2$, then $\yy$ solves a linear equation
\begin{equation}
\dot\yy = J S \yy
\label{linearODE}
\end{equation}
where $S$ is a symmetric matrix. If $\lambda$ is an eigenvalue of $JS$, then $-\lambda$, $\bar{\lambda}$, and $-\bar{\lambda}$ are also eigenvalues, and of the same multiplicity. Thus there are four types of eigenvalues: real pairs $\pm \lambda$, imaginary pairs $\pm i\omega$, complex \emph{Krein} quartets $\lambda =\pm \mu \pm i \omega$, and zeros of even multiplicity. These restrictions imply that the stability of a fixed point can change in only a few ways as parameters in an equation vary, as shown in Figure~\ref{fig:eigsH}: in a pitchfork bifurcation, imaginary eigenvalues collide at the origin and emerge as a pair of real eigenvalues, transitioning from arrangement (a) to arrangement (b) or from (b) to (d), and in a Hamiltonian Hopf bifurcation two pairs collide on the imaginary axis and become a quartet, transitioning from arrangement (a) to (c). Zero eigenvalues that persist as parameters are varied generally indicate the presence of symmetries. The ODE system studied in this paper has such a symmetry, which we use to reduce the dimension of the problem. The reduced system has  eigenvalues equal to zero at only a few bifurcation points in parameter space.

\subsubsection{Relative Fixed Points and Relative Periodic Orbits}

A fixed point in a symmetry-reduced system corresponds to a type of periodic orbit in the non-reduced system that is known as a \emph{relative fixed point} (RFP): it is a fixed point modulo the action of a symmetry group, in our case the group $S^1$ corresponding to gauge invariance. The standing wave~\eqref{stationary} is an RFP of the NLS system~\eqref{NLS}. A periodic orbit in the symmetry reduced system corresponds generically to a quasiperiodic orbit in the original system known also as a \emph{relative periodic orbit} (RPO). 

While our main concern will be RFPs and RFO's, in the reduced system we can consider simpler fixed points and periodic orbits.  Periodic orbits in a neighborhood of a fixed point $\yy^*$ in a finite-dimensional Hamiltonian system are described by the \emph{Lyapunov Center Theorem}~\cite[Ch. 9.2]{Meyer:2010}, which states that if the linearization about $\yy^*$ has spectrum $\left\{\pm i\omega, \lambda_3, \lambda_4,\ldots, \lambda_{2n}\right\}$, where $i\omega\neq 0$ is pure imaginary and $\lambda_j/i\omega$ is never an integer for $j=3,\ldots 2n$, then there exists a one-parameter family of periodic orbits emanating from $\yy^*$. Further, the period of these orbits approaches $2\pi/\omega$  as the periodic orbits approach $\yy^*$. We refer to these as \emph{Lyapunov families} of periodic orbits.

The Lyapunov Center theorem applies in two situations we will encounter when considering our two degree-of-freedom reduced system.  If the spectrum about $\yy^*$ is of the form $\left\{\pm i\omega, \pm \lambda\right\}$ with two real and two imaginary eigenvalues, then there is a single Lyapunov family of periodic orbits connected to $\yy^*$. If the spectrum consists of two pairs of imaginary eigenvalues $\left\{\pm i \w_1, \pm i \w_2\right\}$, with $0< \w_1 < \w_2$, then there is always a Lyapunov family with limiting frequency $\w_2$. If $\w_2/\w_1 \notin \ZZ$, then there is a second Lyapunov family with limiting frequency $\w_1$.

\subsubsection{Resonances}
Suppose that the Hamiltonian is quadratic and can be written
$
H_2 = \sum_{j=1}^n \omega_j\left( p_j^2 + q_j^2 \right)/2.
$
Then if $\kk\in \ZZ^{n}$ is a nonzero vector of integers such that $\innerProd{\kk}{\ww} = 0$, the Hamiltonian is \emph{resonant}. The order of the resonance $|\kk|=\sum{\abs{k_i}}$. If a system has multiple eigenvalues, the leading-order Hamiltonian may not be expressible in this form, but the resonance is defined similarly.

We consider the semisimple -1:1 resonance in Section~\ref{sec:ChowKim}. This is defined by a Hamiltonian whose leading-order linear part is of the above form with $\w_2 = -\w_1$ and 
which has resonance vector $\kk = (1,1)$ of order 2. The associated linear system has eigenvalues $\pm i \Omega$, each of multiplicity two. Since the matrix for this system is diagonalizable, it is referred to as the semisimple form of the resonance. A linear Hamiltonian system with the same eigenvalues but whose matrix is not diagonalizable corresponds to a \emph{non-semisimple} -1:1 resonance which is the generic case and which arises in Section~\ref{sec:HH} below. A zero eigenvalue is always resonant. The Hamiltonian $0^2i\omega$ resonance, with a pair of imaginary eigenvalues and a double zero, is considered in Section~\ref{sec:saddlenode}

\subsubsection{Normal Forms}
\label{sec:normalforms}
The normal form of a given differential equation is another differential equation, obtained by a near-identity change of variables, that is in the ``simplest'' form possible in a given region of phase space. The normal form transformation may lose information. Consider a Hamiltonian of the form
$$
H(\yy,\ep) = H_0(\yy) + \ep H_2(\yy) + H_4(\yy)
$$
where $H_0 \in \cP_2$, $H_2 \in \cP_2$, $H_4 \in \cP_4$. Assuming that $\yy=O(\ep^{1/2})$ makes the term $H_0 = O(\ep)$ and the other terms both $O(\ep^2)$. 

Simplifying the system has two parts. First, choosing coordinates in which $H_0$ is in a standard form, and second, constructing a near-identity change of variables that removes as many terms as possible from $\ep H_2 + H_4$, leaving only the \emph{resonant part} $\ep H_2^{\rm N} + H_4^{\rm N}$. This transformation introduces terms of higher order in $\yy$ and $\epsilon$ that can themselves be simplified by the same procedure, and so on. The sequence of transformations generally does not converge, but useful information can be gleaned from truncating the sequence after a finite number of steps. In particular, if the truncated normal-form system possesses periodic orbits, then, by the implicit function, so does the original system. In our case, we only need to calculate the results of the first change of variables, and we can do this without explicitly determining the change of variables. 

An important step involves determining which terms can be removed and which cannot. A term may be removed from $H_j$ if it is in the range of $\ad{H_0}^{(j)}$. Therefore the normal form of $H_j$ is just the projection of $H_j$ onto some complement of the range. Given the inner product~\eqref{dotPJ}, the Fredholm alternative decomposes $\cP_j$ as
\begin{equation*}
\cP_j = 
\ran{\left(\ad{H_0}^{(j)}\right)} 
\oplus 
\ker {\left(\ad{H_0}^{(j)\rm T}\right)}.
\end{equation*}
In fact there exists an easily constructed quadratic Hamiltonian function $H_0^{\rm T}$ such that 
$
 {\left(\ad{H_0}^{(j)}\right)}^{\rm T} = \ad{H_0^{\rm T}}^{(j)}.
$
Thus 
\begin{equation}
H_j^{\rm N} = {\rm proj}_{\ad{H_0^{\rm T}}^{(j)}} H_j
= \sum_{\vv \in B} \frac{ \innerProd{\vv}{H_j}_{\cP_j}}{ \innerProd{\vv}{\vv}_{\cP_j}}\vv,
\label{Hproject}
\end{equation}
where $B$ is an orthogonal basis for $\ker\left( \ad{H_0^{\rm T}}^{(j)}\right)$. 

It should be further pointed out that the transpose of $\ad{H_0}$ is defined in terms of an inner product on spaces of polynomials. While there are many ways to define such an inner product, the choice~\eqref{dotPJ} is especially elegant and leads to simple formulas.

\section{The model ODE systems}
\label{sec:modelODE}

We consider solutions to equation~\eqref{NLS} with suitable initial data that we assume can be well approximated by a time-dependent linear combination of the eigenfunctions 
\begin{equation}
u(x,t) = \sum_{j=1}^n c_j(t) U_j(x),
\label{ansatz}
\end{equation}
where the $U_j$ are the eigenfunctions of the linearization of equation~\eqref{NLS} and the coefficients $c_j$ satisfy a system of ordinary differential equations given below.

The coefficients $c_j(t)$ satisfy a Hamiltonian system of equations 
\begin{equation*}
i \dot{c}_j = \frac{ \partial  }{\partial \bar{c}_j} \tilde{H}(c,\cbar)
\end{equation*}
with Hamiltonian function
\begin{equation}\begin{split}
\tilde{H}(c,\cbar)=&\W_1 {\abs{c_1}}^2 + \W_2 {\abs{c_2}}^2 + \W_3 {\abs{c_3}}^2 
- \thalf \tilde{a}_{1111} {\abs{c_1}}^4  
- \tilde{a}_{1113}{\abs{c_1}}^2(c_1 \bar c_3 + \bar c_1 c_3) \\
&- \tilde{a}_{1122}\left(\thalf c_1^2 \bar c_2^2 + 2 {\abs{c_1}}^2 {\abs{c_2}}^2 + \thalf \bar c_1^2 c_2^2 \right)
- \tilde{a}_{1133}\left(\thalf c_1^2 \bar c_3^2 + 2 {\abs{c_1}}^2 {\abs{c_3}}^2 + \thalf \bar c_1^2 c_3^2 \right)\\
&- \tilde{a}_{1223} \left(2{\abs{c_2}}^2(c_1 \bar c_3 + \bar c_1 c_3) + c_1 \bar c_2^2 c_3 + \bar c_1 c_2^2 \bar c_3  \right)
- \tilde{a}_{1333} {\abs{c_3}}^2 (c_1 \bar c_3 + \bar c_1 c_3)\\
&- \thalf \tilde{a}_{2222} {\abs{c_2}}^4
- \tilde{a}_{2233} \left( \thalf c_2^2 \bar c_3^2 + 2 {\abs{c_2}}^2 {\abs{c_3}}^2 + \thalf \bar c_2^2 c_3^2 \right)
- \thalf \tilde{a}_{3333} {\abs{c_3}}^4.
\end{split}
\label{Hc}
\end{equation}
where the coefficients are defined by the integrals
$$
\tilde{a}_{jklm}= \int_{-\infty}^{\infty}U_j(x) U_k(x) U_l(x) U_m(x) dx,
$$
which we note are identically zero if $(j+k+l+m)$ is odd. 
A fuller derivation of this system is given in~\cite{Goodman:2011}, along with a careful accounting of the terms ignored in the approximation.
Because the the Hamiltonian satisfies 
$$
\tilde{H}(e^{i \phi_0} \cc) = \tilde{H}(\cc),
$$
Noether's theorem implies the system conserves a discrete version of the power $\cN$ in equation~\eqref{cN},
\begin{equation*}
N = {\abs{c_1}}^2 + {\abs{c_2}}^2 + {\abs{c_3}}^2.
\end{equation*}

For potentials of the form~\eqref{V3}, the coefficients $\tilde{a}_{jklm}$ approach limiting values $a_{jklm}$ as $L\to\infty$, up to exponentially small errors,
\begin{equation*}
\left(
a_{1111},a_{1113},a_{1122},a_{1133},a_{1223},a_{1333},a_{2222},a_{2233},a_{3333}
\right)
=
\left(3,1,2,3,-2,1,4,2,3\right)\cdot A,
\end{equation*}
where 
$$
A =\frac{1}{32} \int_{-\infty}^{\infty} {U_0(x)}^4 dx.
$$
When the potential is given by~\eqref{V0sech}, $A=\tfrac{1}{24}$.
\begin{rem}
Using the approximate values $a_{jklm}$ instead of $\tilde{a}_{jklm}$ simplifies the analysis (and even more the typesetting and reading) of this paper tremendously.
 The effect of this simplification is not completely trivial. The approximation introduces additional symmetry which is reflected in the structure of its solutions, for example in Figure~\ref{fig:all_periodic}.
\end{rem}
Under  these assumptions on the coefficients~$a_{jklm}$, the Hamiltonian~\eqref{Hc} becomes 
\begin{equation}\begin{split}
H(\cc,\ccbar)&= \W_1 {\abs{c_1}}^2 + \W_2 {\abs{c_2}}^2 +\W_3 {\abs{c_3}}^2 - \\
&\phantom{=}A\Big[\tfrac{3}{2} {\left({\abs{c_1}}^2  +{\abs{c_3}}^2\right)}^2
+ \tfrac{3}{2} {\left(c_1 \cbar_3 + \cbar_1 c_3 \right)}^2
+  \left({\abs{c_1}}^2 + {\abs{c_3}}^2\right) \left(c_1 \cbar_3 + \cbar_1 c_3 \right)
+ \\ 
&\phantom{=A\Big[}2 {\abs{c_2}}^4  
+ 4 {\abs{c_2}}^2 \abs{c_3-c_1}^2 
+  {(c_3-c_1)}^2 \cbar_2^2 + {(\cbar_3-\cbar_1)}^2 c_2^2 \Big] 
\label{HcA}
\end{split}\end{equation}

The symmetry of the potential $V(x)$ and of its eigenfunctions is reflected in the Hamiltonian~\eqref{HcA} and the equations of motion. The odd  and even subspaces,
$$
\cBo = \vecspan\left\{\hat{\ee}_2\right\} 
\text{ and } 
\cBe = \vecspan\left\{{\hat{\ee}_1,\hat{\ee}_3}\right\},
$$
where $\hat{\ee}_j$ is the unit vector in coordinate $j$, are both invariant under the flow of $H(\cc,\ccbar)$.

\subsection{Nonlinear Normal Modes}

The normal modes of this system in the linear limit are simply
$$
\cc(t) = e^{-i \Omega_j t}  \hat{\ee}_j.
$$  
These are continued into the nonlinear regime via the linear combinations
\begin{subequations}
\begin{align}
\mop &=
\begin{pmatrix}
0 \\  \sqrt{N}  \\ 0 
\end{pmatrix}
e^{i (\Omega_2- 4 A N)t}, 
\\
\opo& = 
\begin{pmatrix}
 \sqrt{N} \cos{\theta_+} \\ 0 \\  \sqrt{N} \sin{\theta_+}
\end{pmatrix}
e^{i (\Omega_1+\omega_+(N))t}, 
 \quad
\mpm =
\begin{pmatrix}
 \sqrt{N} \sin{\theta_-} \\ 0 \\  \sqrt{N} \cos{\theta_-}
\end{pmatrix}
e^{i (\Omega_3+ \omega_-(N))t},
\label{C1C3}
\end{align}
\end{subequations}
where $\theta_\pm(N)= \pm \tfrac{AN}{2\D}+ O(N^2)$ and $\omega_\pm = -3AN + O(N^2)$.
In addition to the NNM solutions there are ten other solutions (six when symmetries are taken into account) that arise in saddle-node bifurcations as discussed in Section~\ref{sec:allBoundStates}; see~\cite{KapKevChe:06}. 

\subsection{Reduction valid when $c_2\neq 0$}
We make the canonical change of variables from $(c_1,c_2,c_3,i\cbar_1,i\cbar_2,i\cbar_3)$ to $(z_1,\phi,z_3,i\zbar_1,N,i\zbar_3)$ given by 
\begin{equation}
\label{ztoc}
c_1 = z_1 e^{i \phi}; \; 
c_2 = \sqrt{N-\abs{z_1}^2-\abs{z_3}^2}e^{i \phi};  \;
c_3 = z_3 e^{i \phi},
\end{equation}
which is valid when $c_2 \neq 0$. This can be accomplished in steps by, first, converting to canonical polar coordinates $c_j \to \sqrt{\rho_j}e^{i\phi_j}$, then noticing that the Hamiltonian depends on the angles only in the combinations $\phi_1-\phi_2$ and $\phi_3-\phi_2$, using this observation and the conserved quantity $N = {\abs{c_1}}^2 + {\abs{c_2}}^2 + {\abs{c_3}}^2$ to reduce from three degrees of freedom to two, and, finally, converting the system back to complex coordinates.

In the new variables, the Hamiltonian is independent of $\phi$ and contains no square roots because of the quadratic and quartic dependence of Hamiltonian~\eqref{HcA} on $c_2$ and $\cbar_2$:
\begin{equation}
\label{Hreduced}
H(\zz,\zzbar;\ep,N)= H_0(\zz, \zzbar) + \ep H_2(\zz,\zzbar; \ep,N) + H_4(\zz,\zzbar);
\end{equation}
where
\begin{equation}
\begin{split}
H_0  &= -\D \abs{z_1}^2 +  \D \abs{z_3}^2;  \\
\epsilon H_2  &=  \ep\left(\abs{z_1}^2 +   \abs{z_3}^2  \right)
+ A N \left(
4 ( z_1 \zbar_3 +\zbar_1 z_3) -{(z_1-z_3)}^2-{(\zbar_1- \zbar_3)}^2 
\right);\\
H_4  &=  A \left[  \tfrac{1}{2} 
{\left( \abs{z_1}^2 + \abs{z_3}^2 \right)}^2 -
\tfrac{3}{2} {\left(z_1 \zbar_3 +\zbar_1 z_3 \right)}^2 
+   \left( \abs{z_1}^2 + \abs{z_3}^2 \right) 
\left(   {(z_1-z_3)}^2+{(\zbar_1 - \zbar_3)}^2 -5 {(z_1 \zbar_3 + \zbar_1 z_3)}\right)
\right].
\end{split}
\label{H024}
\end{equation}
Fixed points of this system correspond to RFP's of system~\eqref{HcA} and periodic orbits in the reduced system~\eqref{Hreduced} correspond to RPOs in the full system. 

\begin{rem}
Since $c_2$ appears in Hamiltonian~\eqref{HcA} only in even powers, the square root introduced in the change of variables~\eqref{ztoc} disappears from the Hamiltonian~\eqref{Hreduced}. While many of the sources cited in Sections~\ref{sec:priors} and~\ref{sec:DST} perform some sort of reduction, few take advantage of this feature. In order to apply this observation to the DST system~\eqref{DST}, the linear part of the equations must first be diagonalized, which results in a Hamiltonian more closely resembling equation~\eqref{HcA}.
\end{rem}

\subsection{Reduction on the even subspace $\cBe$}

Reduction~\eqref{ztoc} is not valid for representing the two NNM solutions~\eqref{C1C3}, nor for solutions on $\cBe$ more generally, because the angle $\phi$ is ill-defined. The dynamics on $\cBe$ can be simplified  by setting $c_2=0$ and putting $c_1$ and $c_3$ into canonical polar coordinates
$$
c_j = \sqrt{J_j} e^{i \phi_j}
$$
and using the relation~\eqref{W1W2W3}, yielding 
\begin{equation*}
\begin{split}
H &= \Omega_1 J_1 + \Omega_3 J_3 -\frac{3 A}{2} 
\left(  J_1^2 + 2 J_3 J_1+ J_3^2 \right)\\
&\phantom{=} 
-2 A \sqrt{J_1 J_3} \left(J_1+J_3\right) \cos{\left(\phi_1-\phi_3\right)}
-6 A J_1 J_3 \cos^2{\left(\phi_1-\phi_3\right)}
\end{split}
\end{equation*}
The only dependence on the angles comes in terms of the phase difference $\phi_1-\phi_3$, so we make the canonical change of variables
$$
\phi_1= \frac{1}{2} \left(\theta_2-\theta_1\right), \,
\phi_3= \frac{1}{2} \left(\theta_1+\theta_2\right), \,
J_1= \rho_2-\rho_1, \,
J_3= \rho_1+\rho_2.
$$
The angle $\theta_2$ is cyclic, so its conjugate $\rho_2$ is conserved. We note that $\rho_2 = N/2$ and drop the subscripts from $\theta_1$ and $\rho_1$, yielding a Hamiltonian
\begin{equation}
H= 2 \Delta  \rho
-A N  \sqrt{N^2-4 \rho^2}\cos{\theta}
-\frac{3 A}{2}  (N^2-4 \rho^2 )\cos^2{\theta}
-\frac{3 A N^2}{2}+\Delta  N +N \Omega_1;
\label{HEvenPolar}
\end{equation}
and evolution equations
\begin{subequations}
\begin{align}
\dot{\theta} & =
2 \Delta 
+\frac{4 A N \rho  \cos{\theta}}{\sqrt{N^2-4 \rho ^2}}
+12 A \rho  \cos^2{\theta};
\label{thetadot} \\
\dot{\rho} & = - A \left(3 (N^2-4\rho^2) \cos{\theta}+N \sqrt{N^2-4 \rho^2}\right)\sin{\theta}. 
\label{rhodot}
\end{align}
\label{thetarhoEven}
\end{subequations}

\section{Enumerating the nonlinear bound states}
\label{sec:allBoundStates}

The nonlinear bound states become fixed points in the reduced systems corresponding to~\eqref{Hreduced} and~\eqref{HEvenPolar}. We describe the fixed points of the two reductions separately.

\subsection{Fixed points of the reduced Hamiltonian~\eqref{Hreduced}}
\label{sec:asymFix}
\begin{figure}[htbp] 
   \centering
   \includegraphics[width=.3\textwidth]{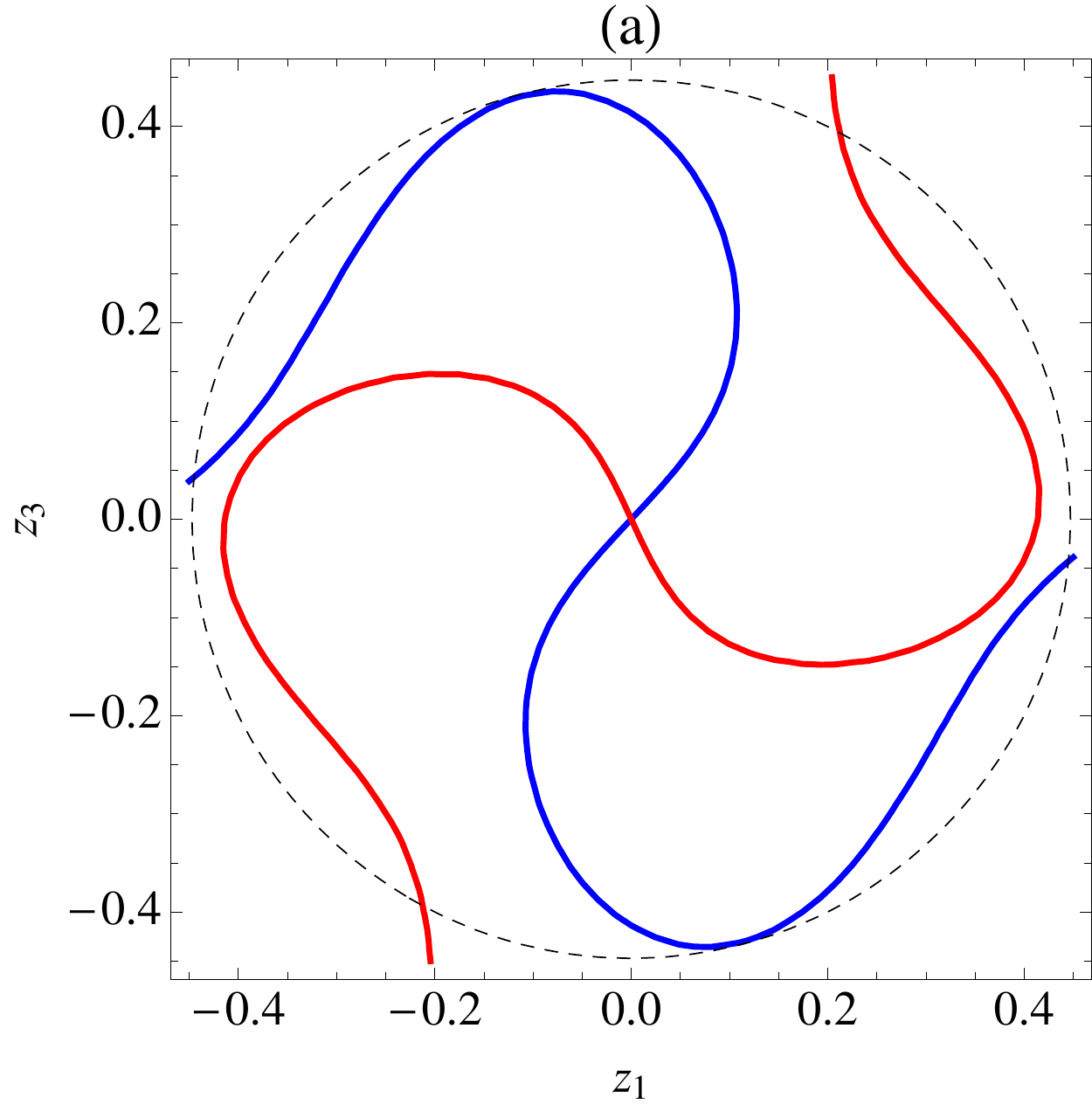}
   \includegraphics[width=.3\textwidth]{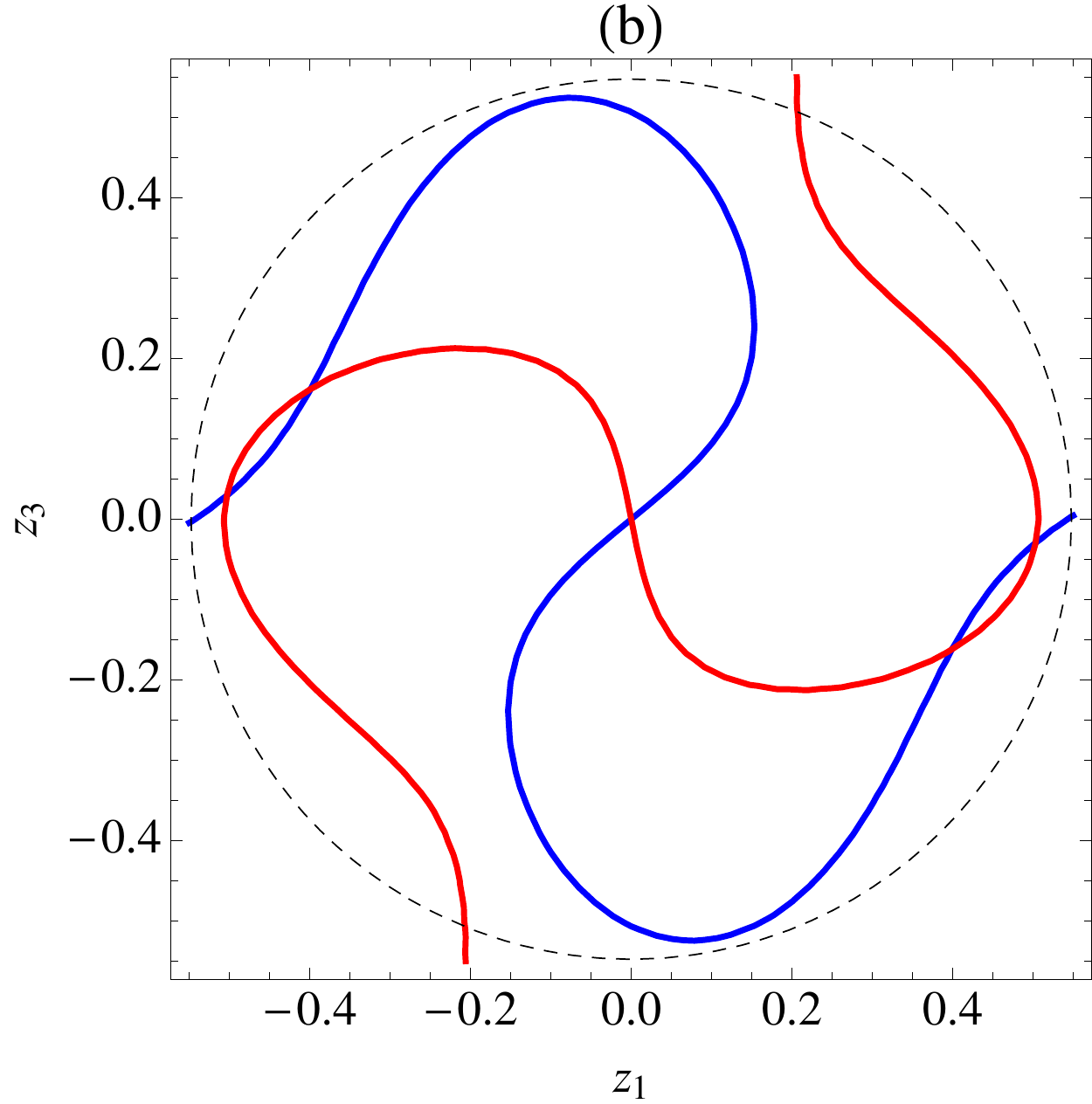}
   \includegraphics[width=.3\textwidth]{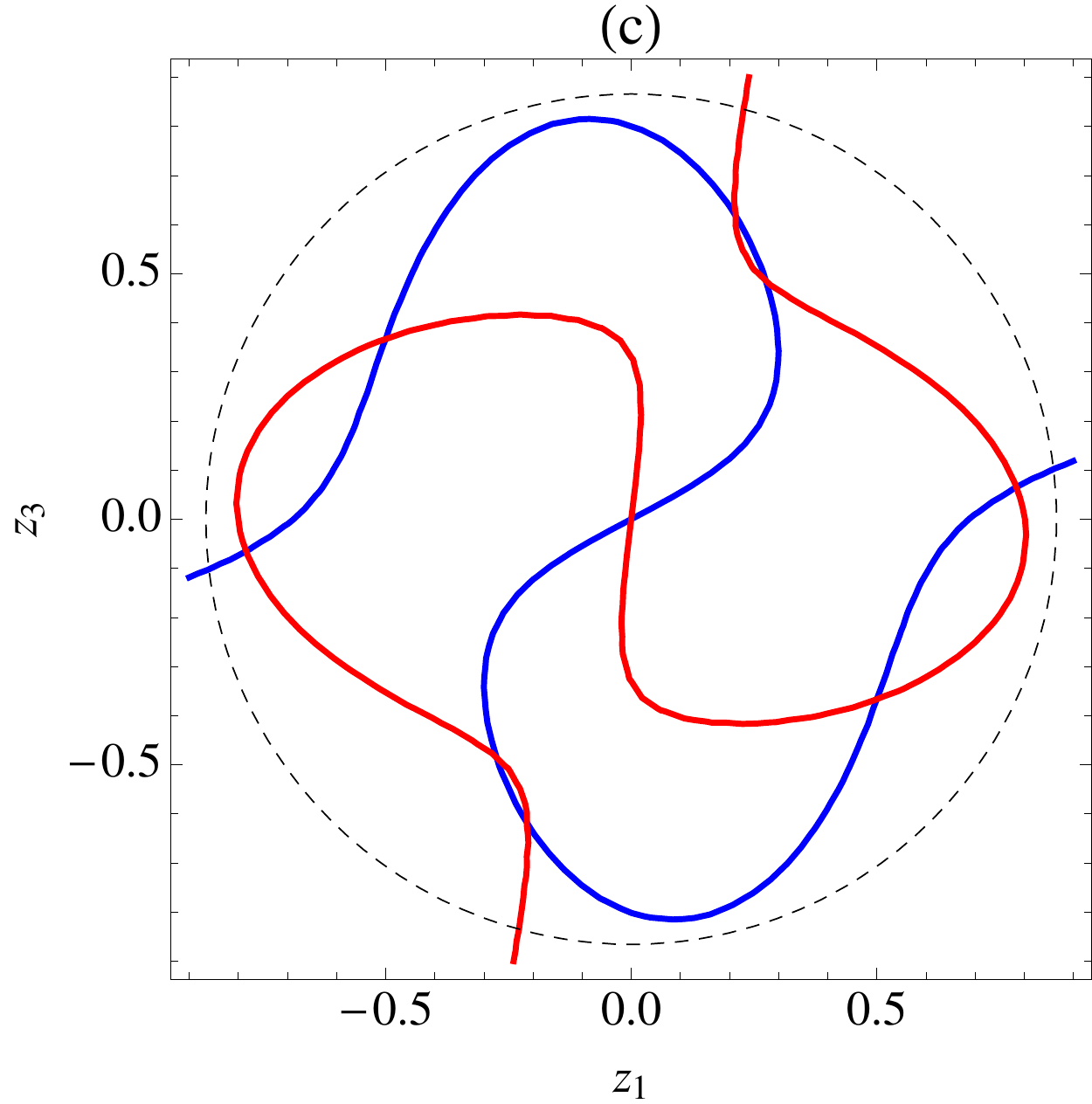}
   \caption{The red and blue curves are, respectively, the zero level-sets of the two equations in system~\eqref{z1z3}, so intersections solve the system. (a) $N=0.2<N_{\rm{a}1}$, (b) $N_{\rm{a}1}<N=0.3<N_{\rm{a}2}$, (c) $N=0.75> N_{\rm{a}2}$.  Since all intersections are inside the circles $z_1^2+z_3^2=N$ (dashed), they correspond to physical solutions. Since all intersections lie near the circle, the PDE solution is dominated by the even eigenfunctions, at least for small $N$. The origin in this figure corresponds to the solution $\mop$ and the boundary circle to the invariant subspace $\cBe$.}
\label{fig:z1z3}
\end{figure}

Since the leading order terms of $H(\zz,\zzbar)$ are quadratic, 
\begin{equation}
\zz=0
\label{z0}
\end{equation}
 is a fixed point. This corresponds to the NNM $\mop$ of system~\eqref{HcA}.  The other fixed points are real, $z_j = \zbar_j \equiv \zeta_j$, and satisfy
\begin{equation}
\label{z1z3}
\begin{split}
(-2 A N-\D+\ep )\zeta_1 +6 A N \zeta_3+5 A \zeta_1^3-21 A  \zeta_1^2 \zeta_3-A  \zeta_1 \zeta_3^2-7 A \zeta_3^3 &= 0 ;\\
6 A N \zeta_1 +(-2 A N+\D+\ep )\zeta_3-7 A \zeta_1^3-A  \zeta_1^2 \zeta_3-21 A \zeta_1 \zeta_3^2 +5 A \zeta_3^3 &=0.
\end{split}
\end{equation}
In order to count its roots, we use Mathematica to eliminate $\zeta_3$, which yields an odd parity ninth-degree polynomial for $\zeta_1$ alone, 
\begin{equation}
b_9 \zeta_1^9 + b_7 \zeta_1^7 + b_5 \zeta_1^5 + b_3 \zeta_1^3 + b_1 \zeta_1 = 0,
\label{z1alone}
\end{equation}
with coefficients depending on the parameters $(A,\D,\ep)$ and on the conserved quantity $N$. This can be factored as $\zeta_1$ times a quartic polynomial in $Z=\zeta_1^2$. We again use Mathematica to compute the discriminant of the $Z$ polynomial. We find numerically the values of $N$ for which the discriminant vanishes, which indicates where bifurcations occur in the solutions of equation~\eqref{z1alone}. For the default values of $(A,\D,\ep)$, bifurcations occur at $N_{\rm{a1}} \approx 0.246$ (creating the RFPs  $\ppo$ and~$\poo$ described in Remark~\ref{rem:naming}) and $N_{\rm{a2}}  \approx 0.667$ (the RFPs $\mpp$ and~$\mpo$).  Figure~\ref{fig:z1z3} shows numerically calculated fixed points $(\zeta_1,\zeta_3)$ of equation~\eqref{z1z3}, at three values of $N$ confirming the existence of two saddle-node bifurcations. Note that $(-\zeta_1,-\zeta_3)$ is the same relative fixed point as $(\zeta_1,\zeta_3)$.

\subsection{Fixed points on $\cBe$}
\label{sec:evenFixed}

Setting the right side of the $\dot\rho$ equation~\eqref{rhodot} to zero implies that either $\sin\theta=0$ or the term in parentheses vanishes. The latter is incompatible with $\dot\theta=0$. Therefore $\theta=0$ or $\theta=\pi$. Inserting $\cos\theta=\pm 1$ into equation~\eqref{thetadot} gives, after some algebra,
\begin{equation}
144 A^2 \rho^4
+48 A \Delta  \rho^3
+4 \left(\Delta^2-8 A^2 N^2\right) \rho^2 
-12 A \Delta  N^2 \rho 
-\Delta^2 N^2
= 0.
\label{quartic}
\end{equation}

\begin{figure}[htbp] 
   \centering
   \includegraphics[width=0.32\textwidth]{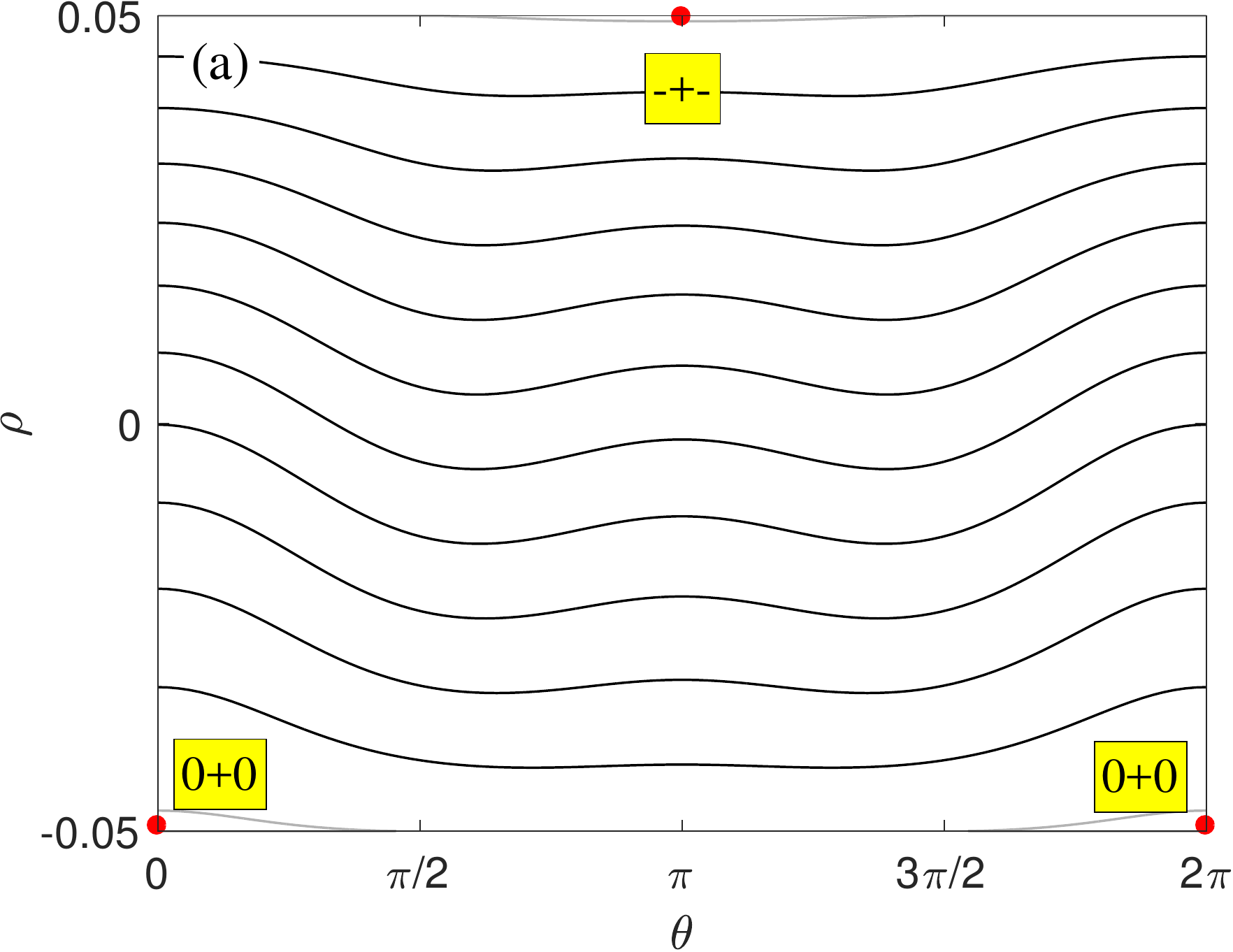} 
   \includegraphics[width=0.32\textwidth]{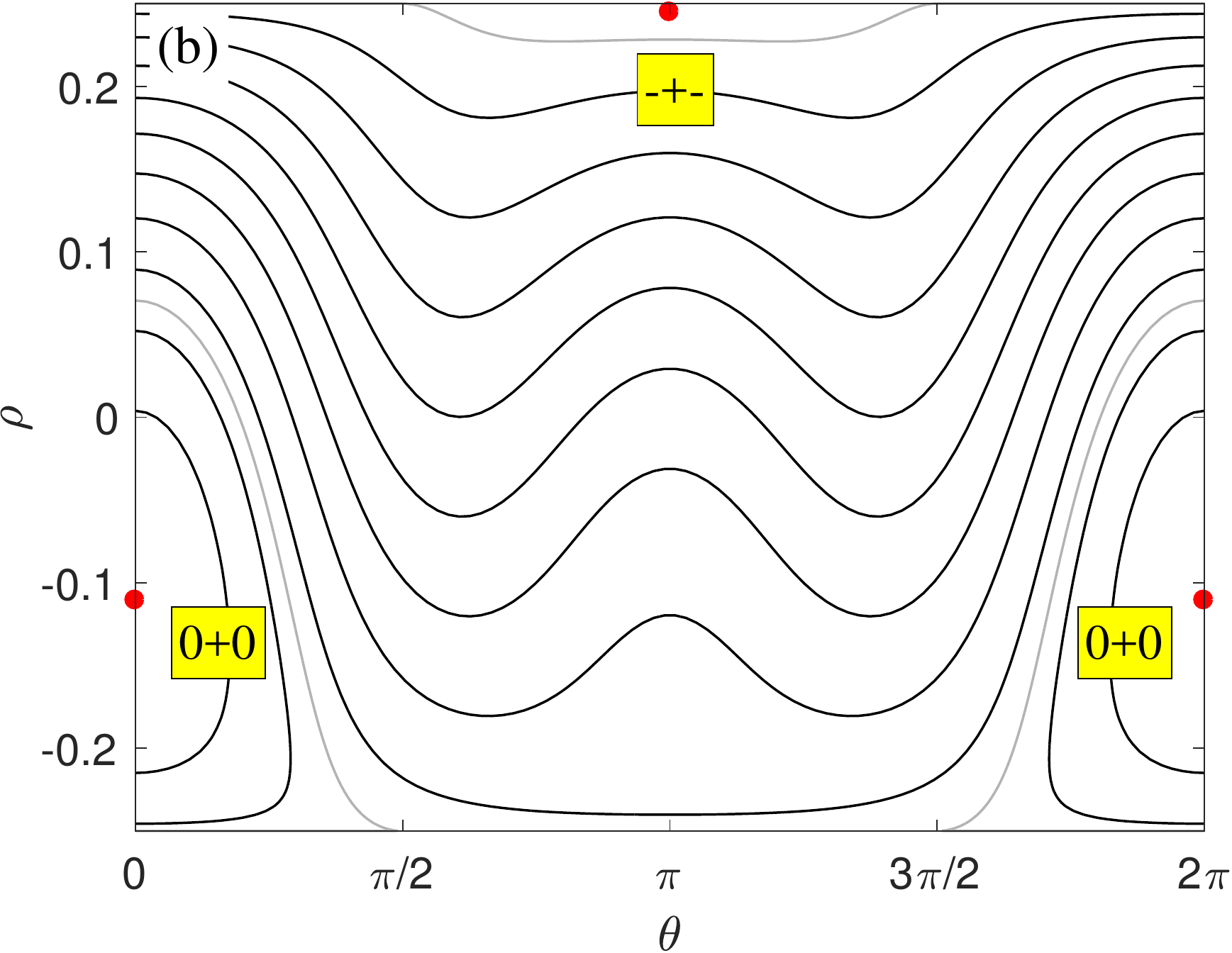} 
   \includegraphics[width=0.32\textwidth]{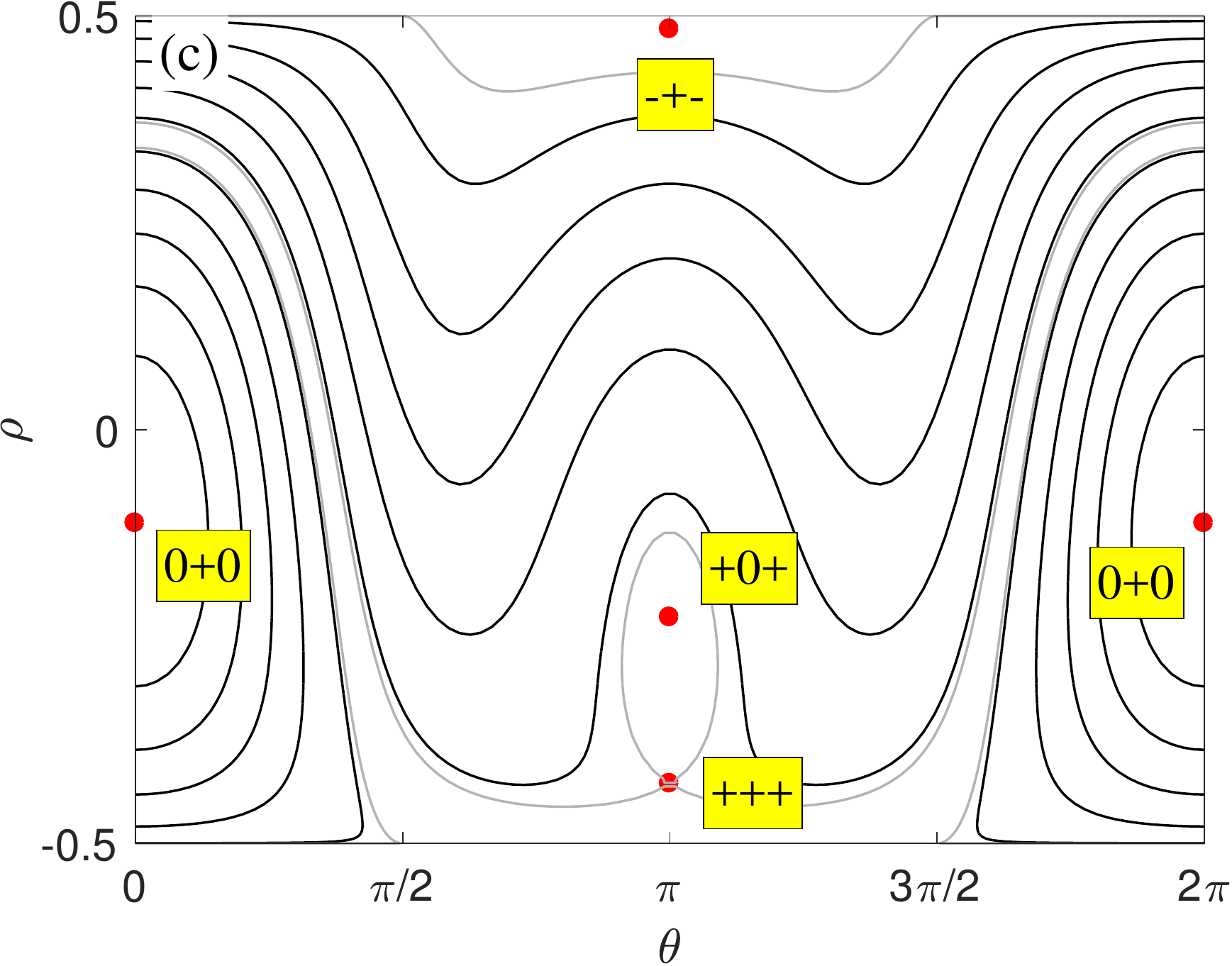} 
   \caption{The phase plane of the even subsystem~\eqref{thetarhoEven} with (a) $N=0.1$, (b) $N = 0.5$ (subcritical), (c) $N=1$ (supercritical). Solutions near the bottom edge are close to the nonlinear ground state $\opo$, while those near the top edge are close to second nonlinear excited state $\mpm$.}
\label{fig:evenSubspace}
\end{figure}

At small values of $N$, this system has two fixed points, one with $\theta = 0$ corresponding to the NNM $\opo$, and the other with $\theta=\pi$, corresponding to $\mpm$. Both are stable to perturbations within~$\cBe$.
A saddle-node bifurcation occurs when the discriminant of equation~\eqref{quartic}, considered as a function of $\rho$, vanishes, which occurs for exactly one real, positive value of $N$,
\begin{equation}
\NSNe = \frac{\sqrt{2 \left(145+99 \cdot 3^{1/3}+57 \cdot 3^{2/3}\right)} \D }{ 32 A} \approx 0.8907 \frac{\D}{A} \approx 0.8125.
\label{SNeven}
\end{equation}
For $N>\NSNe$, there are two additional solutions with $\theta=\pi$. Numerical examples are shown in Figure~\ref{fig:evenSubspace} for $N = 0.1 < \NSNe$, for $N=0.5< \NSNe$, and for $N=1 > \NSNe$.

For $N \ll 1$ the two solutions remain very close to the upper and lower boundaries of the phase space, which, are, in the $(c_1,c_3)$ coordinates
\begin{equation*}
\opo = \sqrt{N} \binom{1 - O(N^2)}{O(N)}
\text{ and }
\mpm = \sqrt{N} \binom{O(N)}{1 - O(N^2)}.
\end{equation*}
However, for large $N$, the behaviors of the two NNM solutions are quite different. The $\rho$ coordinate of solution $\opo$ approaches $\rho=0$, meaning that $\rho_1-\rho_3\to0$. Since $\theta=0$, the two solutions are in phase. Using this in ansatz~\eqref{ansatz} to construct the approximate PDE solution implies that the modes $U_1$ and $U_3$ interfere destructively near $z=\pm L$ but constructively near $z=0$, so that for large $N$, the solution concentrates in the middle well. 

By contrast, the solution $\mpm$ stays near the upper boundary of the region $\rho \approx N/2$, so that the the shape of the PDE standing wave is not greatly altered for large $N$, so this mode is labeled $\mpm$; again, see Remark~\ref{rem:naming}. Similarly for the two modes that appear in the saddle-node bifurcation, the one labeled $\ppp$ stays near the top edge, with a profile that resembles $U_1(x)$, while the one labeled $\pop$ has destructive interference near $z=0$ and constructive interference near $z=\pm L$.

\begin{figure}[htbp] 
   \centering
   \includegraphics[width=3in]{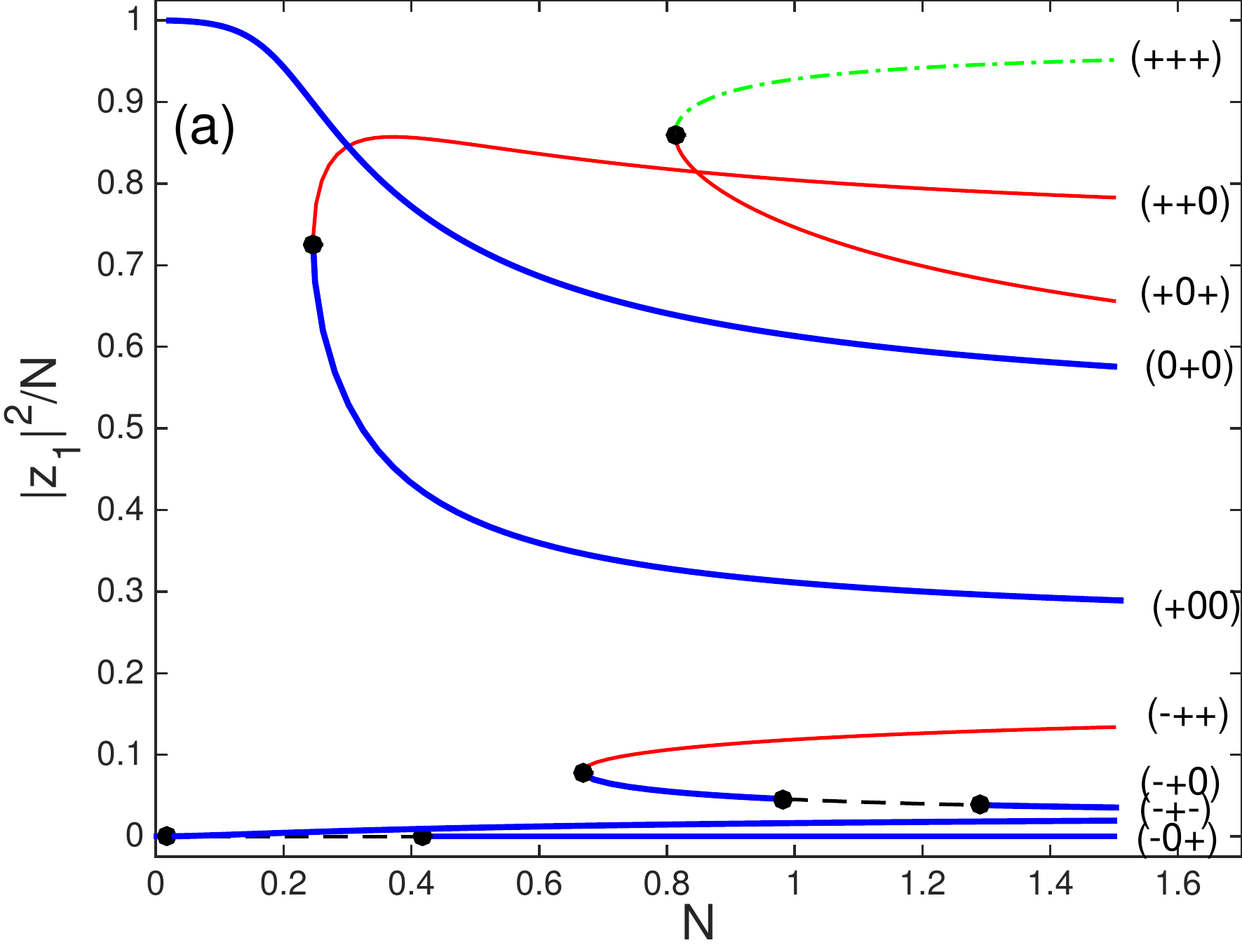} 
   \includegraphics[width=3in]{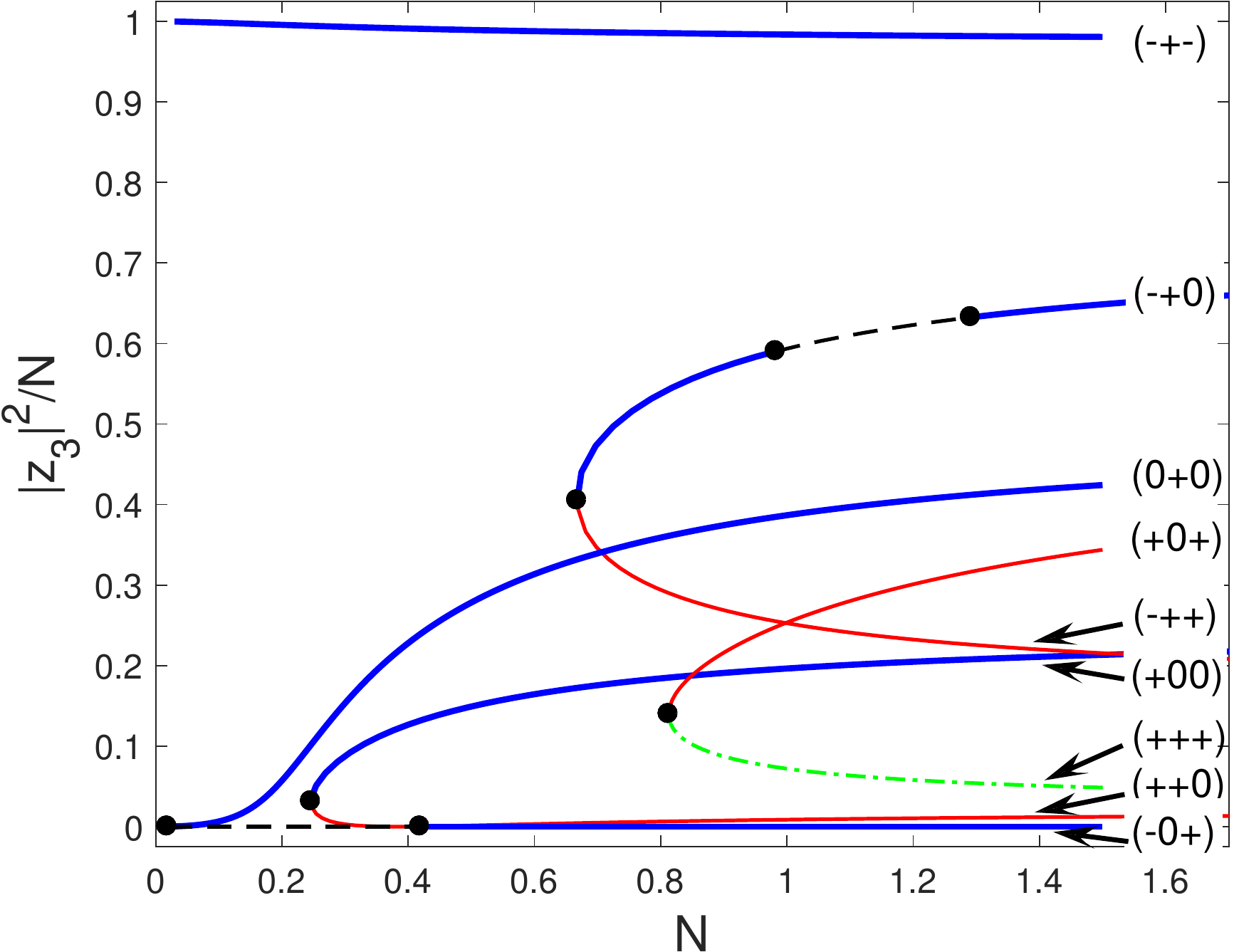} 
   \caption{(Color online) The nine branches of nonlinear bound states, showing (a) the fraction of $N$ given by $\abs{z_1}^2$ and (b) the fraction of $N$ given by $\abs{z_3}^2$. The stability of each branch is denoted by line style. Thick blue: four imaginary eigenvalues. Thin red: two real, two imaginary. Dashed black: Krein quartet. Dash-dot green: four real eigenvalues.}
\label{fig:9Modes}
\end{figure}

Thus far we have described nine nonlinear bound states: three NNM's and six more that come into existence in three saddle node bifurcations. This is summarized in Figure~\ref{fig:9Modes}.

\section{Linear stability of the fixed points}
\label{sec:linear}
\subsection{Stability of the solution $\opo$}
The stability of the $\opo$ solution, the trivial solution~\eqref{z0} in the reduced dynamics, can be found by examining the quadratic part of the reduced Hamiltonian. Separating the ODE 
$$ i \dot{z}_j = \frac{\partial H}{\partial \zbar_j}$$ 
into real and imaginary parts $z_j = x_j + i y_j$ gives linearized evolution equations
$
\frac{d}{dt} \vv = M_0 \vv
$
where 
\begin{equation}
\label{vvM0}
\vv =  \begin{pmatrix} x_1 \\ x_3 \\ y_1 \\ y_3 \end{pmatrix} 
\text{ and } 
M_0 = \begin{pmatrix} 
0 & 0 & n -\D + \ep & n \\
0 & 0 & n & n+\D+ \ep \\
n+\D-\ep & -3 n & 0 & 0 \\
-3n & n-\D-\ep & 0 & 0
\end{pmatrix},
\end{equation}
with $n = 2 A N$.
The trivial solution may change stability when it has multiplicity-two eigenvalues on the imaginary axis, in which case the characteristic polynomial
$$
P(\lambda;n,\ep,\D) = \lambda^4
+\left(4 n^2+2 \D^2+2 \ep^2\right) \lambda^2 
+\left(8 n^2 \D^2-12 n^2 \ep^2+\D^4-2 \D^2 \ep^2+\ep^4 -16 n^3 \ep \right)
 $$
   has double roots. Because $P$ contains only even powers of $\lambda$, this is equivalent to the condition that $P(\sqrt{q};n,\ep,\D)$ have double roots, which is that its discriminant vanishes. The discriminant of this polynomial is
\begin{equation}
   D(n;\D,\ep) = n^4+4  \ep  n^3+ \left(4 \ep^2-\D^2\right)n^2+\D^2 \ep^2.
\label{discrim}
\end{equation}
   Since $\D\gg\ep$, when $n=O(\ep)$, the two leading terms are $\D^2(\ep^2-n^2)$, and the leading order solution is $n=\ep +O(\ep^2)$, i.e.
\begin{equation}
   \NHHa = \frac{n}{2A} = \frac{\ep}{2A} + O(\ep^2) 
\label{Ncrit}
\end{equation}
which is equivalent to the condition found in~\cite{Goodman:2011}. Using $\D\gg\ep$, when $n= O(\D)$ (large) one finds the largest positive root of the discriminant is 
 \begin{equation*}
   \NHHb = \frac{\D-2 \ep}{2A} + O(\ep^2),
 \end{equation*}
so that this solution is unstable if $\NHHa \lessapprox N \lessapprox \NHHb$,  as is shown numerically in~\cite{KapKevChe:06} and~\cite{Goodman:2011}. Directly solving equation~\eqref{discrim} using Remark~\ref{numeric_values}, we find
\begin{equation*}
\NHHa \approx  0.0185 \text{ and } \NHHb \approx 0.419,
\end{equation*}
which is within about one percent of the asymptotic values. By comparison, we may calculate the standing waves by directly solving system~\eqref{stationary} numerically and then numerically calculate the spectrum the NLS/GP equations linearized around these stationary solutions.  We find that the bifurcation values are 
\begin{equation*}
\cNHHa \approx 0.0182 \text{ and } \cNHHb\approx 0.436,
\end{equation*}
in reasonable agreement with the values above. This branch is the bottom curve of both branches of Figure~\ref{fig:9Modes}. HH bifurcations are visible at the two computed values.

\subsection{Stability of other fixed points of the reduced Hamiltonian~\eqref{Hreduced}}

To find the stability of the other fixed points $\zz^*=(\zeta_1,\zeta_3)$, we linearize, letting $z_j = \zeta_j+ (x_j+i y_j)$ and $\zbar_j = \zeta_j + (x_j- i y_j)$. The linearized equations are 
$\tfrac{d}{dt}\vv = (M_0 + A M_1(\zz^*)) \vv$
where $M_0$ and $\vv$ are given in equation~\eqref{vvM0} and 
$$
M_1 = \begin{pmatrix}
 0 & 0 & \zeta_1^2-14 \zeta_1 \zeta_3+\zeta_3^2 & -3 \zeta_1^2-6 \zeta_1 \zeta_3-3 \zeta_3^2 \\
 0 & 0 & -3 \zeta_1^2-6 \zeta_1 \zeta_3-3 \zeta_3^2 & \zeta_1^2-14 \zeta_1 \zeta_3+\zeta_3^2 \\
 -15 \zeta_1^2+42 \zeta_1 \zeta_3+\zeta_3^2 & 21 \zeta_1^2+2 \zeta_1 \zeta_3+21 \zeta_3^2 & 0 & 0 \\
 21 \zeta_1^2+2 \zeta_1 \zeta_3+21 \zeta_3^2 & \zeta_1^2+42 \zeta_1 \zeta_3-15 \zeta_3^2 & 0 & 0
\end{pmatrix}.
$$
The stability is calculated numerically and shown in Figure~\ref{fig:9Modes}. The branches $\ppo$ and $\mpp$ always have two real and two purely imaginary eigenvalues. When created in saddle node bifurcations, the branches $\mpo$ and $\poo$ both have four imaginary eigenvalues and are stable. While branch $\poo$ is stable for all $N$, the branch $\mpo$ loses and regains stability in a pair of HH bifurcations.

\subsection{Stability of solutions on $\cBe$}

The reduced system~\eqref{Hreduced} is ill-defined on $\cBe$, and system~\eqref{HEvenPolar} describes only the motion within $\cBe$ and cannot describe the stability of perturbations in directions complementary to $\cBe$. To determine the stability of the solutions on $\cBe$ requires a reduction similar to equation~\eqref{ztoc}, except one that eliminates $z_1$ or $z_3$ instead of $z_2$. The matrix describing the linearization about the solutions on $\cBe$ is reducible to two smaller systems: one describing the motion of perturbations within $\cBe$ and one for perturbations orthogonal to $\cBe$. 
It is clear from Figure~\ref{fig:evenSubspace} that the eigenvalues corresponding to in-plane motion for modes~$\mpm$, $\opo$, and~$\pop$ are imaginary while those for $\ppp$ are real. The remaining two eigenvalues, corresponding to perturbations out of the plane $\cBe$, come from the linearized equations of motion for $z_2 = r + i s$ in a neighborhood of the fixed points,
$$
\begin{pmatrix} \dot{r} \\ \dot{s} \end{pmatrix} =
\begin{pmatrix} 0 &
 -\D-\ep +A (N+6 \rho)+ \frac{A  (7 N-6 \rho ) \sqrt{\rho } \cos{\theta}}{\sqrt{N-\rho }}
\\
\D+\ep+ A(3 N-6 \rho) -\frac{A (15 N-14 \rho )  \sqrt{\rho } \cos{\theta}}{\sqrt{N-\rho }}
& 0 \end{pmatrix} 
\begin{pmatrix} r \\ s \end{pmatrix}.
$$
If the product of the off-diagonal terms is positive, then the associated fixed point has real eigenvalues, and if it is negative, these eigenvalues are imaginary. This is found to be negative for~$\opo$ and~$\mpm$ and positive for~$\pop$ and~$\ppp$. See~\cite{KapKevChe:06} for a more thorough discussion.

\section{Normal form for $N = O(\ep)$}
\label{sec:ChowKim}
At the HH bifurcation for $N=\NHHa$, the number of Lyapunov families of periodic orbits attached to the RFP $\mop$ drops from two to zero, and we would like to know their fate. Then, at the second HH bifurcation at $\NHHb=O(\D)$, the number of Lyapunov families jumps back up to two. The interval between these two bifurcations is marked by the dashed black curve segment in Figure~\ref{fig:9Modes}. Later we will address the question of how and whether the branches that exist for $N>\NHHb$ relate to those that exist for $N<\NHHa$ and will discover via numerical experiments that they are entirely separate.

We turn first to the HH bifurcation at $N=\NHHa$. We apply a result due to
Chow and Kim~\cite{Chow:1988} which allows us to find the periodic orbits of Hamiltonian, and which requires knowledge only of the truncated normal form of equation~\eqref{Hreduced}
\begin{equation}
\Htrunc(\zz,\zzbar; \ep) = 
H_0(\zz,\zzbar) + 
H_2^{\rm N} (\zz,\zzbar;\ep,N) + 
H_4^{\rm N}(\zz,\zzbar).
\label{Htrunc}
\end{equation}
We proceed in three steps. First, we state the the theorem to be applied. Second, we calculate the terms in the truncated normal form~\eqref{Htrunc}. Third, we we apply the theorem to the normal form, thereby enumerating the periodic orbits. This leads to further insights into the bifurcation that occurs at $N = \NHHa$ from equation~\eqref{Ncrit}, which we explore via further normal-form analysis.

Chow and Kim point out that, according the Moser-Weinstein reduction theorem~\cite{Moser:1976},  there exists for sufficiently small $\ep$, a $C^{\infty}$ normal form $E(\zz,\zzbar;\ep)$ that depends smoothly on $\ep$ for $0<\abs{z}\ll1$ and coincides with $H^{\rm N}(\zz,\zzbar;\ep)$ up to any finite order in~$\ep$. Hence, they conclude, it is sufficient to look for periodic orbits in the truncated normal form system~$\Htrunc$.

\begin{theorem}
\label{thm:CK}
Consider a Hamiltonian of the form~\eqref{Hreduced} whose leading order term $H_0$ is in semisimple -1:1 resonance and which has truncated normal form $\Htrunc$ in equation~\eqref{Htrunc}. Let $\zz = (z_1,z_3,\zbar_1,\zbar_3)$ and define the matrices $L$ and  $B$ by
\begin{equation*}
H_0(\zz,\zzbar) = \innerProd{\zz}{L \zz} 
\text{ and } 
H_2^{\rm N}(\zz,\zzbar) = \innerProd{\zz}{B \zz} 
\end{equation*}
Let $\zz^*$ be a critical point of $\Htrunc$, subject to the constraint $H_0=h$, i.e.\ assume that 
\begin{equation}
H_0(\zz^*) = h
\text{  and  }
\nabla \left.\left(\Htrunc - \eta H_0 \right)\right\rvert_{\zz = \zz^*}=0,
\label{constrainedextremum}
\end{equation}
then system~\eqref{Hreduced} has a $2\pi/\D\eta$ periodic orbit on the level set $H_0=h$
\begin{equation*}
\zz(t) = e^{\eta J L t} \zz^*, 
\end{equation*}
where the Lagrange multiplier $\eta$ is given by
\begin{equation}
\label{eta}
\eta = 1+
 \left.\frac{\left\langle L \zz, \ep B\zz +\nabla H_4^{\rm N} \right\rangle}
{\abs{{\zz}}^2}\right\rvert_{\zz=\zz^*}
\end{equation}
for sufficiently small values of $\ep$, and $\abs{\zz^*}$.
\end{theorem}
In~\cite{Goodman:2011}, we carry out this reduction using von Zeipel averaging, and the periodic orbits are given as fixed points of the corresponding averaged equations.

\subsection{Calculating the truncated normal form}
\label{sec:truncatedForm}

The leading order Hamiltonian~$H_0$ given by equation~\eqref{H024} is in a suitable form to apply Theorem~\ref{thm:CK}, so we can skip the first step described in Section~\ref{sec:normalforms} and move right on to construct the adjoint operator $\ad{H_0}$. In the $(\zz,\zzbar)$ coordinates, the Poisson bracket is 
$$\PB{F}{G} = 
i \sum_{j\in\{1,3\}} 
\left(
F_{\zbar_j} G_{z_j} - F_{z_j}G_{\zbar_j}
\right).
$$
so the adjoint operator of $H_0$ is 
$$
\ad{H_0}{G}= i \D \left( G_{z_1} - G_{z_3} - G_{\zbar_1} + G_{\zbar_3} \right).
$$
When applied to a monomial $\zz^\aalpha \zzbar^\bbeta \in \cP_j$,
$$
\ad{H_0} { \zz^{\aalpha} \zzbar^{\bbeta} }= 
i \D \left(\alpha_1-  \alpha_3-\beta_1 +\beta_3\right)\cdot \zz^{\aalpha} \zzbar^{\bbeta}.
$$
The operator $\ad{H_0}^{(j)}$ acts diagonally on the basis of monomials, and so is symmetric and
$$
\ker{\ad{H_0^T}^{(j)}} = \ker{\ad{H_0}^{(j)}} = \left\{  \left.
\zz^\aalpha \zzbar^\bbeta \in \cP_j \right\rvert
\alpha_1-  \alpha_3-\beta_1 +\beta_3 = 0
 \right\}.
$$
Thus the leading order normal form is simply the projection of equation~\eqref{Hreduced} onto the span of the monomials whose exponents are positive integer solutions of
\begin{equation*}
\begin{pmatrix} 1 & 1 & 1 & 1 \\ -1 & 1 & 1 &-1 \end{pmatrix}
\begin{pmatrix} \alpha_1 \\ \alpha_3 \\ \beta_1 \\ \beta_3 \end{pmatrix} = 
\binom{j}{0},
\end{equation*}
with general solution
$$
\alpha_1 + \beta_3 = \frac{j}{2} \text{ and } \alpha_3 + \beta_1 = \frac{j}{2}.
$$
This has positive integer-valued solutions only for $j\in2\ZZ^+$.  The resonant monomials can be enumerated by specifying $\alpha_1$ and $\alpha_3$ to be integers drawn from the set $\{0,\ldots,j/2\}$.  The resonant quadratic and quartic monomials are tabulated in Table~\ref{table:resonant}. 
\begin{table}[h]
\centering
\subfloat[Degree Two]{%
\begin{tabular}{|c||c|c|}
\hline
$\vspace{0.2cm} \alpha_1\Big\backslash \vspace{-0.2cm} \alpha_3$& 0 & 1\\ 
\hline
0 & $\zbar_1 \zbar_3$ & $\abs{z_3}^2$ \\
1 & $\abs{z_1}^2$ & $z_1 z_3$ \\
\hline
\end{tabular}
}
\subfloat[Degree Four]{%
\begin{tabular}{|c||c|c|c|}
\hline
$\vspace{0.2cm} \alpha_1\Big\backslash \vspace{-0.2cm} \alpha_3$& 0 & 1 &2 \\ \hline
0 & $\zbar_1^2 \zbar_3^2 $ & $\abs{z_3}^2 \zbar_1 \zbar_3$ & $\abs{z_3}^4$ \\
1 & $\abs{z_1}^2 \zbar_1 \zbar_3$ & $\abs{z_1}^2 \abs{z_3}^2$ & $\abs{z_3}^2 z_1 z_3$ \\
2 & $\abs{z_1}^4$ & $\abs{z_1}^2 z_1 z_3$ & $z_1^2 z_3^2$ \\
\hline
\end{tabular}
}
\caption{Resonant monomials of degree two and four}
\label{table:resonant}
\end{table}

Thus, we can  read off the resonant terms that make up the normal form of the Hamiltonian~\eqref{Hreduced}
\begin{subequations}
\begin{align}
H_2^{\rm N}(\zz,\zzbar;\ep,N) &= 
\ep \left(\abs{z_1}^2 +  \abs{z_3}^2\right)  
+ 2A N(z_1 z_3 + \zbar_1 \zbar_3); \\
H_4^{\rm N}(\zz,\zzbar) & = 
 A \left[ 
  \tfrac{1}{2} \abs{z_1}^4 - 2 \abs{z_1}^2 \abs{z_3}^2 
  + \tfrac{1}{2} \abs{z_3}^4  
-2  \left( \abs{z_1}^2 + \abs{z_3}^2 \right)  
     (z_1 z_3 + \zbar_1 \zbar_3) 
\right].
\end{align}
\label{H2NH4N}
\end{subequations}

\subsection{Finding the periodic orbits}
We find it convenient to eliminate the Lagrange multiplier from system~\eqref{constrainedextremum}, and to present the results in canonical polar coordinates $z_j = \sqrt{J_j}e^{i \theta_j}$.

 The constrained critical points satisfy the equations 
\begin{subequations}
\label{ChowKim}
\begin{align}
 \sqrt{J_1 J_3} \left(2\ep- A \left(J_1+ J_3\right)\right) + 2 A  \left(N \left( J_1+  J_3\right)-J_1^2-6 J_1 J_3-J_3^2\right)\cos{\Theta} &=0\label{CK1};\\
\sqrt{J_1 J_3}  \left(N -J_1-J_3\right) \sin{\Theta} &=0,\label{CK2}
 \end{align}
 \end{subequations}
where $\Theta=\theta_1+\theta_3$. 
The physically meaningful solutions lie in a solid torus, with triangular cross-section defined by the simplex
\begin{equation}
\label{triangle}
0 < J_1 ;\; 0 < J_3;\; J_1+J_3 < N;\; \Th \in \SS^1.
\end{equation}
On the boundaries of the triangular cross-section  one or both of the angles $\theta_1$, $\theta_3$ is undefined, as is~$\Theta$.  Therefore some solutions of system~\eqref{ChowKim} do not correspond to periodic orbits of system~\eqref{Hreduced}.

Since equation~\eqref{CK2} is factored, it is convenient to solve this first, and to use its factors to classify the periodic orbits. The equations are equivariant to the transformation that interchanges $J_1$ and $J_3$, so the diagram containing the solution curves of this system is symmetric across the plane $J_1 = J_3$. The factor $(N-J_1-J_3)$ lies on the boundary of the open domain of definition~\eqref{triangle}, and solutions in which this term is zero do not correspond to periodic orbits of the system~\eqref{Hc}.

\subsubsection{Trivial solutions: continuations of linear bound states}
\label{sec:trivial}

We find that setting $J_1=0$ or $J_3=0$ in system~\eqref{ChowKim} gives rise to three solutions
\begin{equation}
\textbf{(a) } (J_1,J_3)=(0,0),
\textbf{(b) } (J_1,J_3)=(N,0),
\text{ and {\bf(c)} }(J_1,J_3)=(0,N)
\label{easy_ones}
\end{equation}
for any value of $\Th$. We note that $\Th$ is ill-defined for these solutions, but that these solutions are exactly the NNMs. Solution~(a) corresponds to $\mop$ using~\eqref{z0}, while solutions (b) and (c) correspond to the NNM solutions $\opo$ and $\mpm$ of equation~\eqref{C1C3}, with which they agree to the order of the truncation. 

\subsubsection{Solutions with $J_1+J_3 <N$}
\label{sec:solutionssin0}
Two continuous branches of solutions are defined by setting $\sin{\Th}=0$ in equation~\eqref{CK2}, so that $\Th = n \pi$, making $\cos{\Th}=\pm 1$ in~\eqref{CK1}.  There are therefore two families of solutions, $\cS_0$ with $\Th = 0 \mod 2\pi$ and $\cS_\pi$ on which $\Th = \pi \mod 2\pi$. It is useful to look at the intersections of these two branches with the symmetry axis $J_1 = J_3 = J$, $0 < J < \tfrac{N}{2}$. On this axis, equation~\eqref{CK1} becomes 
\begin{equation}
\label{bigeasy}
J (\ep - A J + A(2N-8J)\cos{\Theta} ) = 0
\end{equation}

On the family $\cS_0$, this has solution 
$$
J = \frac{\ep + 2AN}{9A}
$$
which is in the interval $(0,\tfrac{N}{2})$ when 
\begin{equation*}
N>\Neven =\tfrac{2\ep}{5A}.
\end{equation*}

On the family  $\cS_\pi$, equation~\eqref{bigeasy} has solution $J=0$ as well as 
$$
J = \frac{\ep-2AN}{7A}
$$
which is positive when $N>\ep/2A$, which we recognize as the right-hand side of $\NHHa$ given by the leading-order term of condition~\eqref{Ncrit} for the instability of the trivial solution. The birth of this solution would appear to be a consequence of the Hamiltonian Hopf bifurcation which is discussed further in Section~\ref{sec:HH}.

\begin{figure}
\includegraphics[height=.3\textwidth]{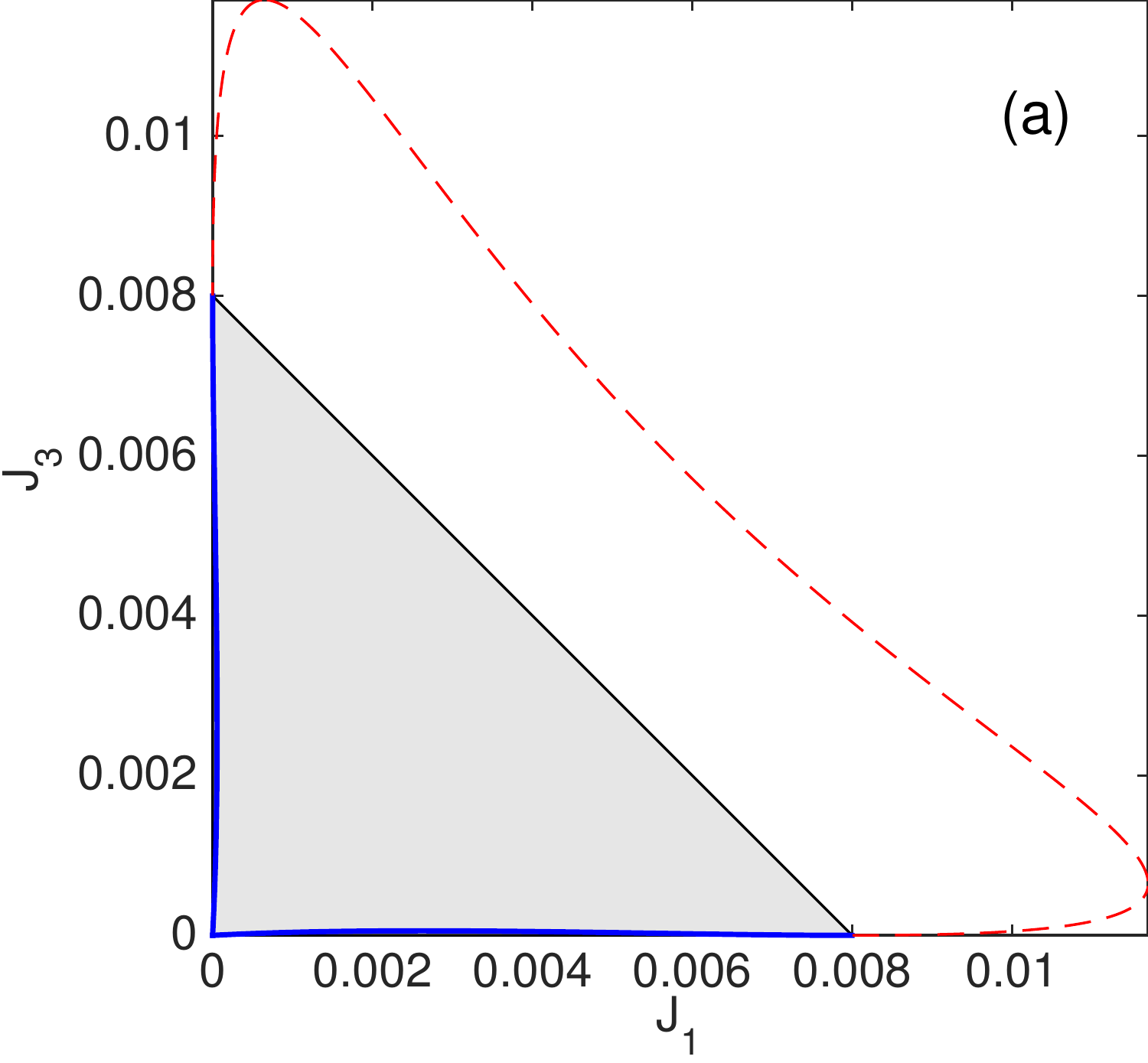}
\includegraphics[height=.3\textwidth]{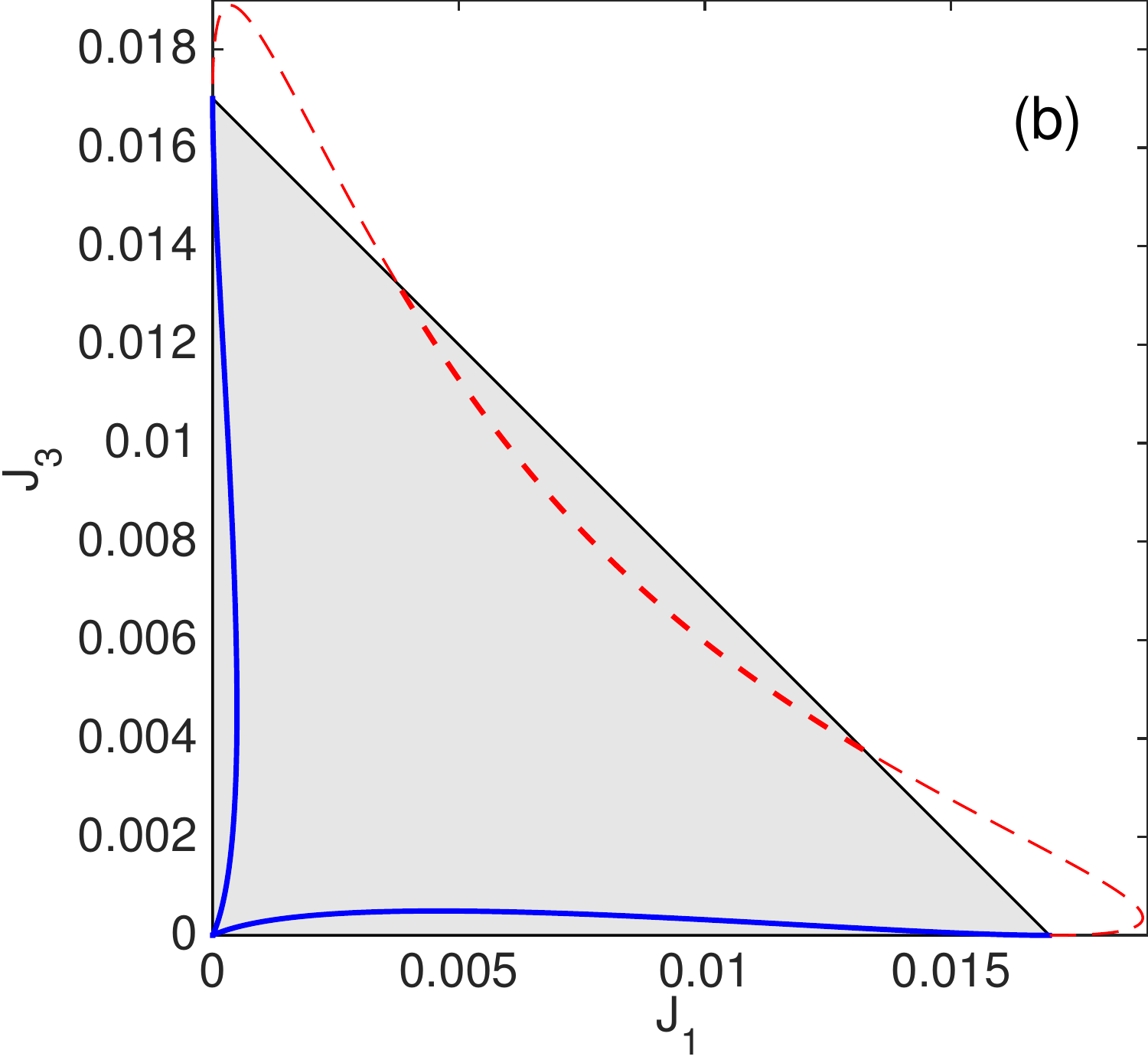}
\includegraphics[height=.3\textwidth]{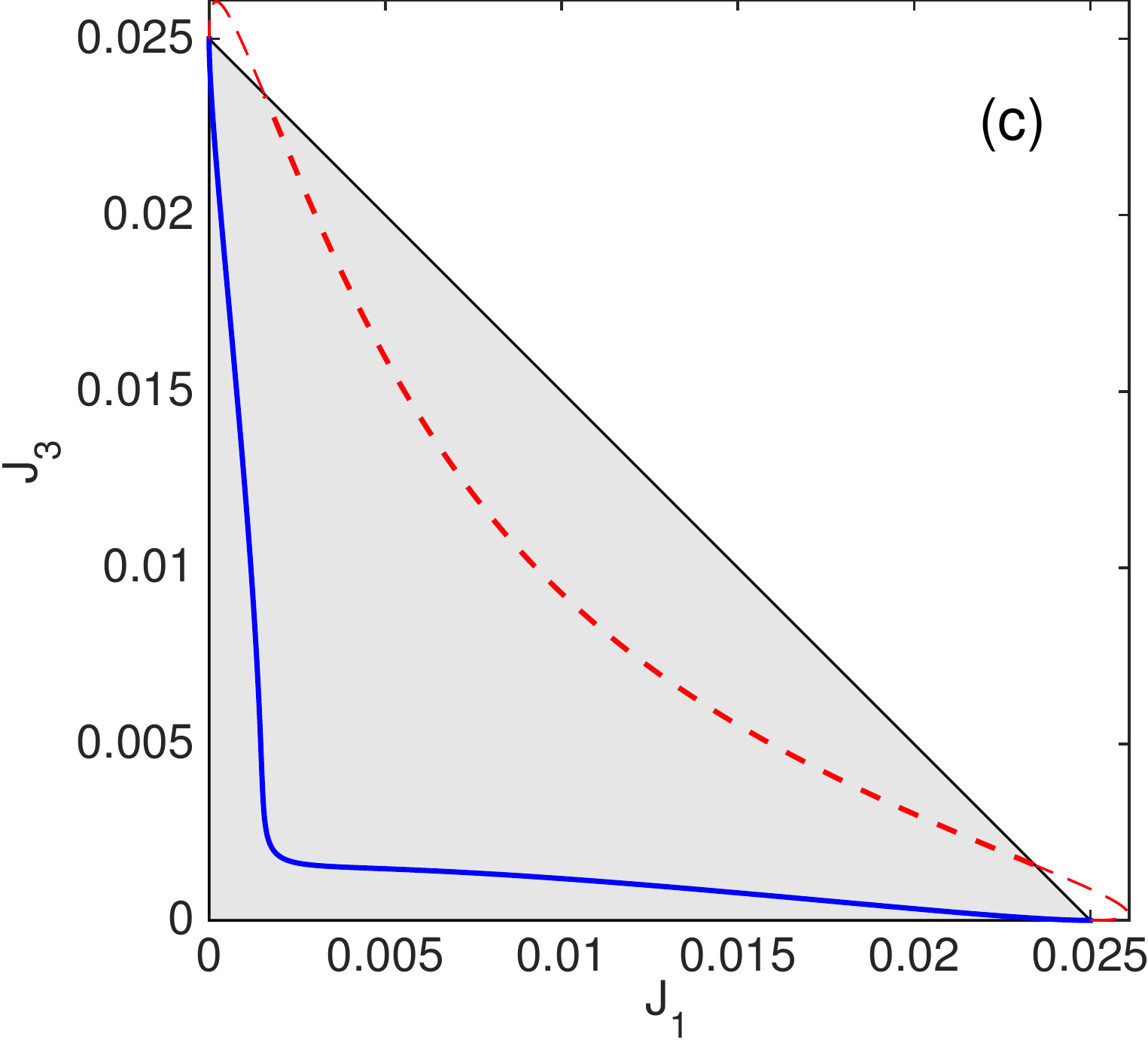}
\caption{The locations in $(J_1,J_3,\Theta)$ space of periodic orbits of system~$\Htrunc$. Solutions are shown for $N_j \in \{0.008, 0.017, 0.025\}$, chosen such that $N_1 < N_{\rm even} < N_2 < \NHHa < N_3 $. The trivial solutions from Section~\ref{sec:trivial} are at the corners and the solutions $\cS_\pi$ in blue and on $\cS_0$ in dashed red from Section~\ref{sec:solutionssin0}.}
\label{fig:all_periodic}
\end{figure}

The results of these calculations are shown in Figure~\ref{fig:all_periodic}. The triangle of physical solutions~\eqref{triangle} is shaded gray, with the three NNMs~\eqref{easy_ones} at the corners: $\mop$ at the origin, $\opo$ to its right, and $\mpm$ above it. The branch $\cS_\pi$ is shown as a solid line, and the branch $\cS_0$ dashed. All the RPOs on the branch $\cS_0$ are unphysical for $N<\Neven$, and lie outside the triangle in Figure~\ref{fig:all_periodic}(a). For $N>\Neven$, as in images (b) and (c), some of the solutions on $\cS_0$ are physical, having crossed into the triangle. For $N<\NHHa$, as in images (a) and (b), the branch $\cS_\pi$ consists of two pieces, the first interpolating between $\mop$ and $\opo$, and the second between $\mop$ and $\mpm$. At $N=\NHHa$, the two pieces merge with each other and detach from $\mpo$, interpolating directly between $\opo$ and $\mpm$.

% These images generated 5/13/15 using program allbranches.m

\section{The Hamiltonian Hopf bifurcations of $\mop$}
\label{sec:HH}
The goal in this section is to understand the two HH bifurcations of $\mop$ in more detail by focusing on a neighborhood of the bifurcation points. After describing the normal form for a Hamiltonian near a non-semisimple -1:1 resonance, we will further normalize the normal-form Hamiltonian defined by equations~\eqref{Htrunc} and~\eqref{H2NH4N} where it becomes unstable at $N= \ep/2A \approx \NHHa$. In Section~\ref{sec:Numerics}, we numerically check the bifurcation type at $\NHHb$ in terms of this analysis.

\subsection{The general normal form}
It is well known~\cite{Meyer:2010} that a two-degree-of-freedom quadratic Hamiltonian system
with a deficient pair of imaginary eigenvalues $\pm i\Omega$ can be put in the form
\begin{equation}
H_0(\xxi) = 
\W \left(  \xi_2 \eta_1-  \xi_1 \eta_2 \right) 
+ \frac{\sigma}{2}(\xi_1^2+\xi_2^2),
\label{H0xi}
\end{equation}
with $\sigma=\pm 1$.
This leads to a linear evolution equation
$\frac{d}{dt} \xxi = \cB \xxi $ where
\begin{equation}
\cB = \W \cS - \sigma \cJ  , \,
\cS = \begin{pmatrix}
0 & 1  & 0 & 0 \\
- 1  & 0 & 0 & 0 \\
0 & 0 & 0 & 1  \\
 0 & 0 & -1  & 0
\end{pmatrix}, \, \text{ and }
\cJ = \begin{pmatrix}
0 & 0  & 0 & 0 \\
0  & 0 & 0 & 0 \\
1 & 0 & 0 & 0  \\
 0 & 1 & 0  & 0
\end{pmatrix}.
\label{cB}
\end{equation}

Constructing the normal form for the perturbed system
\begin{equation}
H(\xxi) = H_0(\xxi) + \delta H_2(\xxi) + H_4(\xxi)
\label{Hxi}
\end{equation}
requires projecting $H_2(\xxi)$ and $H_4(\xxi)$ onto the null spaces, respectively, of the operators $\ad{H_0^{\rm T}}^{(2)}$ and $\ad{H_0^{\rm T}}^{(4)}$. 
Unlike in Section~\ref{sec:ChowKim}, however, the operator $\ad{H_0}$ does not act diagonally on the monomial basis, so determining its adjoint null-space, and constructing the projection operator~\eqref{Hproject} requires explicit construction of the operators $\ad{H_0}^{(2)}$ and $\ad{H_0}^{(4)}$ in terms of the monomial bases of $\cP^{(2)}(\RR)$ and $\cP^{(4)}(\RR)$. These are $10\times10$ and $35\times35$ matrices, so computer algebra is very helpful.

Defining the quadratic combinations
\begin{equation*}
\Gamma_1 = \xi_2 \eta_1-\xi_1 \eta_2, \,
\Gamma_2 = \frac{\xi_1^2+\xi_2^2}{2}, 
\text{ and }
\Gamma_3 = \frac{\eta_1^2+\eta_2^2}{2}
\end{equation*}
gives
$$
H_0(\xxi)= \W \Gamma_1 + \sigma \Gamma_2
$$ 
and resonant terms of the form
$$
\ker {\left(\ad{H_0^{\rm T}}^{(2)}\right)}= \vecspan{\left\{\Gamma_1,\Gamma_3\right\}}
\text{ and }
\ker {\left(\ad{H_0^{\rm T}}^{(4)}\right)} = 
\vecspan{\left\{\Gamma_1^2,\Gamma_1\Gamma_3,\Gamma_3^2\right\}}.
$$
Projecting terms from the Hamiltonian~\eqref{Hxi} onto these bases yields the normal form:
\begin{equation}
H^{\rm N} = H_0(\Gamma_1,\Gamma_2) + \d H_2^{\rm N}(\Gamma_1,\Gamma_3) + H_4^{\rm N}(\Gamma_1,\Gamma_3)
\label{HHnormal}
\end{equation}
where
$$
H_2^{\rm N} = \k_1 \Gamma_1 + \k_3 \Gamma_3,  \text{ and }
H_4^{\rm N} =  \k_{1,1} \Gamma_1^2 + \k_{1,3} \Gamma_1 \Gamma_3 + \k_{3,3} \Gamma_3^2.
$$                                                                      

After making a scaling (which is symplectic with multiplier $\mu^3$) with $\mu\ll 1$:
$$
\xi_j \to \mu^2 \xi_j;   \;
\eta_j \to \mu \eta_j;   \;
\delta \to \mu^2 \ss \text{ (with $\ss = \pm 1$)}; \;
H \to H/\mu^3,
$$
the Hamiltonian becomes 
$$
H = \W \Gamma_1 + 
\mu H_{\rm pert}+ O{(\mu^3)} \text{ where } H_{\rm pert} =  \sigma \Gamma_2 + \ss\k_3 \Gamma_3 + \k_{3,3} \Gamma_3^2.
$$
Then we can use the Poincar\'e-Lindstedt method to look for solutions of the form
$$
\uu(t) = \uu^0(\tau) + \mu \uu^1(\tau) + O{(\mu^2)}
$$
where 
$\tau = (\w_0 + \mu \w_1 + O{(\mu^2)}) t$ 
and 
$\uu^j(\tau+2\pi) = \uu^j(\tau)$.
the sequence of equations is
\begin{align}
O(1): \; \omega_0 \frac{d \xxi_0}{d\tau} - \W\cS \xxi_0 & = 0\label{O1} \\
O(\mu): \; \omega_0 \frac{d \xxi_1}{d\tau} - \W\cS \xxi_1 & = -  \omega_1 \frac{d \xxi_0}{d\tau} + J\nabla H_{\rm pert}.\label{Omu}
\end{align}
The sequence continues for higher orders in $\mu$, but the essential information, i.e.\ the topology of the periodic orbits is determined by the expansion to this order. 

The solution to equation~\eqref{O1} is 
$$\xxi_0 = 
\begin{pmatrix}
 \cos{\tau} & \sin{\tau} & 0 & 0 \\
 -\sin{\tau} & \cos{\tau} & 0 & 0 \\
0 & 0 & \cos{\tau} & \sin{\tau} \\
 0& 0&-\sin{\tau} & \cos{\tau}
\end{pmatrix}
\begin{pmatrix}
\Xi_1 \\ \Xi_2 \\ Y_1\\ Y_2
\end{pmatrix} \text{ and } \omega_0 = \W.$$
 The conditions to avoid secular terms in equation~\eqref{Omu} are $\Xi_1 = -\omega_1 Y_2$, $\Xi_2 = \omega_1 Y_1$,  and  
\begin{equation}
\sigma \w_1^2 + 2\k_{3,3}  r^2 = \ss \k_3.
\label{hyperbola}
\end{equation}

\begin{figure}[htbp] 
   \centering
   \includegraphics[width=.3\textwidth]{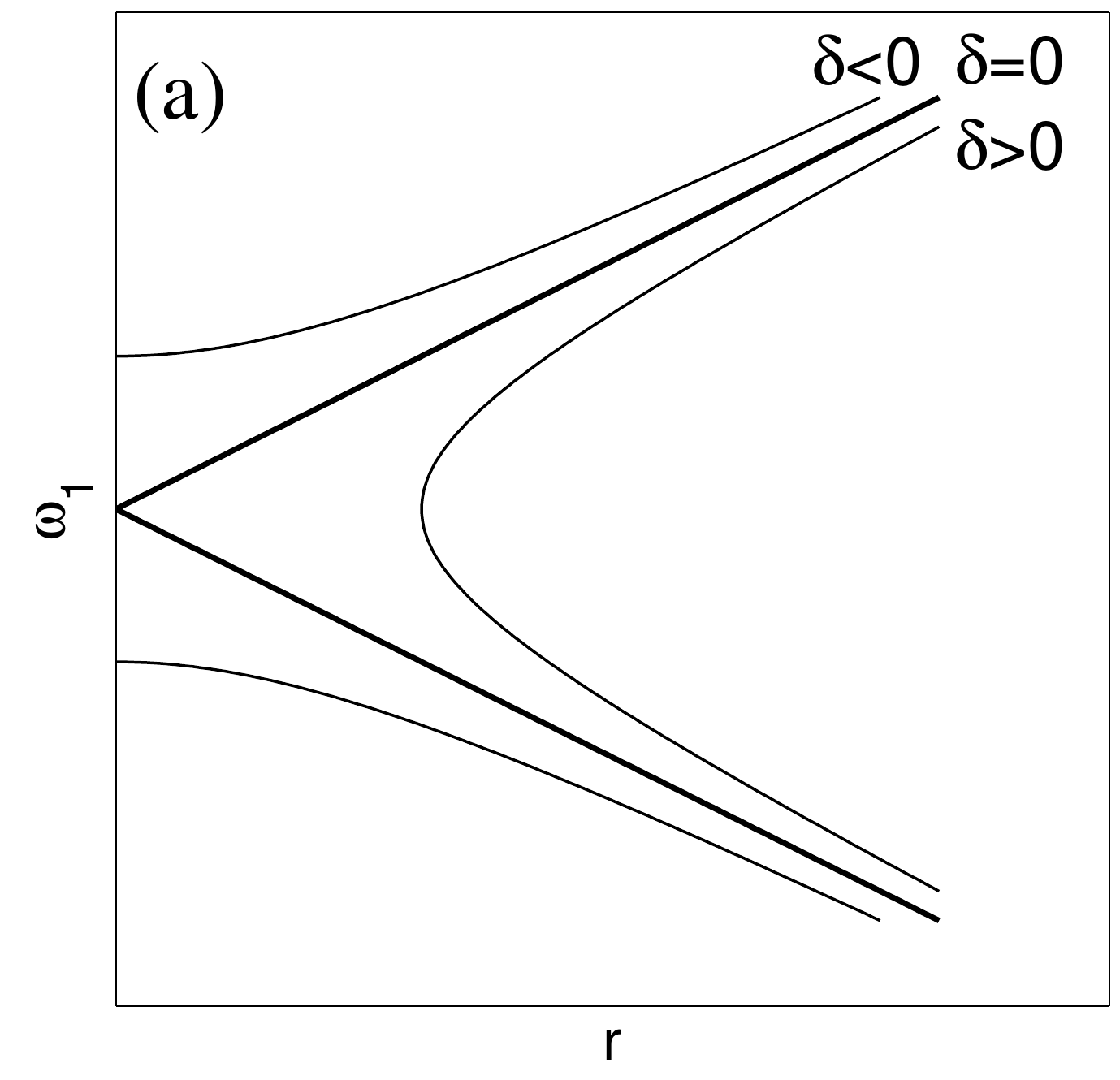} 
   \includegraphics[width=.3\textwidth]{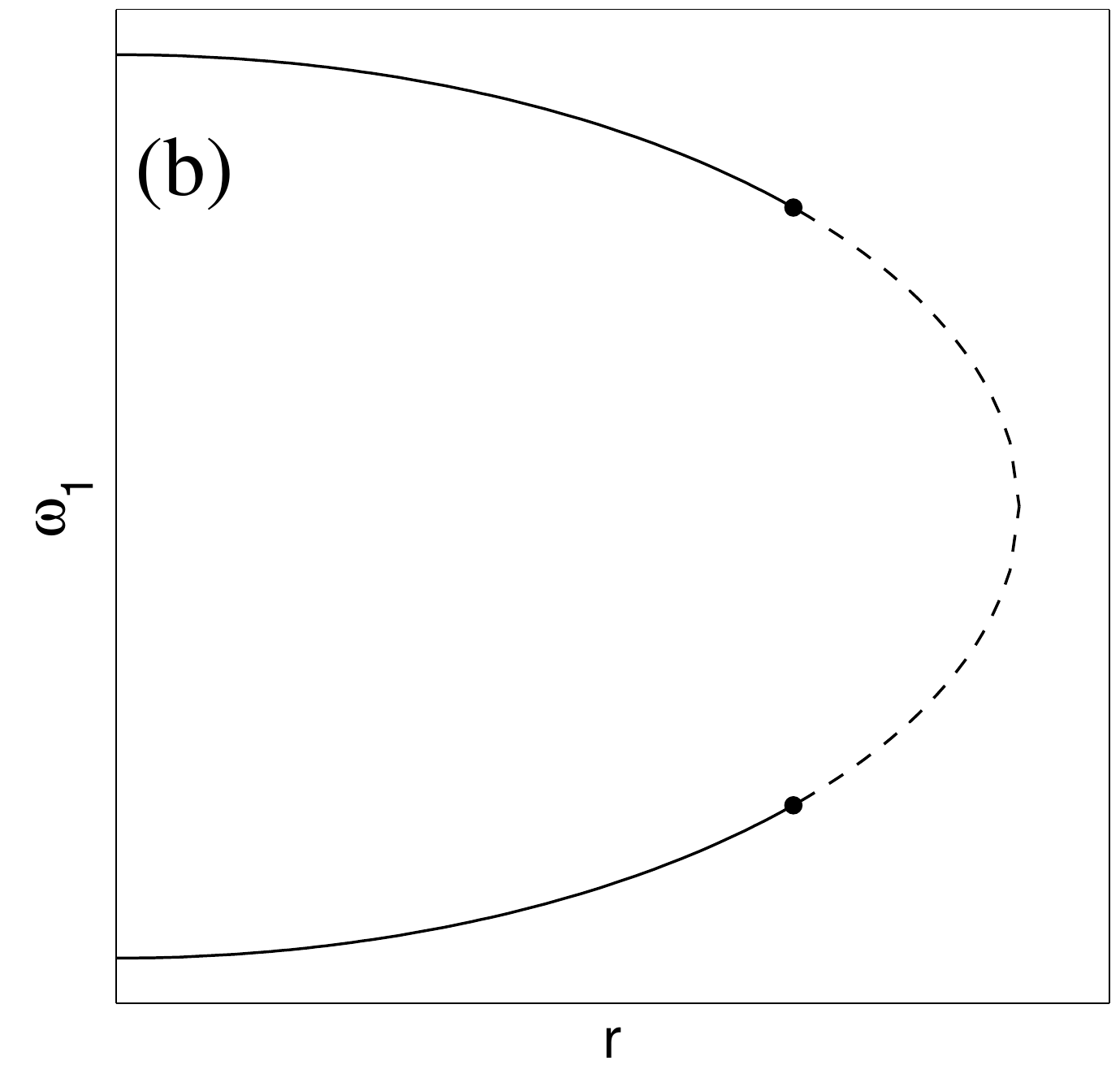} 
   \caption{The branches of time-harmonic solutions in a neighborhood of the origin. (a) Hyperbolic HH (b) Elliptical HH.}
\label{fig:hyperbola}
\end{figure}

The bifurcation can thus take two forms depending on the sign of $\sigma \k_{3,3}$. Assume that the origin is stable for $\delta<0$ and unstable for $\delta>0$. Define the \emph{hyperbolic HH bifurcation} to occur when $\sigma \k_{3,3}<0$. In this case, the two Lyapunov families of periodic orbits lie on separate branches of a hyperbola for $\delta<0$; for $\delta=0$ these two branches collide at the point $(r,\w_1)=(0,0)$; and for $\delta>0$, the two branches have re-attached to form a single branch of a hyperbola that is bounded away from $r=0$. See Figure~\ref{fig:hyperbola}(a). If $\sigma \k_{3,3}>0$, then the two Lyapunov families lie on a single, continuous, approximately elliptical branch for $\delta<0$. This shrinks to a point at $\delta=0$ and disappears for $\delta>0$; see Figure~\ref{fig:hyperbola}(b). This will be denoted the \emph{elliptical HH bifurcation}.

\subsection{The HH bifurcation at $\NHHa$}
In this section, we  apply the above analysis to the normal form Hamiltonian~\eqref{Htrunc} with
\begin{equation*}
N = \ep/2A + \delta \text{ with } \delta \ll \epsilon
\end{equation*}
using real coordinates $x_j+i y_j=z_j$. At $\delta=0$, the linearized evolution equations under Hamiltonian~$\Htrunc$ are
$$
\frac{d}{dt} \begin{pmatrix} x_1 \\ x_3 \\ y_1 \\ y_3 \end{pmatrix} =
\begin{pmatrix}
 0 & 0 & \ep -\D  & -\ep  \\
 0 & 0 & -\ep  & \D +\ep  \\
 \D -\ep  & -\ep  & 0 & 0 \\
 -\ep  & -\D -\ep  & 0 & 0 \\
\end{pmatrix}
 \begin{pmatrix} x_1 \\ x_3 \\ y_1 \\ y_3 \end{pmatrix} \equiv \cA \xx
$$
This matrix has multiplicity-two eigenvalues $\pm i \D$ and is non-diagonalizable.
To put it in the form~\eqref{H0xi}, an algorithm of Burgoyne and Cushman~\cite{Burgoyne:1974} is used to construct a symplectic matrix $P$---one satisfying $P^{\rm T} J P = J$---such that in the new coordinates 
$$
\xxi = \begin{pmatrix} \xi_1 \\ \xi_2 \\ \eta_1 \\ \eta_2 \end{pmatrix} = P^{-1}\xx,
$$
the leading order linear part is given by $\cB = P^{-1}\cA P$ in equation~\eqref{cB}. The sign $\sigma$ is determined by the algorithm. The Burgoyne algorithm gives
$$
P = \begin{pmatrix}
\frac{1}{2 \sqrt{\ep}} & 0 & 0 & \sqrt{\ep} \\
 \frac{1}{2 \sqrt{\ep}} & 0 & 0 & -\sqrt{\ep} \\
 0 & -\frac{1}{2 \sqrt{\ep}} & \sqrt{\ep} & 0 \\
 0 & \frac{1}{2 \sqrt{\ep}} & \sqrt{\ep} & 0
\end{pmatrix}
$$
and the normal form algorithm gives
$$
\sigma=1, \,
\k_1 = 0, \, 
\k_3 = -4 A   \delta, \,
\k_{1,1}= \frac{2}{3} A, \,
\k_{1,3}= 0, \, \text{and}\
\k_{3,3}= 7 A \ep^2
$$
Since $\sigma \k_{3,3}=7 A \epsilon^2>0$, the HH bifurcation is hyperbolic.

\begin{figure}[tb]
\center
\includegraphics[height=.3\textwidth]{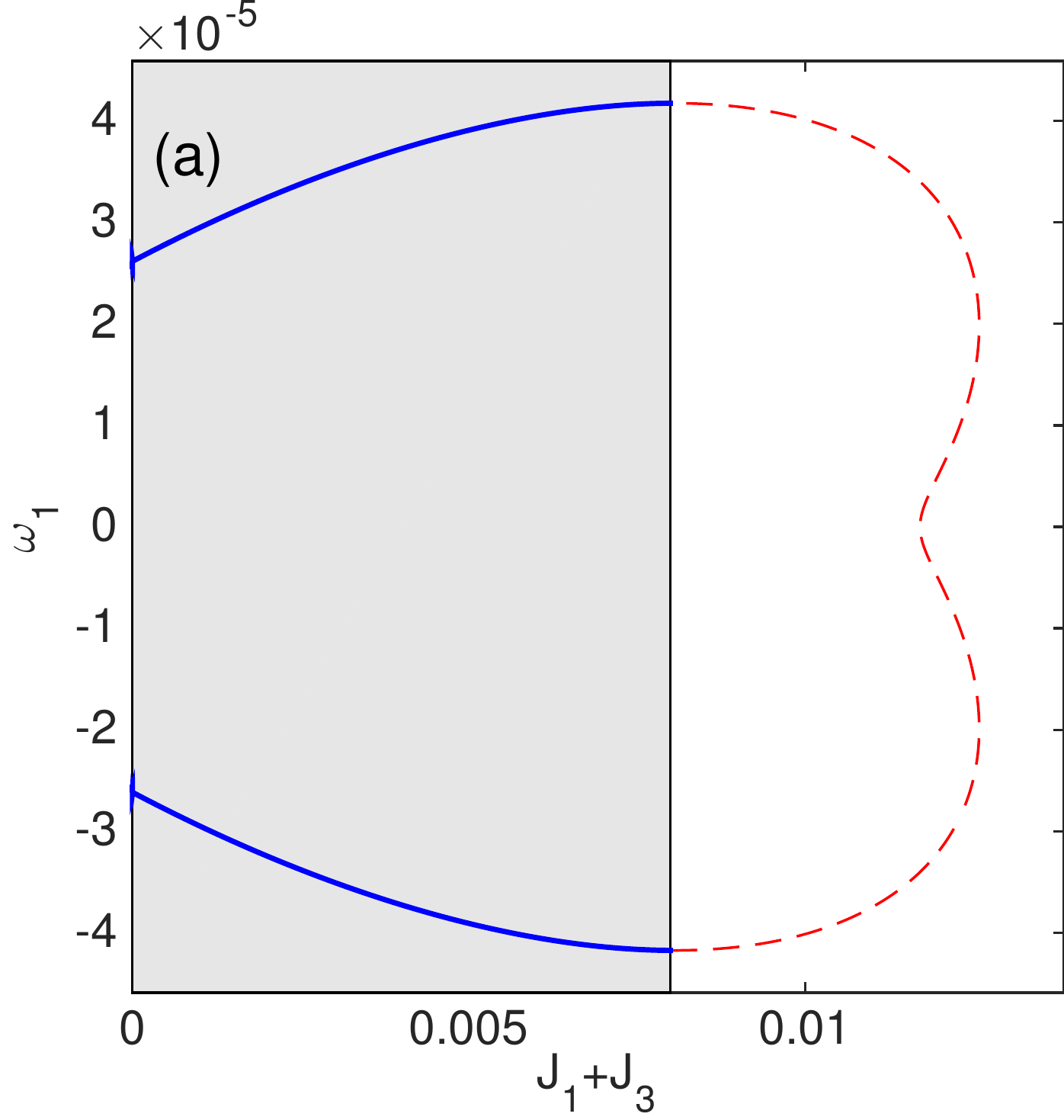}
\includegraphics[height=.3\textwidth]{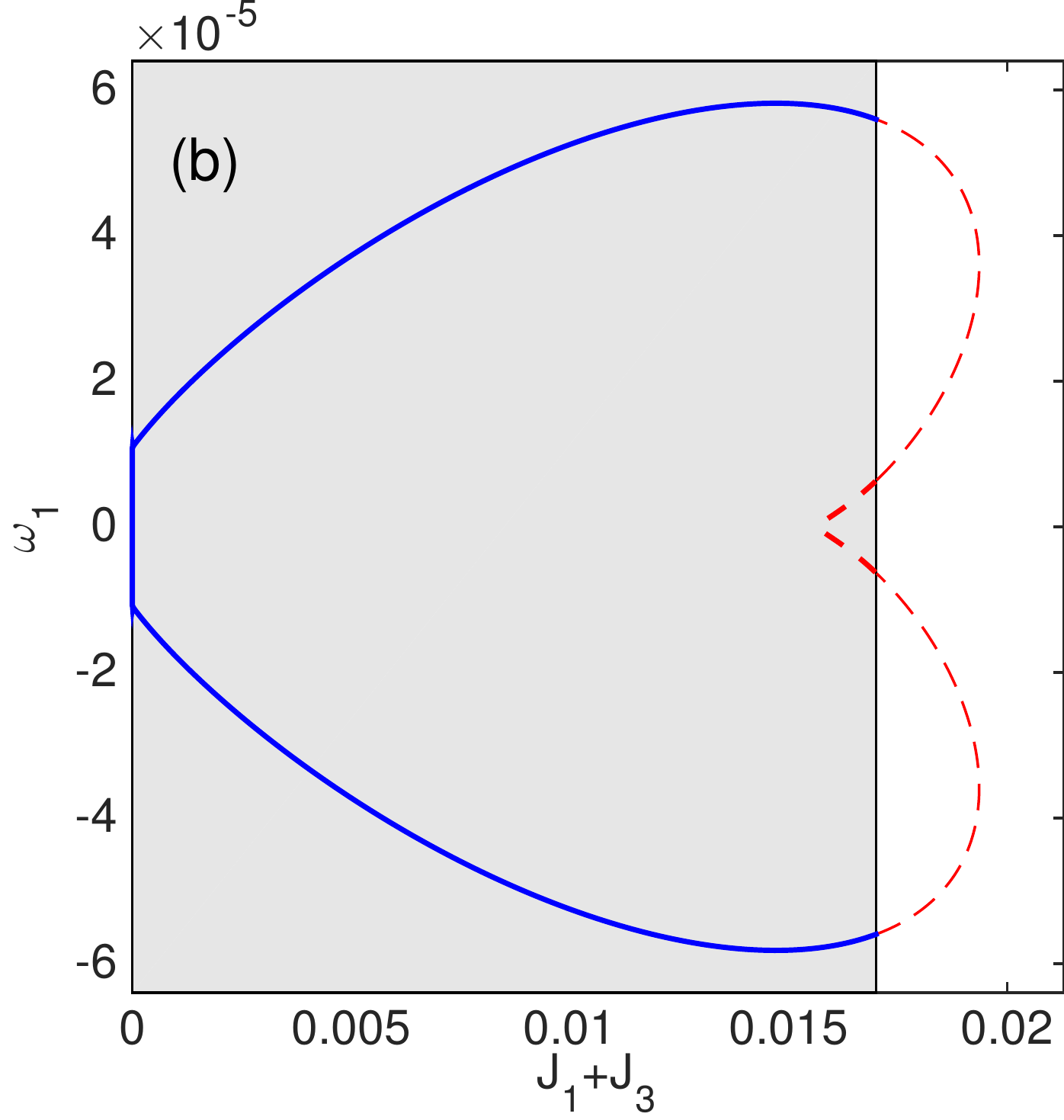}
\includegraphics[height=.3\textwidth]{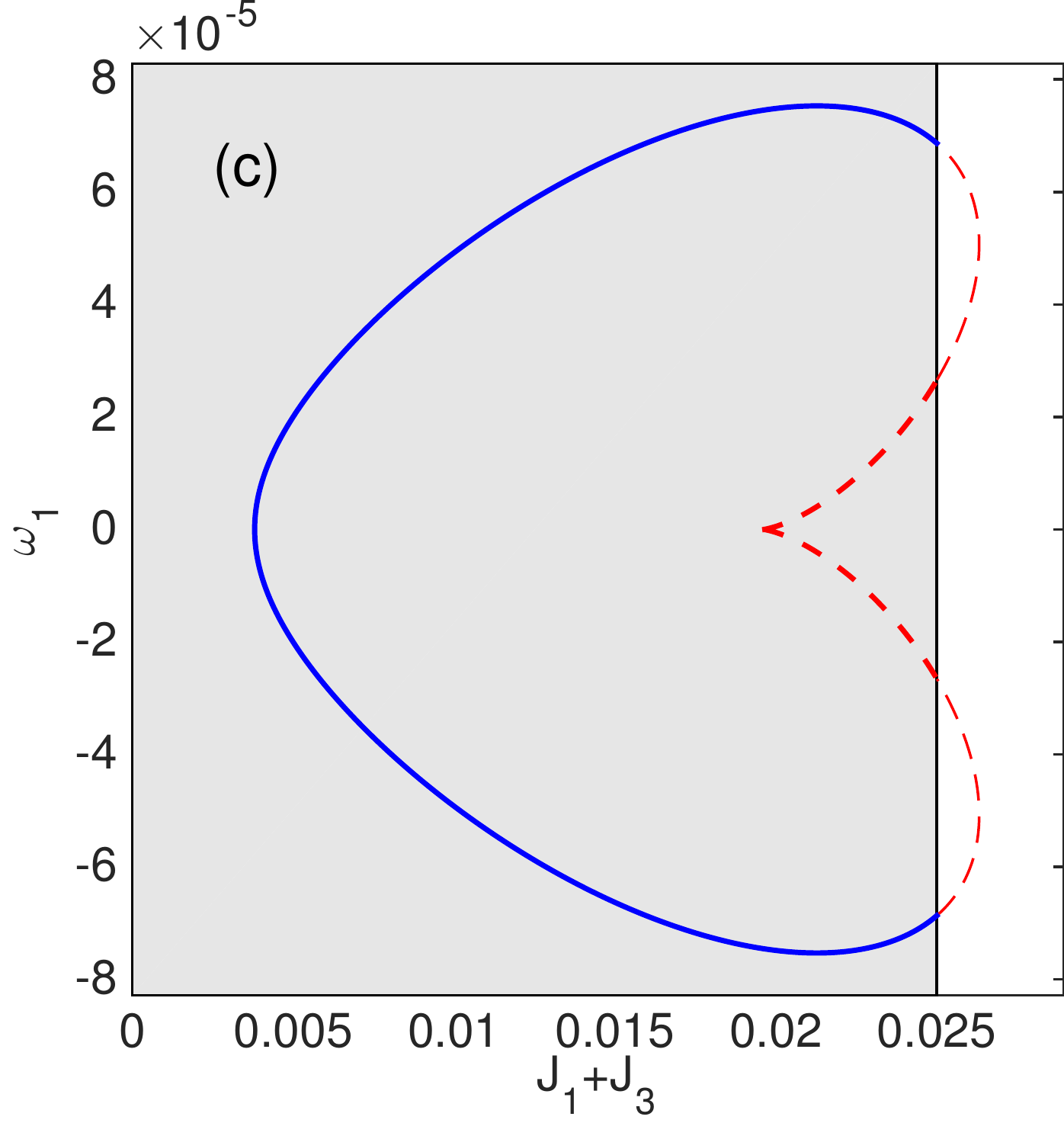}
\caption{Graphs of the solution amplitude vs.\ the frequency correction, with the family $\cS_0$ solid and the family $\cS_\pi$ dashed, reproducing the bifurcation picture in Figure~\ref{fig:hyperbola}.}
\label{fig:r_w1}
\end{figure}

The solid curve in Figure~\ref{fig:r_w1} is a numerical analog to Figure~\ref{fig:hyperbola} and contains three plots generated from the same data as Figure~\ref{fig:all_periodic}. On the $x$-axis is $r=\rho_1+\rho_3$ and on the $y$-axis we plot $(\eta-1)$ from equation~\eqref{eta}, which in these coordinates is 
$$
-\frac{ \D  \left(\rho_1-\rho_3\right) \left(-8 A \sqrt{\rho_1\rho_3} \cos{\Theta}
+2 A (\rho_1+ \rho_3) +\ep \right)}{2(\rho_1+\rho_3)}.
$$
This figure clearly shows the pinch-off associated with the hyperbolic HH bifurcation.

% Figures generated on 5/13/15 using allbranches.m

We have analyzed the bifurcation at $\NHHa$ by applying a second normal form reduction to the truncated normal form Hamiltonian derived in the previous section. We do not have this available at the other HH bifurcation that occurs at $N=\NHHb$. In Section~\ref{sec:HH2} we determine the type of this bifurcation numerically.

\section{Normal form of an SN in two degrees of freedom}
\label{sec:saddlenode}
The saddle-node bifurcations resulting in the pairs of RFPs $\ppo$/$\poo$ and $\mpo$/$\mpp$, depicted in Figures~\ref{fig:z1z3} and~\ref{fig:9Modes}, both show the creation of a pair of fixed points, one with two pairs of imaginary eigenvalues, and one with one pair of imaginary and one pair of real eigenvalues. This bifurcation is also known at the Hamiltonian $0^2i\omega$ bifurcation~\cite{Broer:1993ch,Gelfreich:2014fn}. In this section we consider a normal form for such behavior. The system undergoes the bifurcation as the parameter $\mu$ crosses zero:
\begin{equation}
H = \left(\frac{q_1^2}{2} + \frac{p_1^2}{2} \right)
  + \a \left(\frac{p_2^2}{2} + \mu q_2 - \frac{q_2^3}{3}\right)
  + \b I q_2
  + \Hhigh(q_2,I)
  + R_{\infty}(q_1,q_2,p_1,p_2)
\label{H02iw}
\end{equation}
where $I=(q_1^2+p_1^2)/2$ is the action variable in the first degree of freedom. Without loss of generality we may take $\alpha>0$, but $\beta$ may have either sign. This normal form can be formally deduced in the same manner that we determined the resonant terms in Sections~\ref{sec:truncatedForm} and~\ref{sec:HH}, where $H_0 = \tfrac{1}{2}(q_1^2+p_1^2+p_2^2)$ and the remaining terms lie in $\ker {\left(\ad{H_0^{\rm T}}\right)}$.

The term $\Hhigh$ represents a formal power series expansion and contains only resonance terms. The term $R_{\infty}$ is a beyond-all-orders small term. Because $\Hhigh$ is independent of the angle $\theta_1$ in the $(q_1,p_1)$ plane the action $I$ is conserved in any truncation of the system to finite degree. Thus, the act of computing a normal form for this bifurcation introduces a new conserved quantity, thus rendering the normal-form equations integrable when the underlying dynamics may not be. This will be important in the numerical calculations of Section~\ref{sec:what}.

In this preliminary analysis, we consider the truncated system where the terms $ \Hhigh(q_2,I)$ and $R_{\infty}(q_1,q_2,p_1,p_2)$ are ignored. The equations of motion are
\begin{subequations}
\begin{align}
\dot{q}_1 & = (1 + \beta q_2) p_1;&  \dot{q}_2 & = \alpha p_2; \\
\dot{p}_1 & = -  (1 + \beta q_2) q_1;&  \dot{p}_2 & = - \alpha (\mu  - q_2^2) - \beta I.
\end{align}
\label{zero2iwODE}
\end{subequations}
When $\mu>0$ the system has fixed points at 
\begin{equation}
\qmupm =(q_1,q_2,p_1,p_2)\approx(0,\pm\sqrt\mu,0,0),
\label{qdelta}
\end{equation}
with $\qmum$ stable and $\qmup$ unstable. Each has a pair of imaginary eigenvalues $\lambda=\pm i \omega_1 \approx \pm i$ coming from the $(q_1,p_1)$ directions. The stable fixed point has an additional pair of imaginary eigenvalues $\lambda = \pm i \omega_2 = \pm O(\sqrt{\mu}) i$. Therefore both fixed points have a  family of \emph{short}-periodic orbits coming from $\omega_1$ and the stable fixed point $\qmum$ has an additional family of  \emph{long}-periodic orbits coming from $\omega_2$ when $\w_2^{-1}\notin \ZZ$. A third family of periodic orbits that is not Lyapunov plays an important role in this bifurcation. We describe the global character of the three families.

The subspace defined by $(q_1,p_1)=(0,0)$ is invariant. Off of this subspace, we introduce canonical action-angle coordinates in the $(q_1,p_1)$ degree of freedom
$$
q_1 = \sqrt{2 I} \cos{\theta}, \, p_1 = \sqrt{2 I} \cos{\theta}.
$$
We find that $I$ and $\theta$ satisfy
$$
\dot{I} = 0 \text{ and } 
\dot{\theta} = -(1+\beta q_2),
$$
which has solution
 \begin{equation}
 \begin{split}
 I & = \frac{\nu^2 Q_0^2}{2};  \;
 \theta(t)-\theta(0) = - \int_{0}^{t} \left(1 + \beta q_2(t') \right) dt.
 \end{split}
	\label{DeltaTheta}
 \end{equation}
 
Let
$$ 
\mu = \nu^2 s \text{ with } \nu>0 \text{ and } s = \pm 1, 
q_2 = \nu x, \, 
\text{and } 
\gamma = \frac{\beta Q_0^2}{2 \alpha} + s.
$$
Then $x(t)$ solves
 \begin{equation}
 \ddot x + \nu \alpha^2 \left( \gamma - x^2 \right) =  0.
\label{ddotx}
\end{equation}
When $\gamma>0$ (i.e.\ when $s=1$ and $\mu>0$)  and for any $b$ satisfying $b^2 < \gamma$ this system has a solution
\begin{equation}
x(t) = c + (b-c) \sn^2{(\omega t, k)},
\label{xsn}
\end{equation}
where 
\begin{equation}\begin{split}
a = \frac{-b + \sqrt{12\gamma-3b^2}}{2} ; \; 
c = \frac{-b - \sqrt{12\gamma-3b^2}}{2}; \\
k = \sqrt{\frac{b-c}{a-c}}; \;
\omega = \sqrt{\frac{(a-c)\alpha^2 \nu}{6}}.
\end{split}
\label{auxiliary}
\end{equation}

The above calculation shows that equations~\eqref{DeltaTheta} and~\eqref{xsn} are the general solution to system~\eqref{zero2iwODE}. Among these orbits are three families of periodic orbits, which we describe next.

\subsection{The short-periodic orbits}
\begin{figure}[htbp] 
   \centering
   \includegraphics[width=.24\textwidth,valign=t]{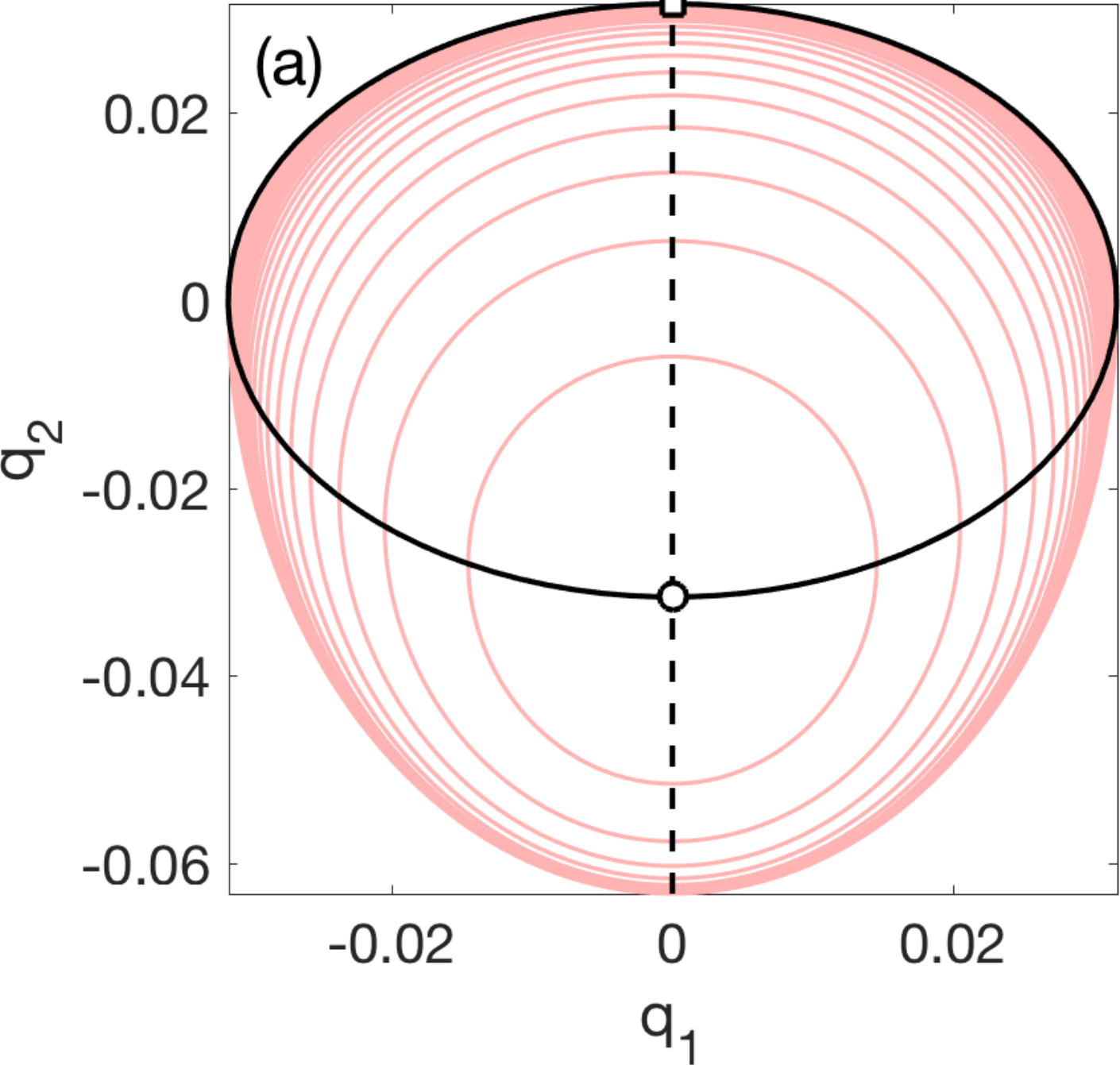} 
   \includegraphics[width=.24\textwidth,valign=t]{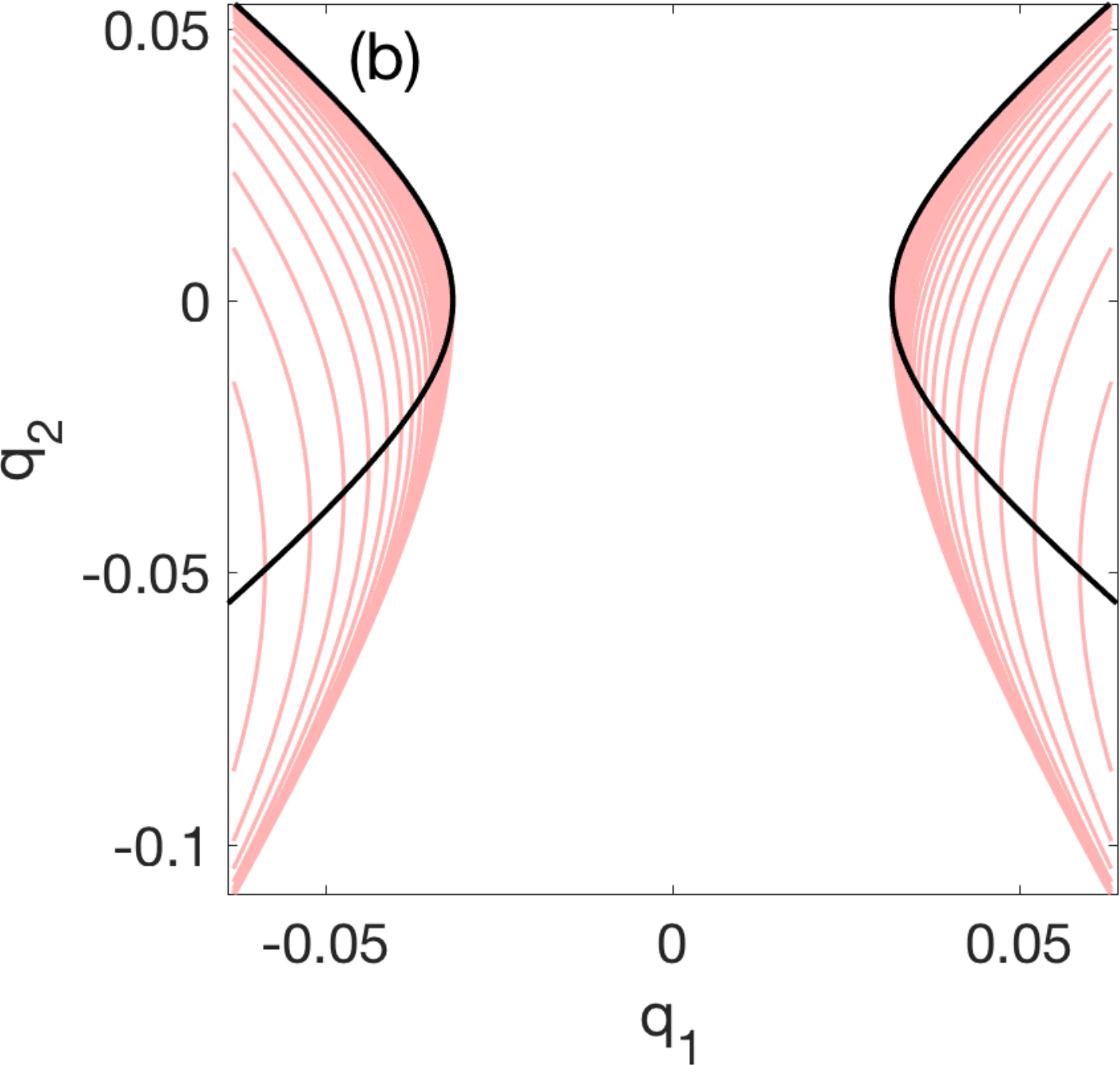} 
   \includegraphics[width=.24\textwidth,valign=t]{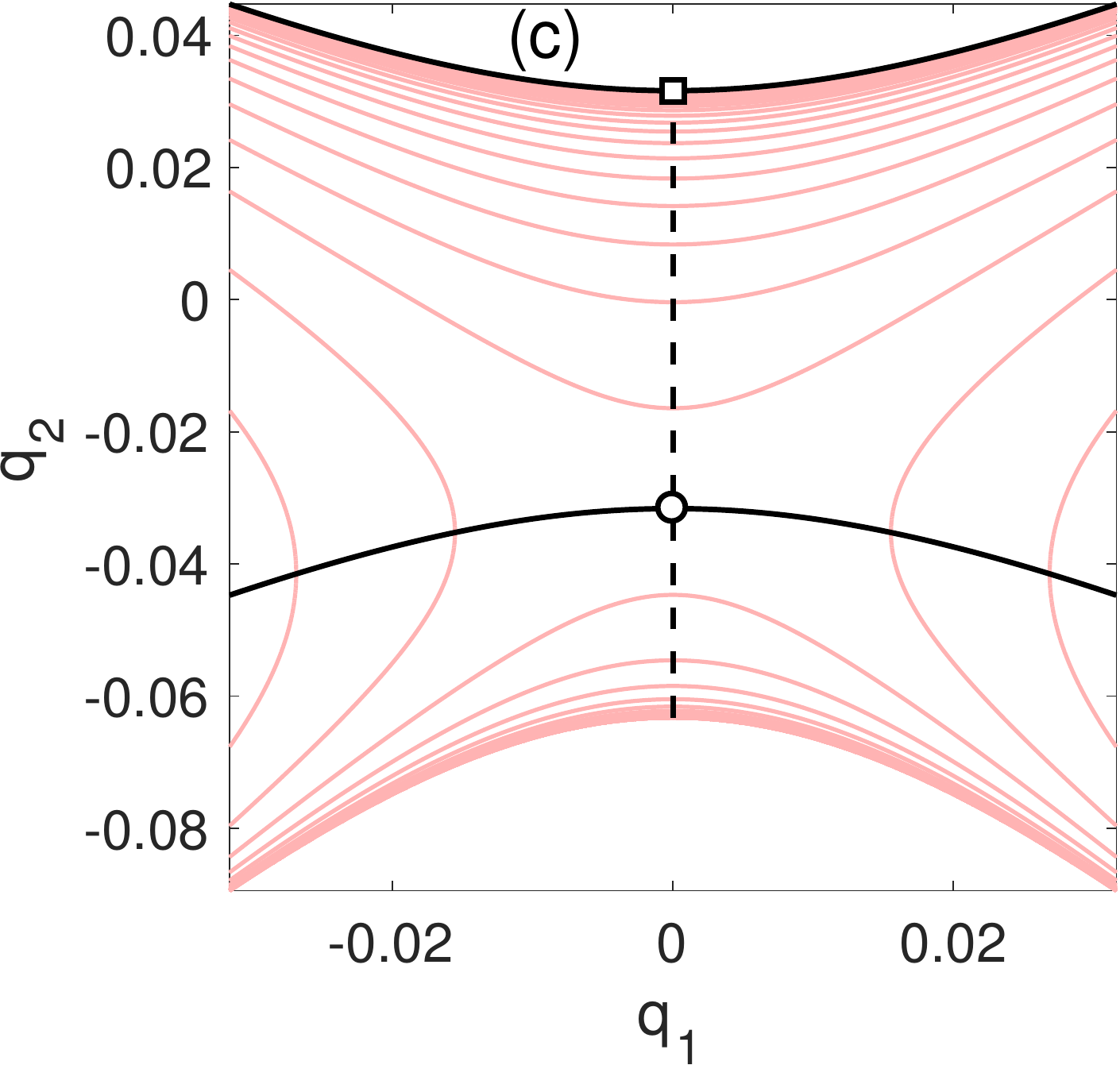} 
   \includegraphics[width=.22\textwidth,valign=t]{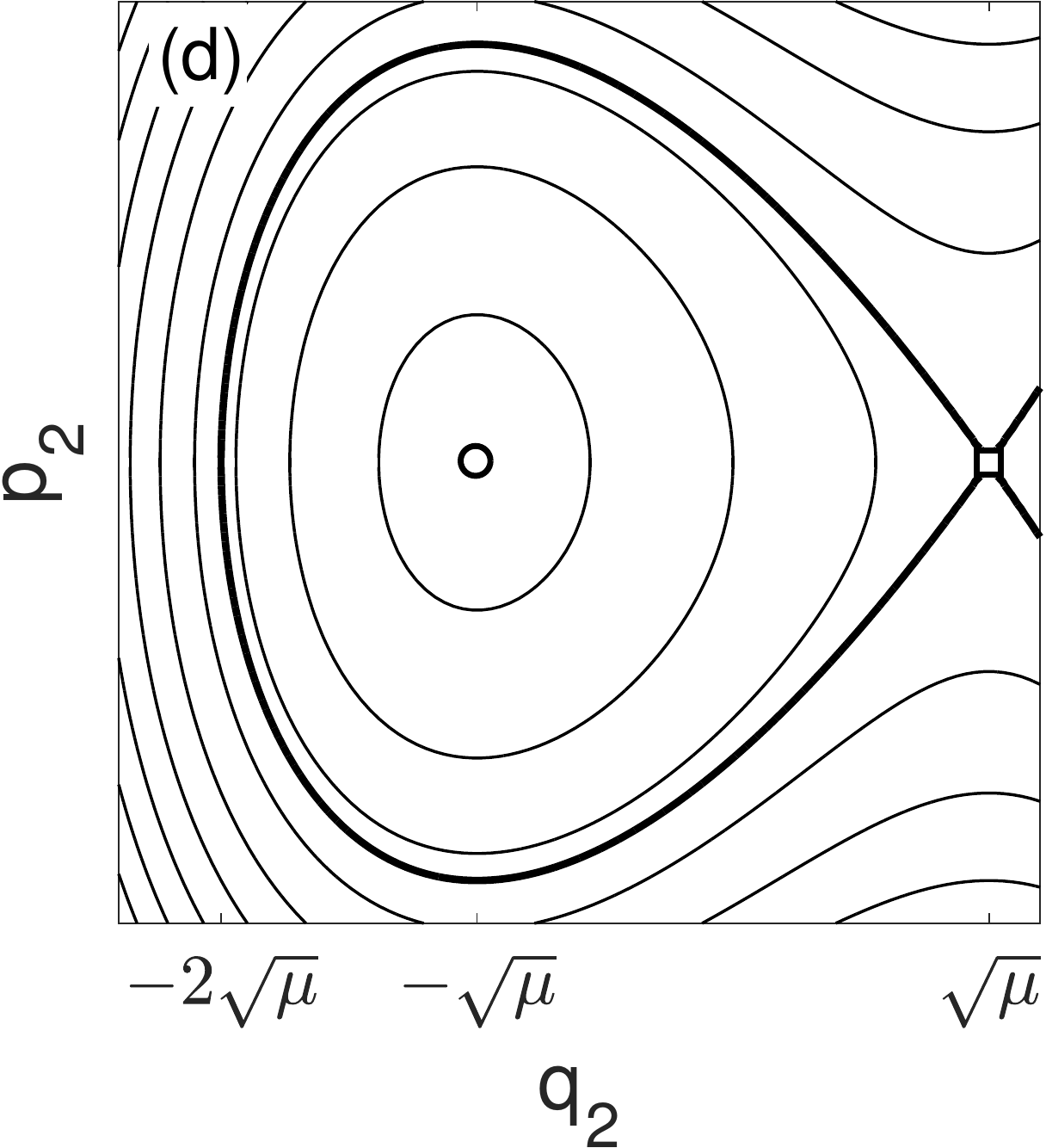} 
   \caption{System~\eqref{H02iw} has three families of periodic orbits---short, long, and mixed, represented by heavy solid, dashed, and light red curves, respectively---and two types---elliptic and hyperbolic. In the elliptic case, the branches only exist for $\mu>0$, shown in~\textbf{(a)} for $(\alpha,\beta,\mu) = (1,-2,10^{-3})$. In the hyperbolic case, the short and mixed branches exist regardless of the sign of $\mu$, with the short-periodic branches undergoing a reconnection bifurcation at $\mu=0$ and the mixed branches reconnecting at a sequence of resonant values $\mu_n$. The long periodic branch only exists for $\mu>0$. This case is shown  in \textbf{(b)} and~\textbf{(c)} for $(\alpha,\beta,\mu) = (1,2,-10^{-3})$ and $(\alpha,\beta,\mu) = (1,2,10^{-3})$, respectively. \textbf{(d)}~The long-period dynamics due to the saddle-node bifurcation. }
\label{fig:SN}
\end{figure}

These orbits correspond to fixed points of the ODE~\eqref{ddotx}, i.e.\ $x^2(t) = X_0^2 = \gamma$ or
\begin{equation}
 X_0^2-\frac{\b}{2\a} Q_0^2 =  s.
\label{SNtype}
\end{equation}
This is reminiscent of equation~\eqref{hyperbola}, showing that this bifurcation comes in two types. If $\beta<0$, then equation~\eqref{SNtype} describes the \emph{elliptical $0^2 i \omega$ bifurcation}. There are no periodic orbits when $s=-1$ and a closed curve of such orbits when $s = 1$, which is depicted as the heavy solid curve in Figure~\ref{fig:SN}(a). Note that $s=1$ means $\mu>0$, the condition  for the existence of fixed points, and that these are just degenerate periodic orbits along this curve. The curve of periodic orbits shrinks to a point as $\nu \to 0$. 

The condition $\beta>0$ describes the \emph{hyperbolic $0^2 i \omega$ bifurcation}. Regardless of whether $s=\pm 1$, the periodic orbits lie along the two branches of a hyperbola, oriented vertically when $s=-1$ and horizontally when $s=1$. Only in the case $s=1$ do these branches contain the two fixed points. In the limit $\mu=0$, the two branches merge into a pair of crossed lines, and re-connect. These are shown as the heavy solid curves in Figure~\ref{fig:SN}(b) and~(c).

\subsection{The long-periodic orbits}
The Lyapunov family of long-period  orbits exists only when $\mu>0$. It lies in the invariant plane $(q_1,p_1)=(0,0)$ and consists of the bounded solutions to equation~\eqref{ddotx} with $Q_0=0$; see Figure~\ref{fig:SN}(d).  Representing each member of a family of periodic orbits by the values of $(q_1,q_2)$ when $p_2=0$ gives a line segment from the hyperbolic fixed point at $(q_1,q_2)=(0,\sqrt{\mu})$ to the furthest point on the homoclinic loop at $(0,-2\sqrt{\mu})$. Each periodic orbit is represented by two points on this line on opposite sides of the elliptic fixed point at $(q_1,q_2)=(0,-\sqrt{\mu})$; these are the marked points and dashed lines in Figure~\ref{fig:SN}(a) and~(c). The period increases to infinity toward the endpoints of the segment. 

\subsection{The mixed periodic orbits}
\label{sec:mixed}
The two families of Lyapunov periodic orbits have trivial dynamics in one or the other degree of freedom. In the third family, both degrees of freedom evolve in time.

The component $x(t)$ given by equation~\eqref{xsn} has period
$$
T = \frac{2 K(k)}{\omega}
$$
where $K(k)$ is a complete elliptic integral of the first kind. It oscillates between $x(0) = c$ and $x(T/2) = b$.

The orbit is periodic if $(q_1,p_1)$ oscillates an integer number of times during one period of $x(t)$ i.e.\ if $\Delta \theta$ is integer multiple of $2\pi$ over one period of $x(t)$. Integrating equation~\eqref{DeltaTheta}, we find
$$
\Delta \theta = -T - \nu \beta \int_{0}^{T} x(t') dt' =
-2 \sqrt{\frac{6}{(a-c) \alpha^2 \nu}} 
\bigl(
K(k) + \nu \beta \left[a K(k) - (a-c) E(k) \right]
\bigr).
$$ 
Thus the condition that an orbit be periodic is that 
\begin{equation}
\sqrt{\frac{6}{(a-c) \alpha^2 \nu}} 
\bigl(
K(k) + \nu \beta \left[a K(k) - (a-c) E(k) \right]
\bigr)
= \pi n
\label{mixedPeriodic}
\end{equation}
for some $n \in \ZZ$. This condition depends on the initial condition when $\dot{q}_2=0$, i.e.\ on $b$ or $c$, and on the value $Q_0$. These orbits are shown in Figure~\ref{fig:SN} for $\a=1$, $\beta = \pm 2$, and $\mu = \pm 10^{-3}$.

The topology of these branches mirrors that of the short-periodic orbits. The upper-limiting envelope satisfies equation~\eqref{SNtype} with $X_0=b=\sqrt{\gamma}$ in equation~\eqref{xsn}. The lower envelope is related to $b$ by the equation for $c$ in equation~\eqref{auxiliary} and is given by $c = -2\sqrt{\gamma}$. Each of the branches of mixed periodic orbits is a $\pi$-times-integer-valued level-set curve of the form~\eqref{mixedPeriodic}.

The most interesting question to ask is how the branches change as $\mu$ changes. The most obvious thing to note is that for $\beta <0$ there are no mixed branches when $\mu<0$ and an infinite number when $\mu>0$. The behavior for they hyperbolic case $\beta>0$ is more subtle. For $\mu<0$ there is an infinite number of such branches, each opening horizontally. For $\mu>0$ there is still an infinite number of branches, with all but a finite number opening vertically. 

The mixed branches undergo bifurcations for values of $\mu$ where the linearization about the fixed point is resonant, i.e.\ where the linearization of system~\eqref{zero2iwODE} about the fixed point $\qq_{\mu-}$ has a long period that is $n$ times its short period, i.e.\ when 
$$
\frac{1}{\alpha\sqrt{2\nu}} = \frac{n}{1-\beta\nu}.
$$
This occurs at 
$$
\nu = \sqrt{\mu} 
= \frac{\alpha^2 n^2 + \beta - \sqrt{\alpha^4 n^4 + 2 n^2 \alpha^2 \beta}}{\beta^2}
=\frac{1}{2 \alpha^2 n^2 } - \frac{\beta}{2\alpha^4 n^4} + O(n^{-6}).
$$
To analyze the bifurcation, we perform a perturbation expansion. We let
$$
\nu = \frac{1}{2 \alpha^2 n^2 } - \frac{\beta}{2\alpha^4 n^4} + \frac{C}{2\alpha^4 n^4},
$$
with $C$ to be determined. We further assume that $Q_0 = \nu^{1/2} \tilde{Q}$ and assume the solution~\eqref{xsn} is small amplitude with $b = -\sqrt{\gamma} + O(\nu)$, in particular choosing
$$
b=-\sqrt{\gamma + \nu {\tilde B}^2}.
$$
Inserting these assumptions into the resonance relation~\eqref{mixedPeriodic}, we find that to eliminate terms of $O(n^{-4})$ requires 
$$
4 \tilde{B}^2 - \frac{3\beta}{\alpha} \tilde{Q}^2 = 48 C.
$$
This quadratic form equation mimics equation~\eqref{SNtype} for the short-periodic orbits. The case $\beta<0$ case has elliptical families of solutions only when $C>0$, where as the case $\beta>0$ has as its solutions hyperbolas which undergo a reconnection at $C=0$. This confirms the elliptical and hyperbolic configurations of the branches in Figure~\ref{fig:SN}.

\begin{rem}
For the orbit to be periodic, it is sufficient that the ratio of the two frequencies be rational. However at the resonant bifurcation the Lyapunov center theorem fails to hold, and this has important consequences for the periodic orbits, as shown by the numerics below.
\end{rem}

\subsection{The effect of the the neglected terms}

The effect of the term $\Hhigh(q_2,I)$ in the Hamiltonian~\eqref{H02iw} is minor, at least for small initial conditions. The dynamics in $(q_1,p_1)$ remains a circular rotation with an $I$-dependent frequency, and the $(q_2,p_2)$ dynamics remain a Newtonian evolution with a potential that depends parametrically on $I$. The effect of the neglected term $R_{\infty}(q_1,q_2,p_1,p_2)$ is more significant as it destroys these qualitative properties and introduces nontrivial interactions between the two degrees of freedom. In particular, the structure of the normal form ignoring $R_{\infty}$ is such that the phase difference between the two oscillators is arbitrary. The simplest effect of $R_{\infty}$ is to select a phase difference. Its other effects will be described in detail in Section~\ref{sec:SNnumerics} where periodic orbits are calculated numerically.

\section{Numerical simulations of the ``full'' ODE and PDE systems}
\label{sec:Numerics}

Our analysis has proceeded as a sequence of simplifications: from PDE to ODE and then to normal-form ODE. Our numerical investigations will proceed in the opposite direction. We have already visualized the bifurcation diagram determined by our analysis of the normal form. We compare these with branches of numerically-computed periodic orbits of the ``full'' two degree-of-freedom ODE system~\eqref{Hreduced}.

\subsection{Numerical Calculations near the bifurcation at $\NHHa$}
\subsubsection*{Numerical continuation of periodic orbits of system~\eqref{Hreduced}}
\label{sec:HH1}

In this section we compare numerically computed periodic orbits of the Hamiltonian system~\eqref{Hreduced}  with the approximate periodic orbits obtained from the normal form equations, satisfying system~\eqref{ChowKim}. The ODE orbits are computed using an algorithm by Viswanath that combines aspects of the Poincar\'e-Lindstedt algorithm and Newton iteration~\cite{Vis:01}. Because the periodic orbits lie on one-parameter families, we use pseudo-arclength continuation to generate them. Due to the system's symmetries, all periodic orbits can be chosen such that $z_1(0)$ and $z_3(0)$ are real.

\begin{figure}[htb] 
   \centering
   \includegraphics[height=.3\textwidth]{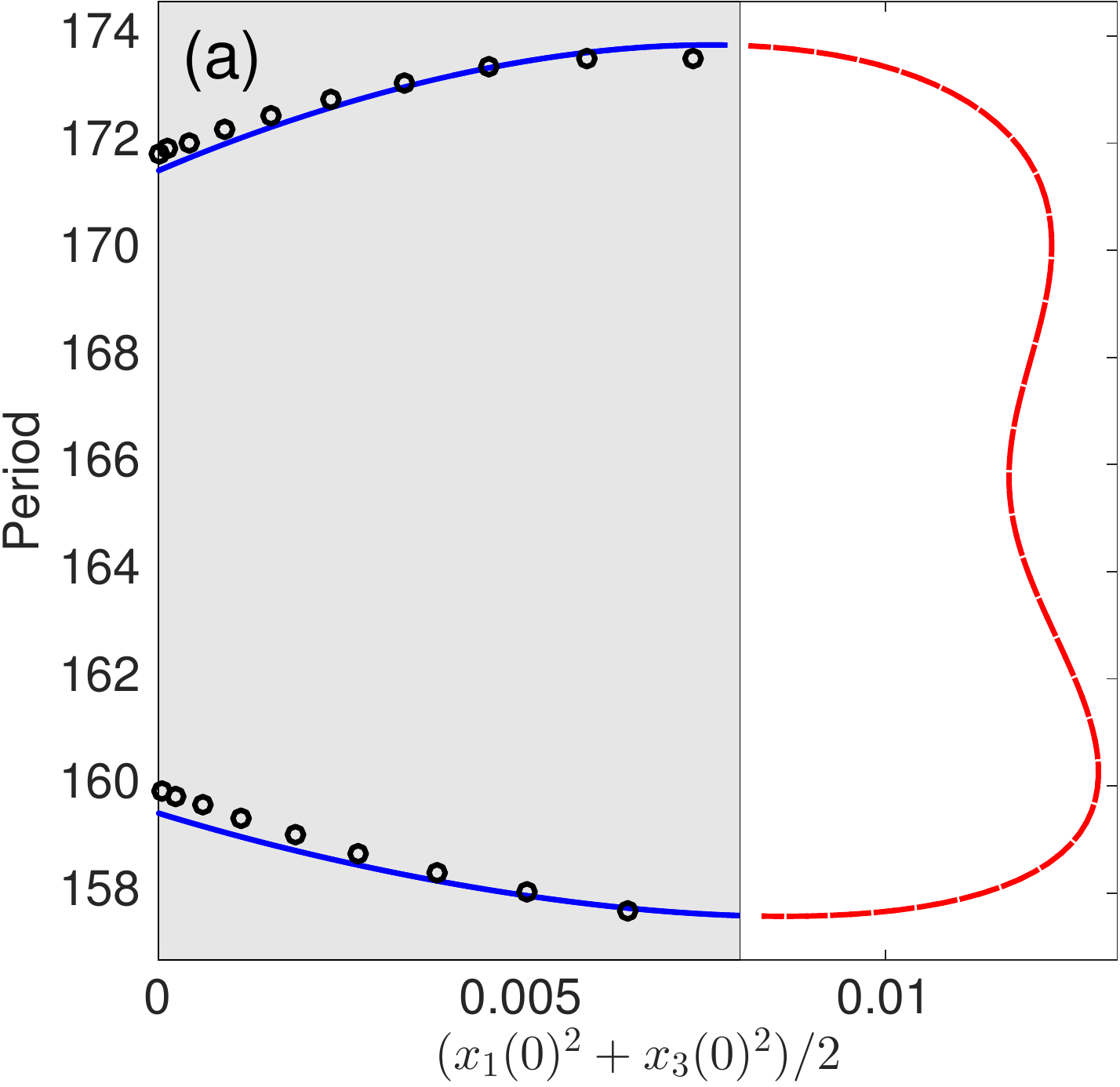}
   \includegraphics[height=.3\textwidth]{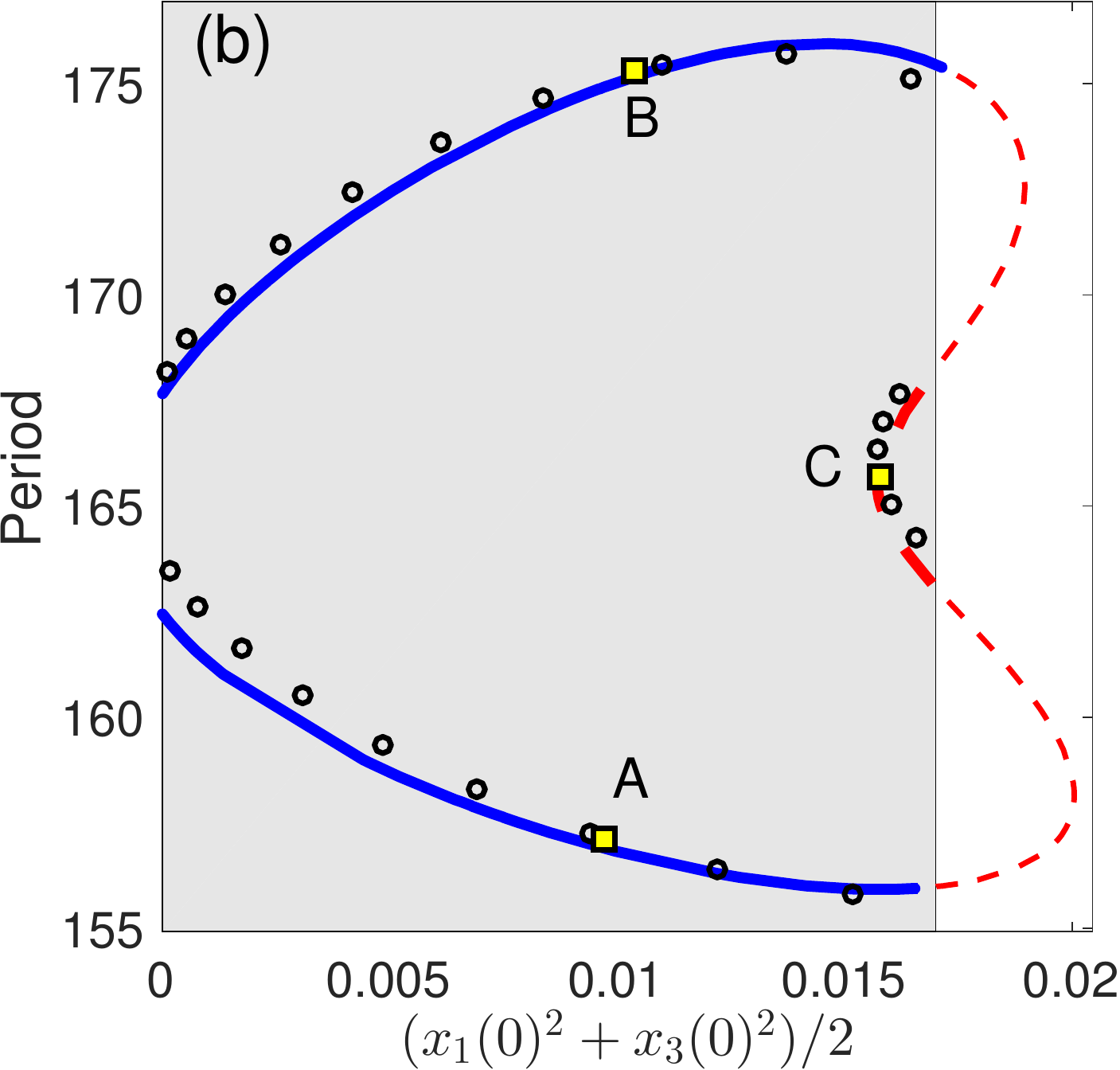}
   \includegraphics[height=.3\textwidth]{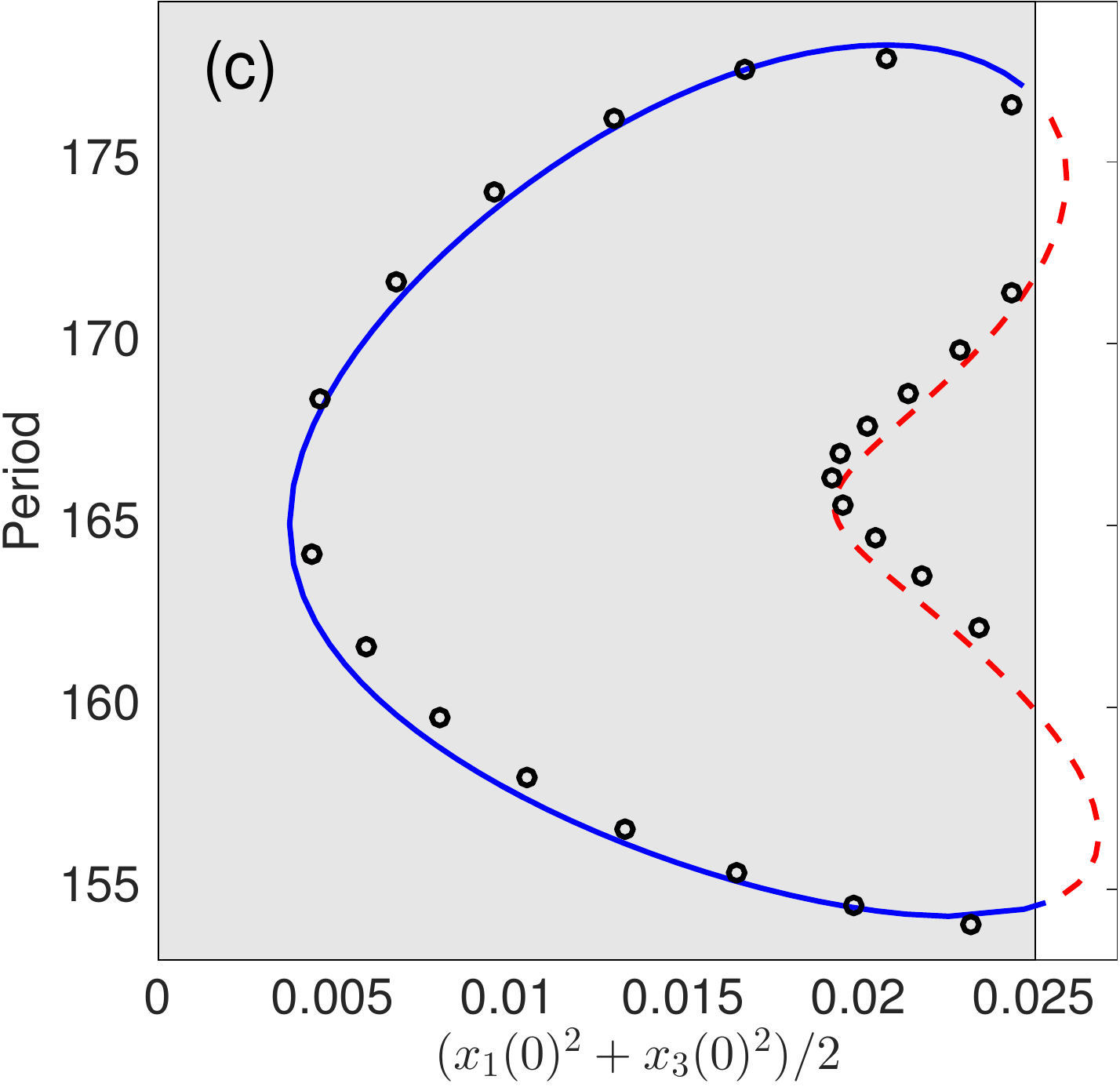}
   \caption{The period of solutions plotted as a function of the maximum magnitude of the numerically computed periodic orbit of system~\eqref{HcA} over one period, with the same parameter values as previous figures. Lines represent solutions of the Hamiltonian system~\eqref{Hreduced}, while black circles denote near-periodic orbits of NLS/GP.}
\label{fig:meanNorm}
\end{figure}

Figure~\ref{fig:meanNorm} shows the period of the numerical solution as a function of the initial condition magnitude,  analogous to Figure~\ref{fig:r_w1}. These solutions confirm the predictions based on the normalized system~\eqref{Htrunc}, but lack the symmetry present in the solutions of the truncated normal form equations.

\subsubsection*{Numerical calculation of near-periodic solutions to NLS/GP}

Exact time-quasiperiodic orbits of system~\eqref{NLS} on the whole real line have been shown not to exist~\cite{Sigal:1993wl}. Numerically, Marzuola find such solutions with a two-well potential slowly lose energy to radiation~\cite{Marzuola:2010}. Nonetheless the decay of energy can be very small, and we are able find solutions that return very close to their initial conditions and match closely the periodic solutions to the reduced system above.

We take initial conditions of the form~\eqref{ansatz}, with $c_j(0)\ll1$, run the PDE solver and then project the numerical solutions onto the eigenfunctions, giving approximate projections $c_j(t)$. Cancelling out the phase of $c_2(t)$ gives slowly evolving envelopes $z_1(t)$ and $z_3(t)$. Our procedure to find near-periodic orbits works entirely with these time series and not with the full solution, and thus relies on the fact that the numerical solution is well-described by the solution ansatz. 

We note that the periodic orbits to the reduced ODE system are all symmetric across both the real and imaginary $z_j$-axes, and that the solution components $z_1$ and $z_3$ are always either in-phase or $180^\circ$ out of phase. Thus at instants $t_0$ when $z_1(t_0)\in \RR$, so is $z_3(t_0)$, and, similarly, both are purely imaginary at the same instants. We use this observation to construct a shooting procedure. Starting with real-valued initial conditions, integrate the PDE until a time $\tau$ such that $z_1(\tau)$ is purely imaginary. Then use a root finding algorithm to adjust the initial conditions such that $z_3(\tau)$ is purely imaginary to some small tolerance. The computed solution is then one quarter of a periodic orbit. Depending on the shape of the periodic orbits, it is sometimes numerically preferable to start with purely imaginary initial data and integrate until the solutions are real, or else to run the simulation until $z_3$ is imaginary and adjust until $z_1$ is also imaginary. Some periodic orbits discussed in later sections have fewer symmetries, so we calculate half-periods instead of quarter-periods.

The program to implement this scheme numerically solves the NLS/GP system using a code written by T. Dohnal. It uses operator splitting to separate the stiff linear terms from the non-stiff nonlinear term, and perfectly matched layers to absorb outgoing radiation at the boundaries~\cite{Dohnal:2007,Kennedy:2003}.

While the PDE system of course loses energy to radiation escaping the computational domain, we find at these amplitudes that the numerical solution $(z_1,z_3)$ returns to within $10^{-8}$ of the initial condition. Using this initial condition for a long numerical integration, we find that the solution remains close to periodic for times of the order of thousands. The black circles in Figure~\ref{fig:meanNorm}  indicate approximate periodic orbits obtained in this way, showing good qualitative agreement.

\begin{figure}[htbp]
   \centering
   \includegraphics[width=2.5in]{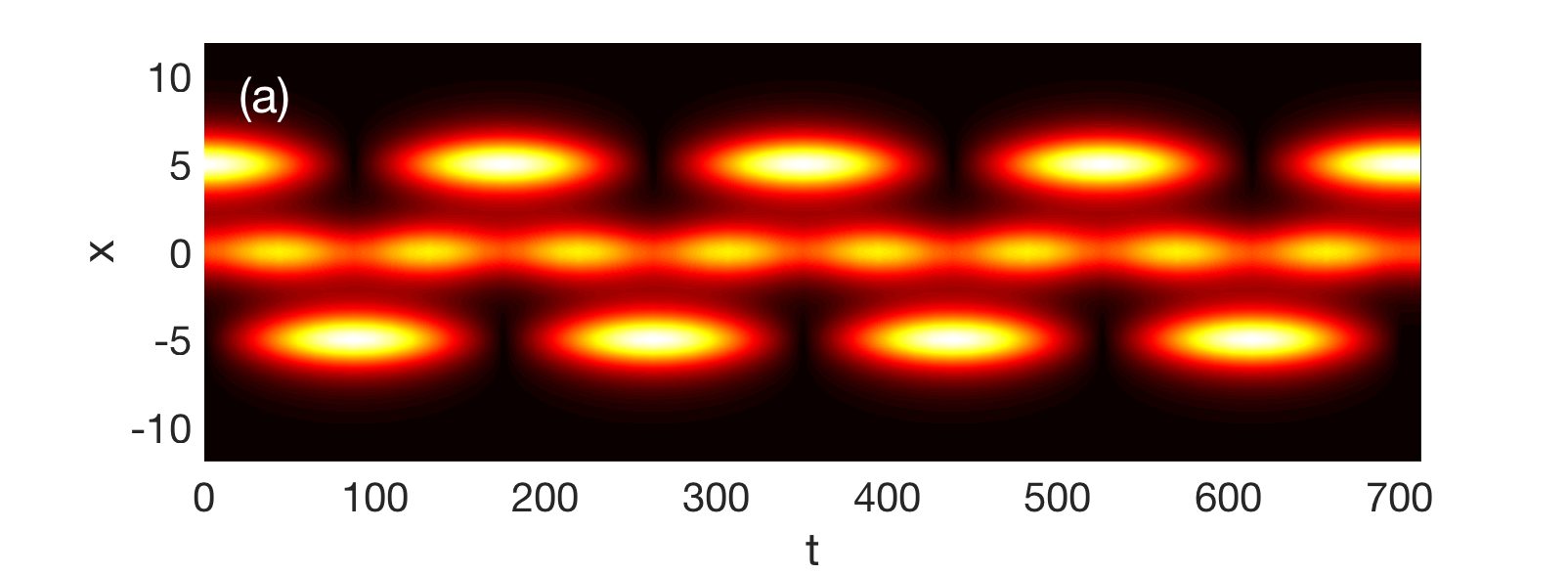}
   \includegraphics[width=2.5in]{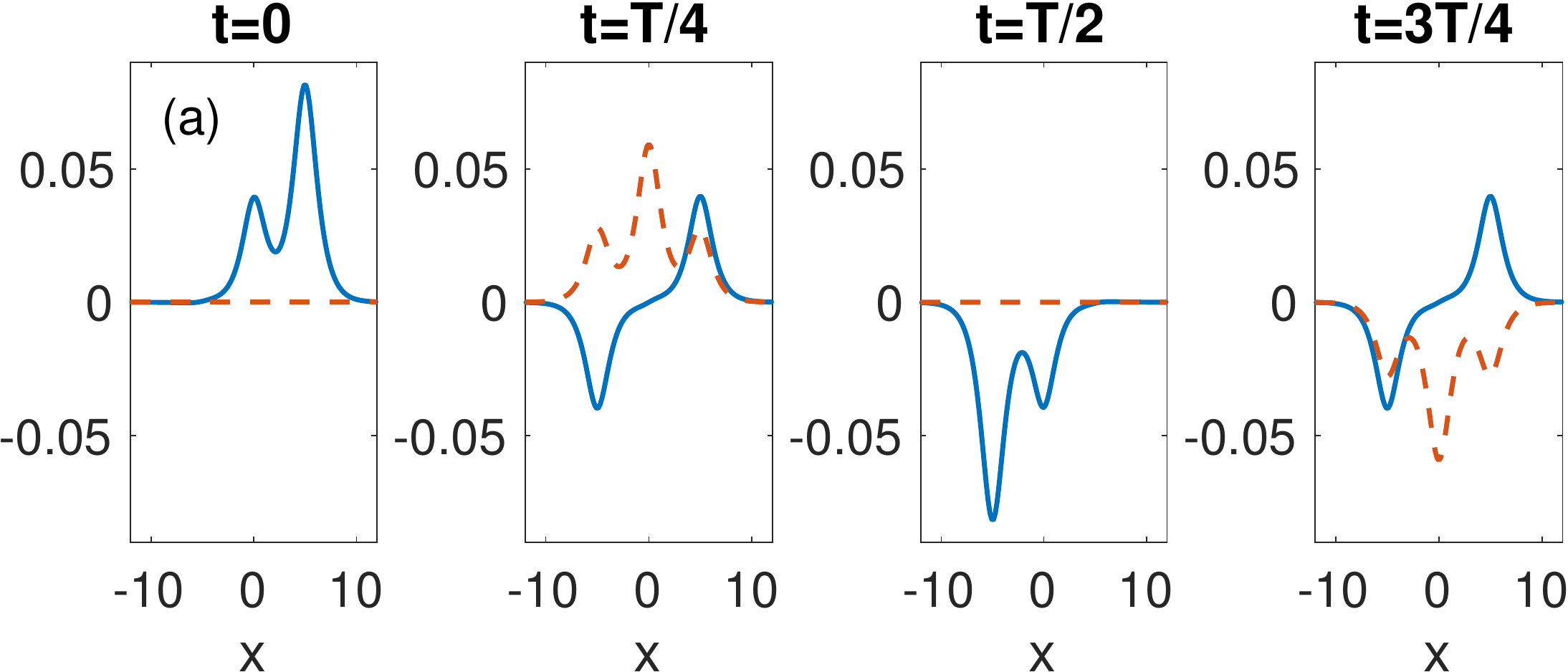}
   \includegraphics[width=1in]{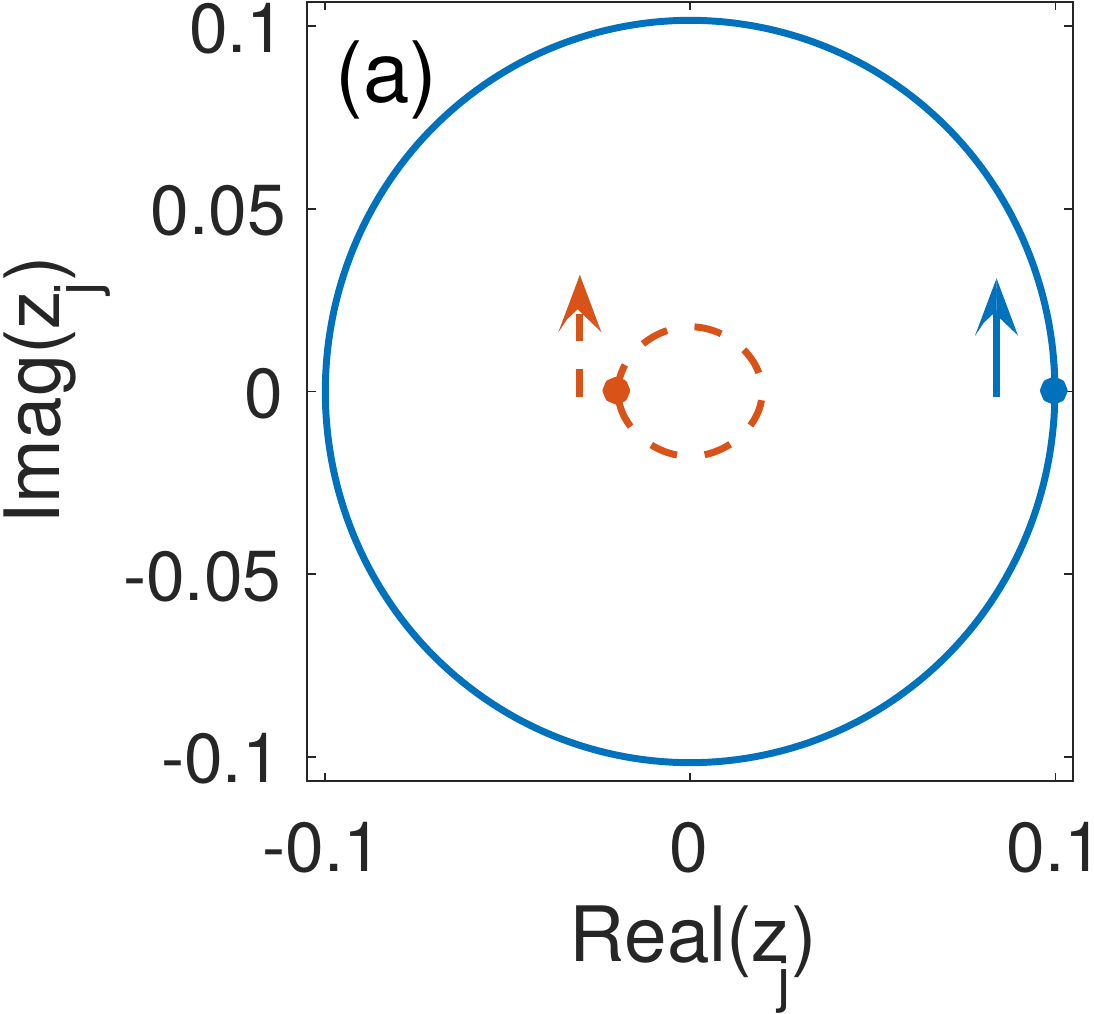} \\
   \includegraphics[width=2.5in]{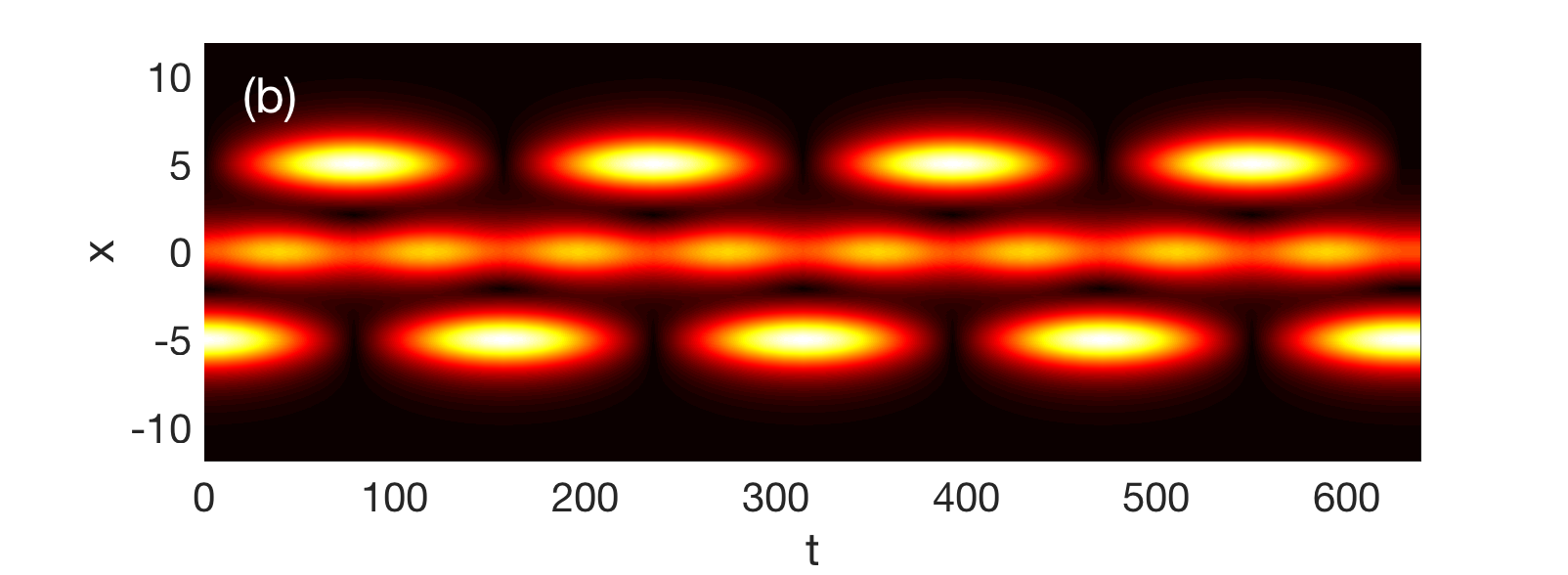}
   \includegraphics[width=2.5in]{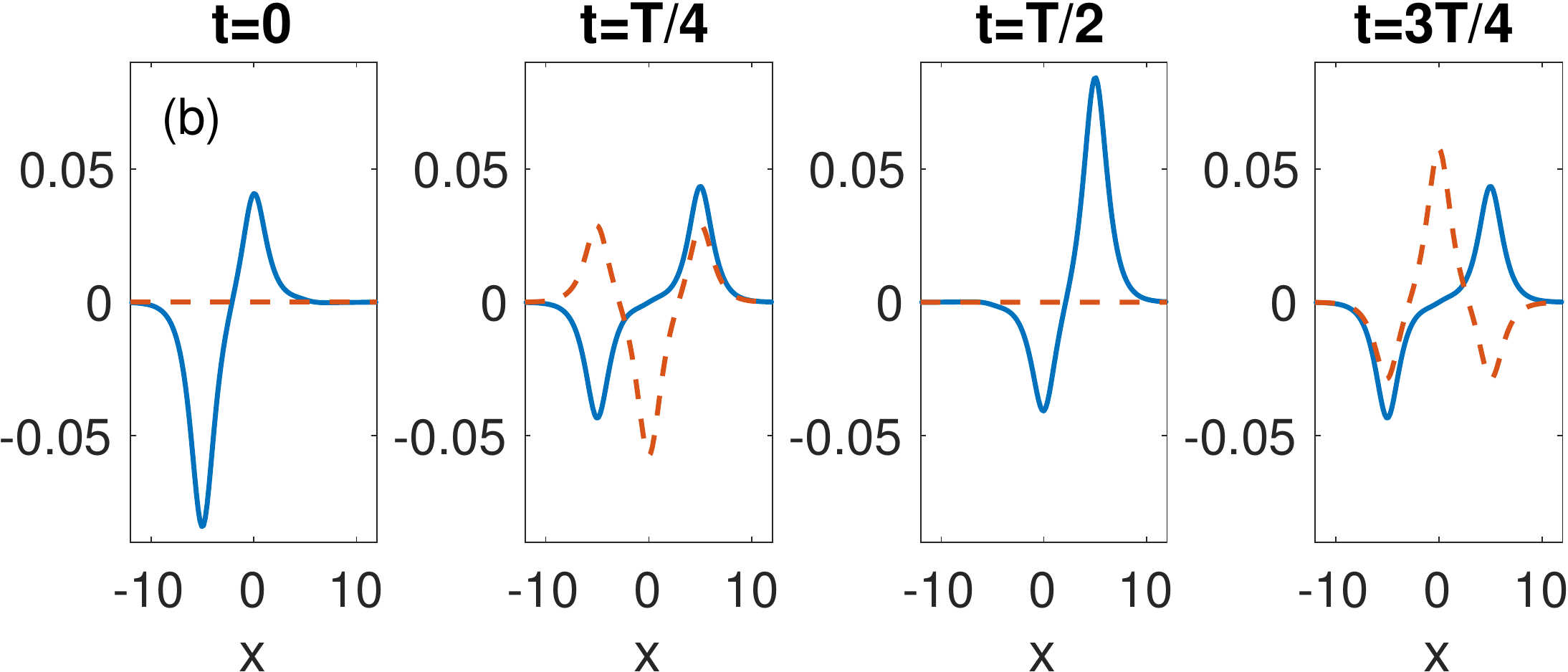}
   \includegraphics[width=1in]{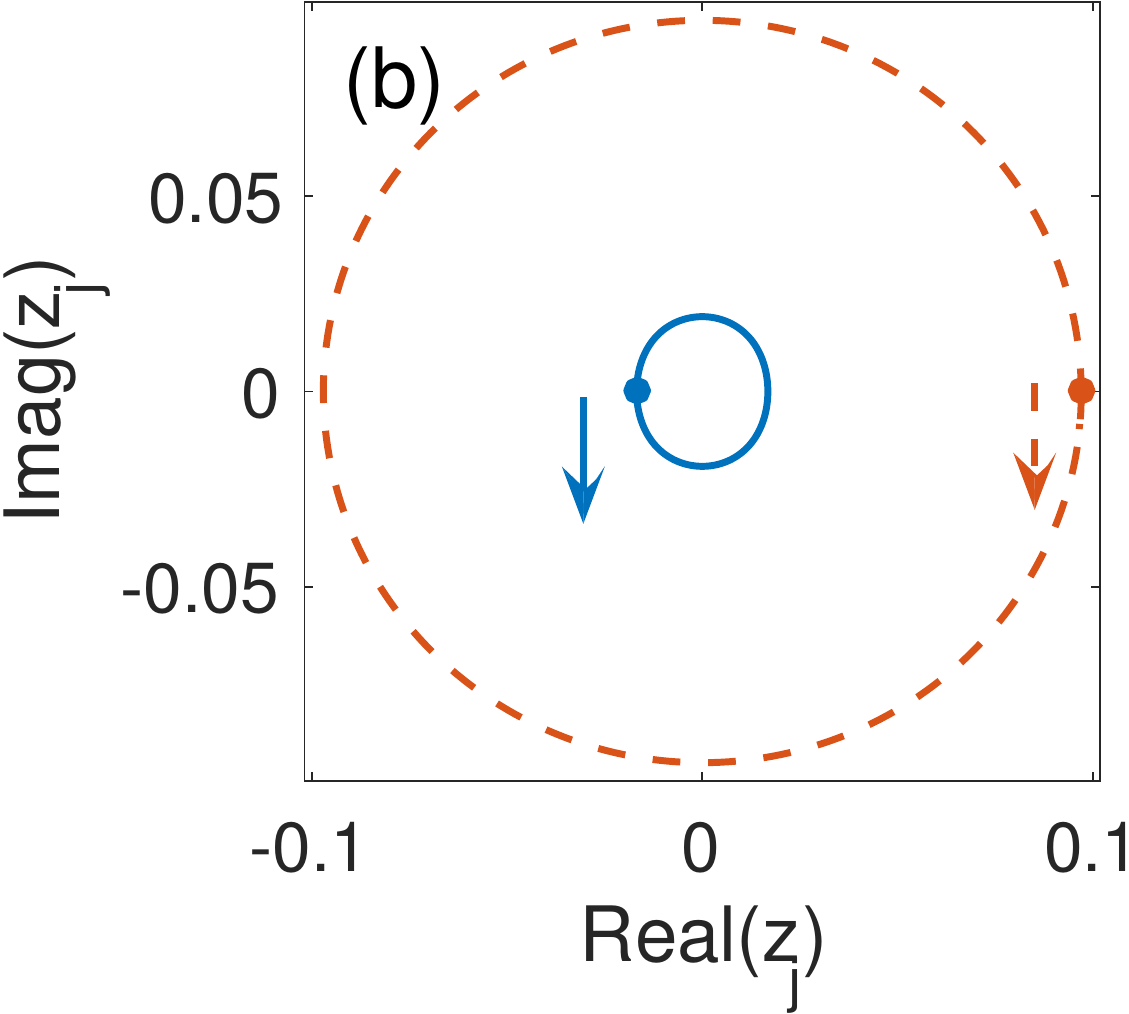} \\
   \includegraphics[width=2.5in]{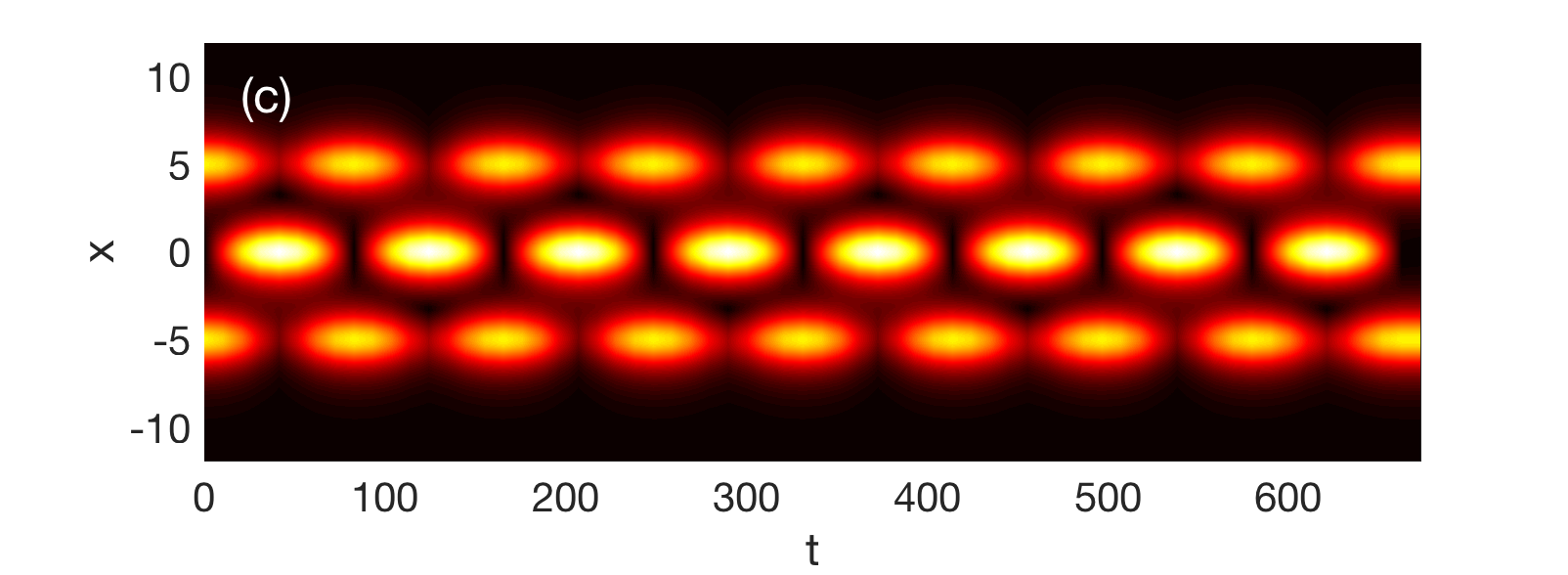}
   \includegraphics[width=2.5in]{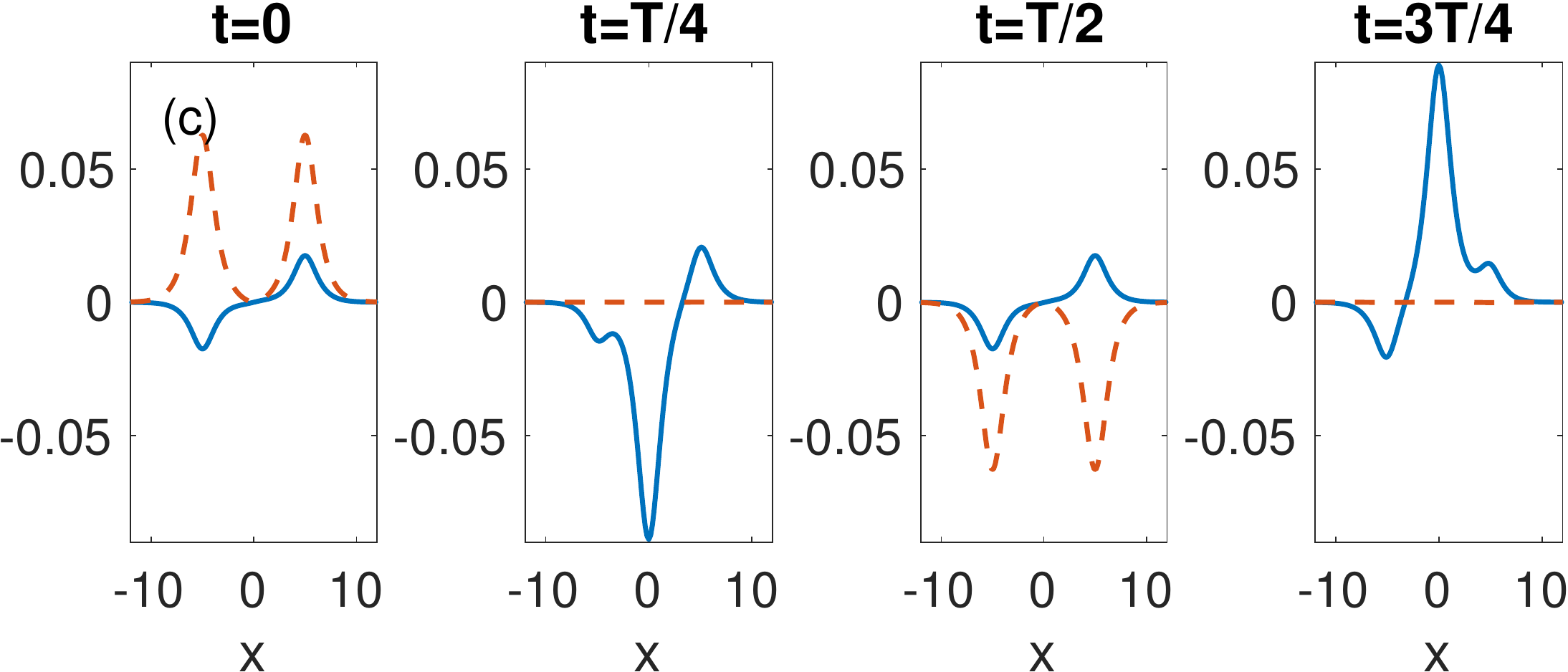}
   \includegraphics[width=1in]{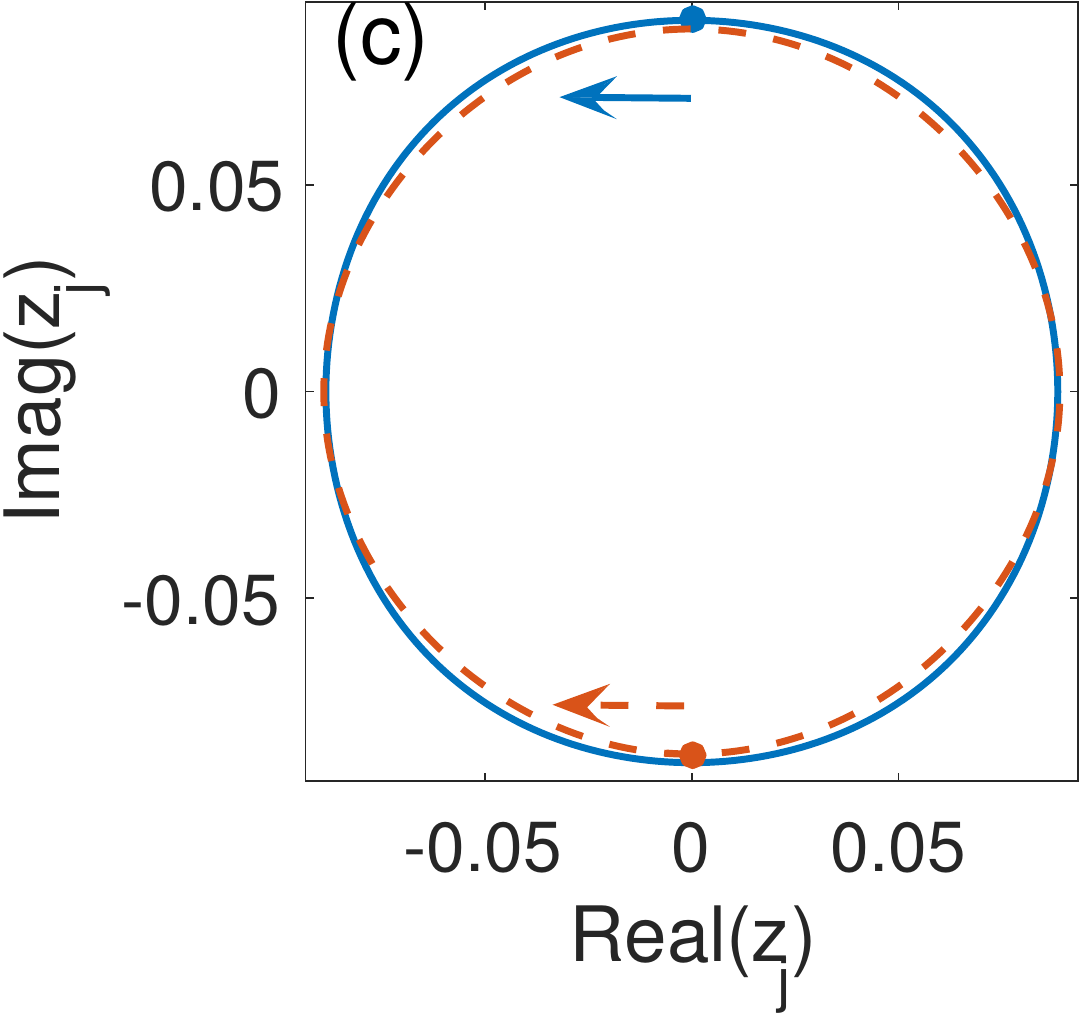} 
   \caption{Left: The illumination patterns of the three periodic orbits marked `A', `B', and `C' in Figure~\ref{fig:meanNorm} over four complete periods. Center: The real and imaginary parts of the solution normalized such that the projection onto $U_2$ is always real and non-negative. Right: The variables $z_1$ (solid, blue) and $z_3$ (dashed, red) calculated from the numerical solution.}
\label{fig:PDEnumerics}
\end{figure}

Figure~\ref{fig:PDEnumerics} shows numerical solutions of NLS/GP corresponding to the solutions marked \textbf{A-C} in Figure~\ref{fig:meanNorm}(b). These three solutions are chosen because they have near complete cancellation leading to dark spots in the PDE solution field. In cases~\textbf{A} and~\textbf{B}, these dark spots occur in the waveguides on the edge whereas in~\textbf{C}, the dark spot occurs in the middle waveguide.

\subsection{Computation of branches arising at $\NHHb$}
\label{sec:HH2}
We turn our attention to the behavior near the bifurcations at $\NHHb$ and $\cNHHb$. Considering first the dynamics of the reduced ODE system with $N - \NHHb$ positive, but small, we find a branch of periodic orbits that shrinks to a point as $N \searrow \NHHb$; an elliptical HH bifurcation. Our numerical experiments show that the periodic orbits along these branches reach their maximum amplitudes when $z_1$ and $z_3$ are purely imaginary. We therefore set the phase of our orbits such that $(z_1(0),z_3(0)) = i (y_1, y_3)$, and use $y_1$ and $y_3$ in the subsequent plots.

\begin{figure}[htbp] 
   \centering
   \includegraphics[width=3in]{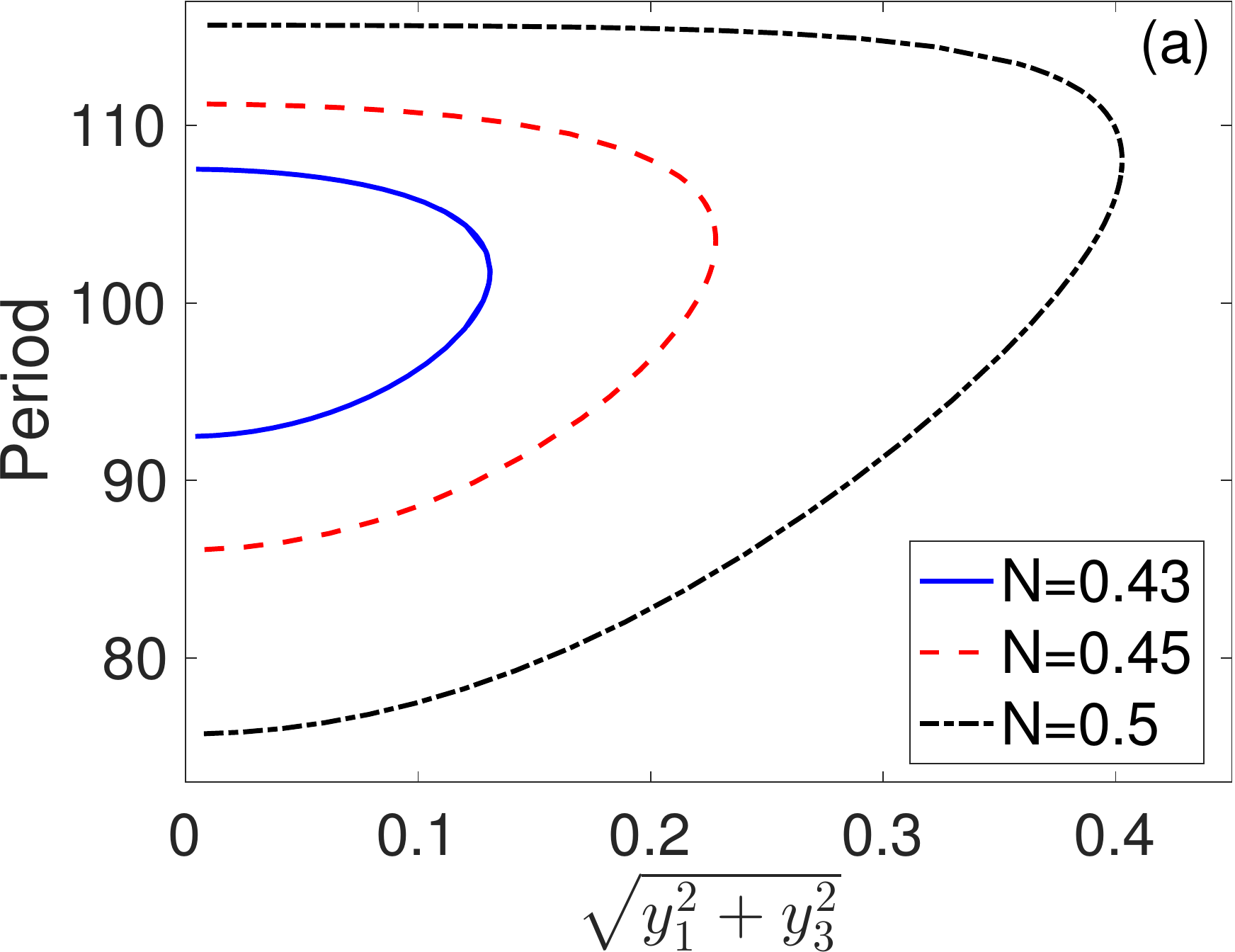} 
   \includegraphics[width=3in]{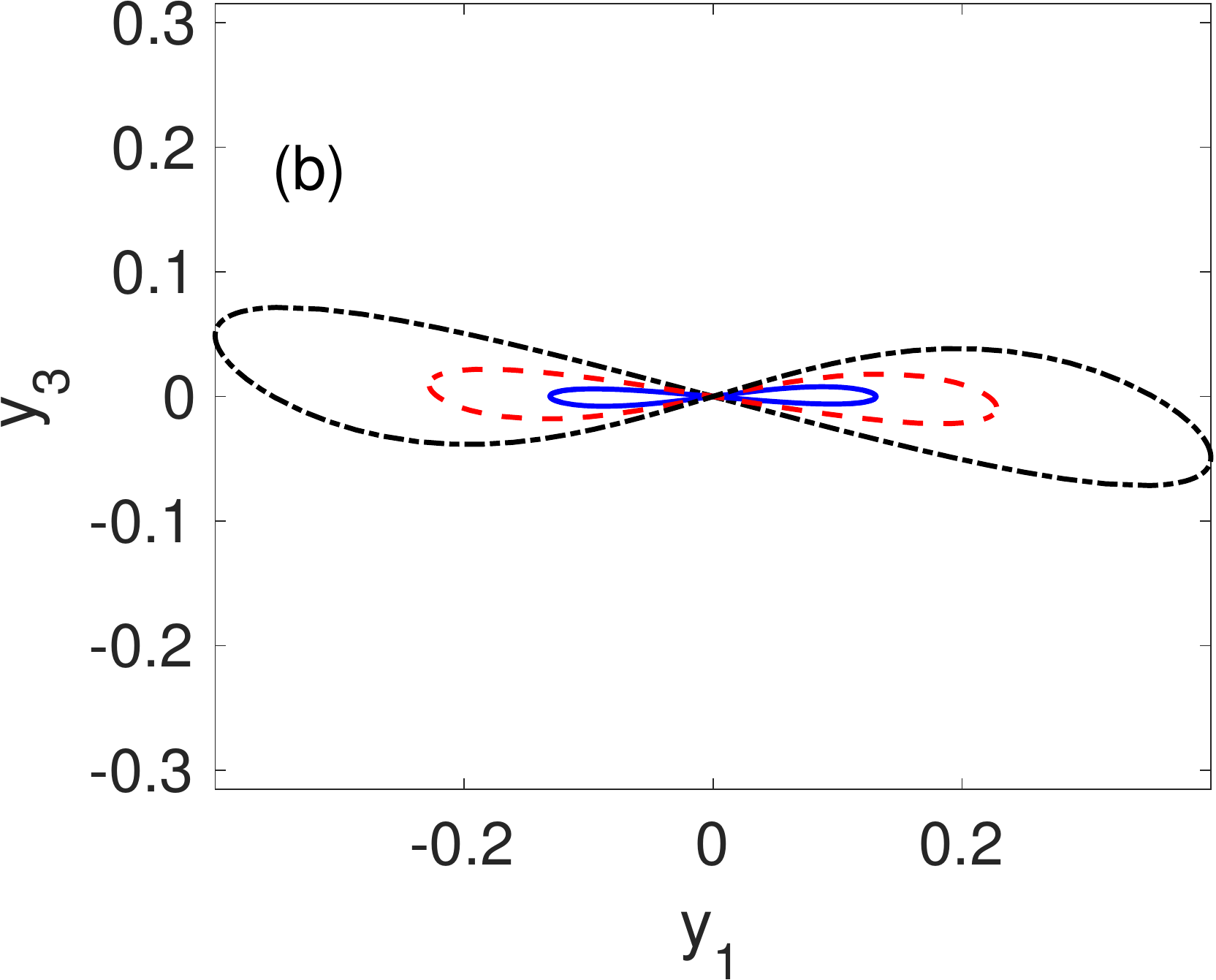} 
   \caption{Branches of periodic orbits of the reduced system that arise in the HH bifurcation at $\NHHb$. Note that because of symmetries, the point on the curve at $(y_1,y_3)$ and $(-y_1,-y_3)$ represent the same periodic orbit.}
\label{fig:HH2ODE}
\end{figure}

\begin{figure}[htbp] 
   \centering
   \includegraphics[width=3in]{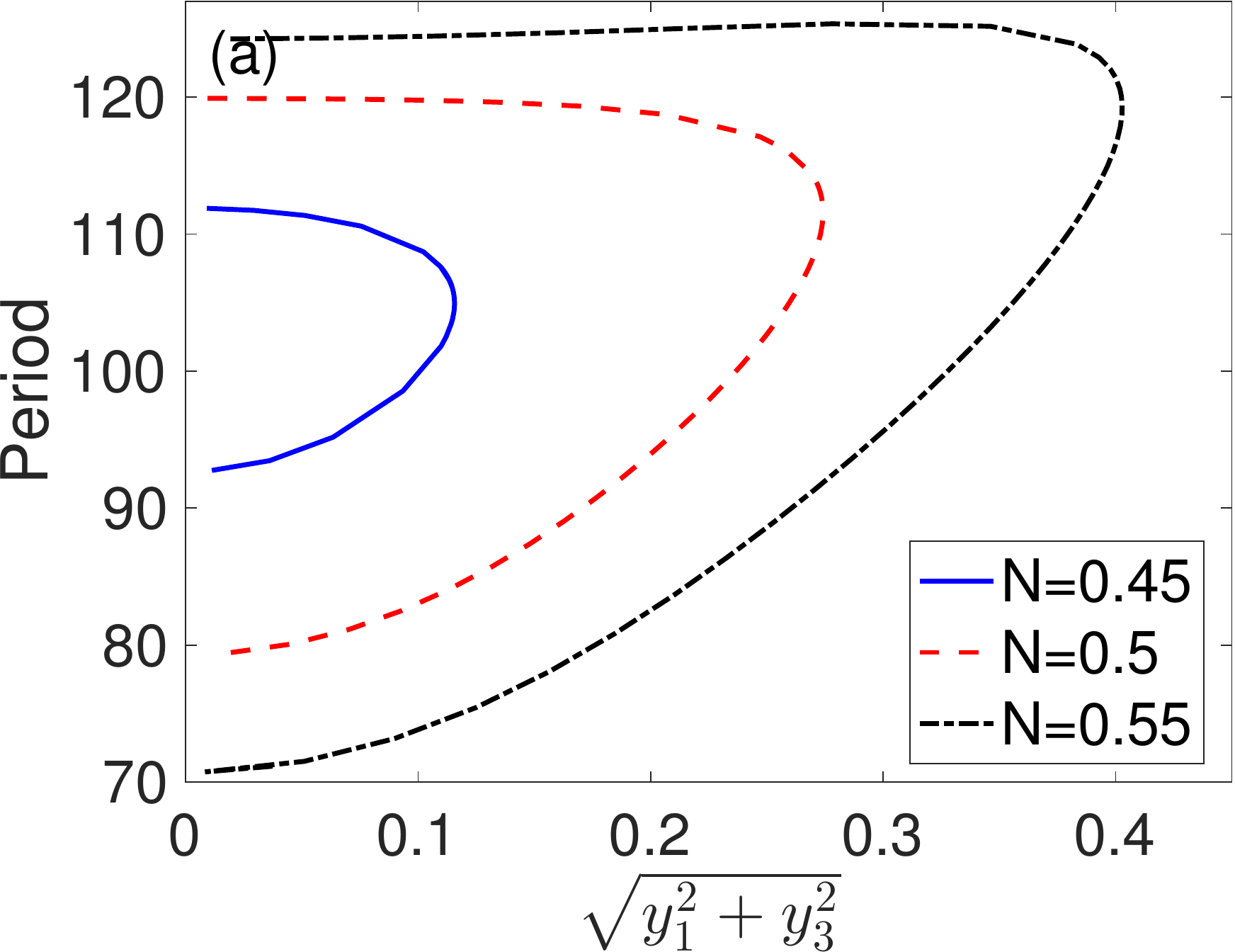} 
   \includegraphics[width=3in]{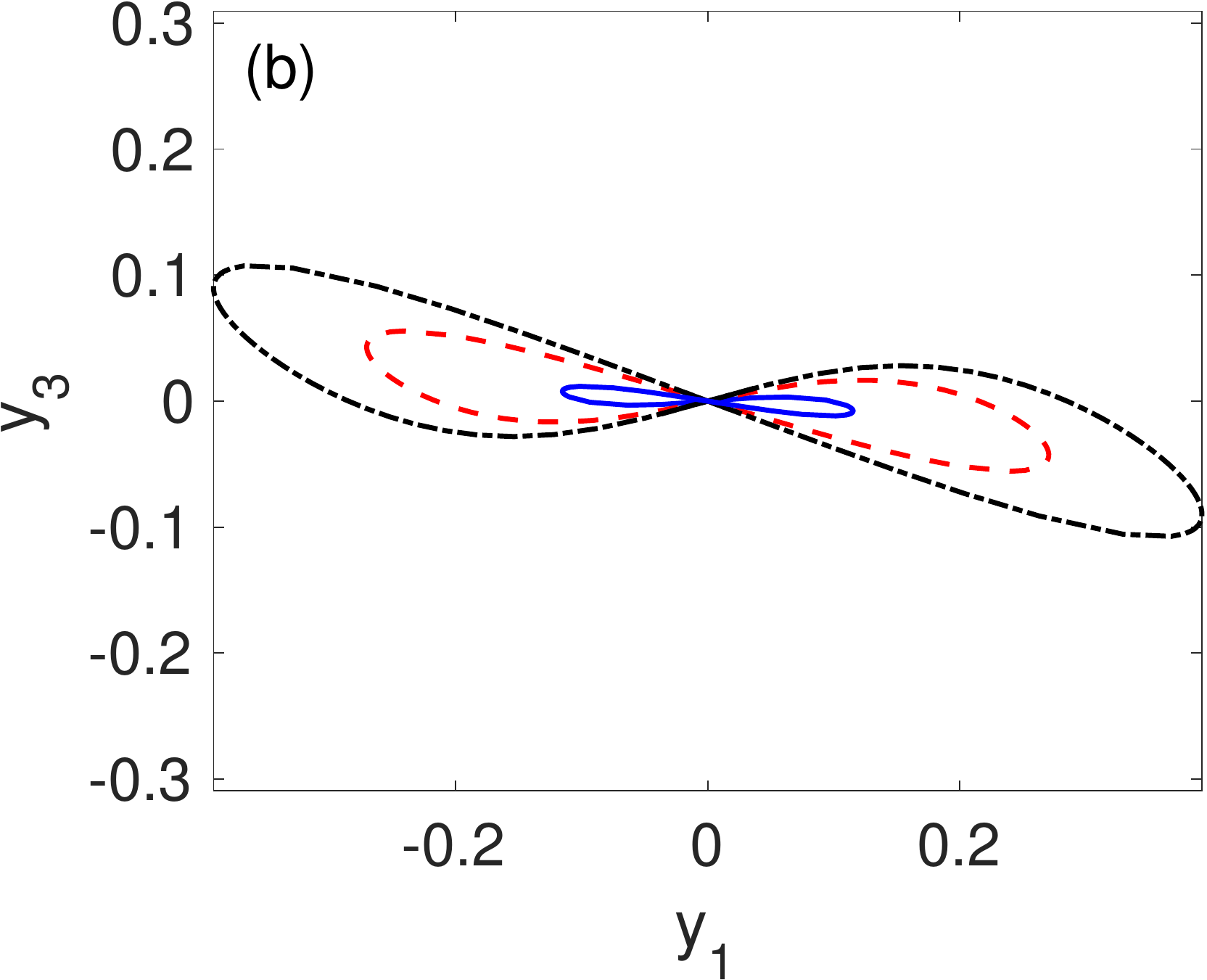} 
   \caption{Branches of periodic orbits of the NLS/GP system that arise in the HH bifurcation at $\cNHHb$.}
\label{fig:HH2PDE}
\end{figure}

Figure~\ref{fig:HH2ODE} shows the new branch of Lyapunov periodic orbits that arises as $N$ increases across the bifurcation value $\NHHb$. We show both the analog of Figure~\ref{fig:hyperbola}b, and also $y_1$ vs.\ $y_3$. Figure~\ref{fig:HH2PDE} is the equivalent for the PDE. Note we use different values of $N$ and $\cN$ in the two figures, since the bifurcation values are not exactly the same.

In the above computations, the total amplitude is close to the bifurcation value, and the values of $\abs{z_1(t)}^2$ and $\abs{z_3(t)}^2$ remain small in comparison with $N$ or $\cN$, meaning that all periodic orbits along these branches are essentially small perturbations of the dipole mode $\mop$. Over the entire branches depicted in these bifurcation diagrams, $z_1$ and $z_3$ vary nearly harmonically as their evolution is only weakly nonlinear.

\subsection{RPOs emanating from RFPs on the $\cBe$}

The two RFPs $\opo(N)$ and $\mpm(N)$ are each stable for all values of $N$ and thus should each possess two families of Lyapunov periodic orbits. In fact, these have already been calculated. Each has one family within the even invariant subspace $\cBe$ itself, as seen in Figure~\ref{fig:evenSubspace}. Each also has a family of Lyapunov periodic orbits that extends out of this subspace, and these families have been computed in Figure~\ref{fig:meanNorm}. The points on these curve at the boundary between the shaded and unshaded regions are precisely the NNM solutions and the branches extending out from them are the other family of Lyapunov periodic orbits.

Of the other two RFPs on the $\cBe$ discussed in Section~\ref{sec:evenFixed}, RFP $\ppp$ has no imaginary eigenvalues and thus no Lyapunov periodic orbits, while RFP $\pop$ has two imaginary eigenvalues and thus one family of Lyapunov periodic orbits, which encircle this point within $\cBe$ as shown in Figure~\ref{fig:evenSubspace}.

\subsection{RPOs emanating from the asymmetric RFPs that arise in saddle-node bifurcations}
\label{sec:SNnumerics}

The remaining four RFPs arise in a pair of saddle-node bifurcations as described in Section~\ref{sec:asymFix}. The normal form analysis near these bifurcations---see Section~\ref{sec:saddlenode}---partially describes the branches of RPOs. The eigenvalue count in Figure~\ref{fig:9Modes} shows that the branches $\poo$ and $\ppo$ should be accomapanied by, respectively, two branches and one branch of Lyapunov periodic orbits. Similarly, the branches $\mpo(N)$ and $\mpp(N)$, have, respectively two branches and one branch of Lyapunov periodic orbits near the saddle-node bifurcation at $N_{\rm{a}2}$. In addition the branch $\mpo(N)$ first destabilizes and then restabilizes in a pair of HH bifurcations. 

The normal form argument of Section~\ref{sec:saddlenode} misses an important feature of the dynamics which we describe below.

\subsubsection{Lyapunov families through $\poo$ and $\ppo$}
\label{sec:ellipticSN}
\begin{figure}[htbp] %  figure placement: here, top, bottom, or page
   \centering
   \includegraphics[width=3in]{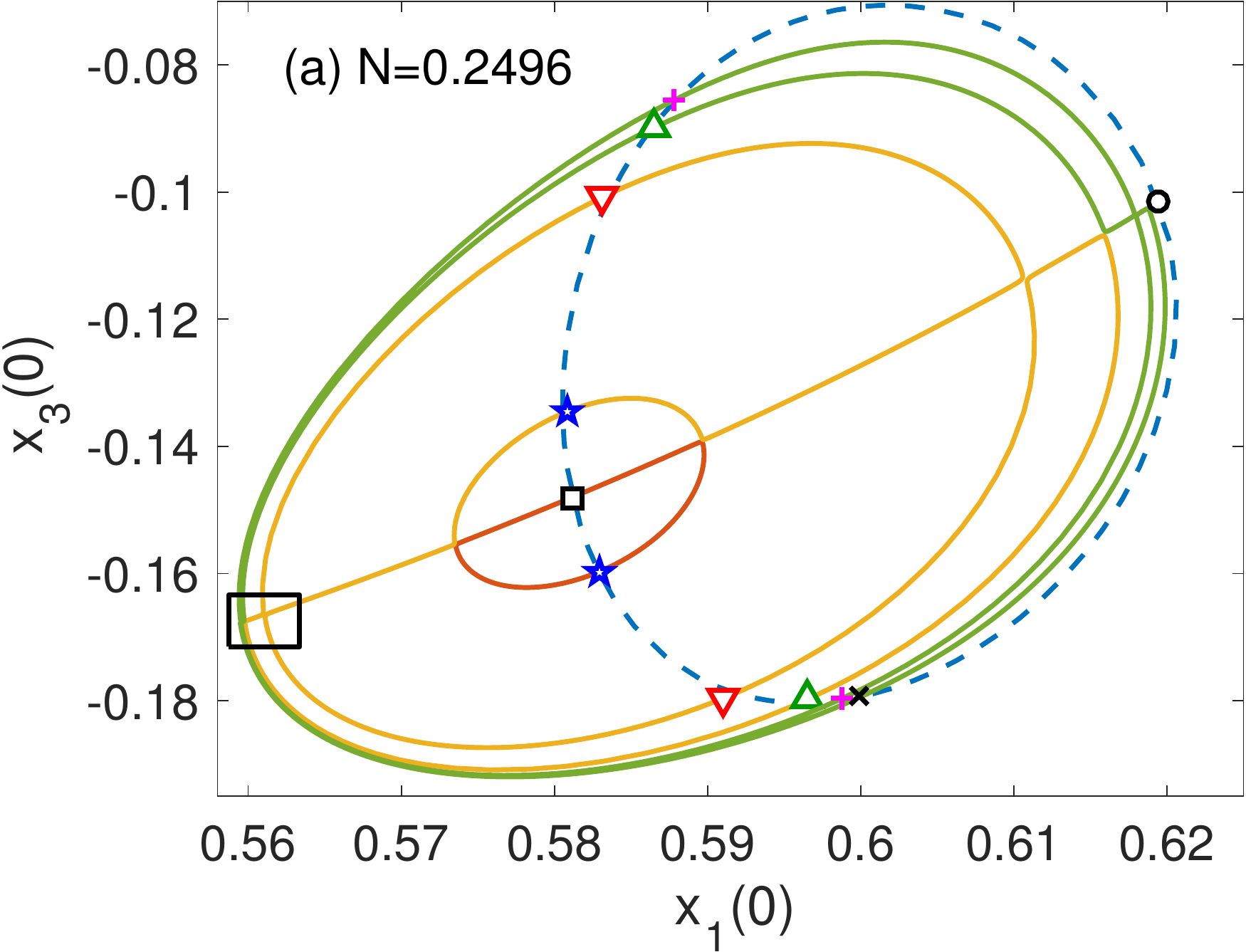} 
   \includegraphics[width=3in]{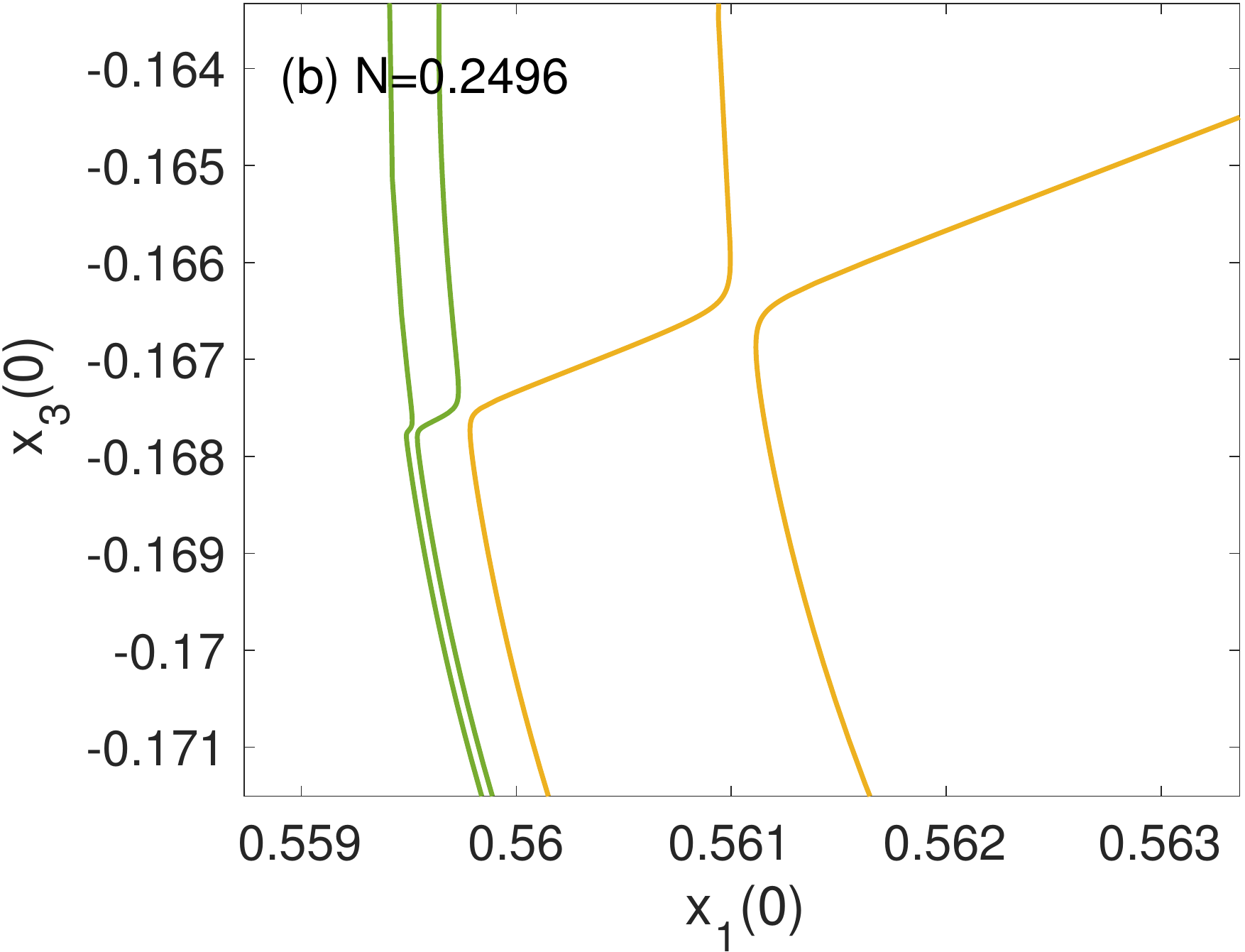}\\ 
   \includegraphics[width=3in]{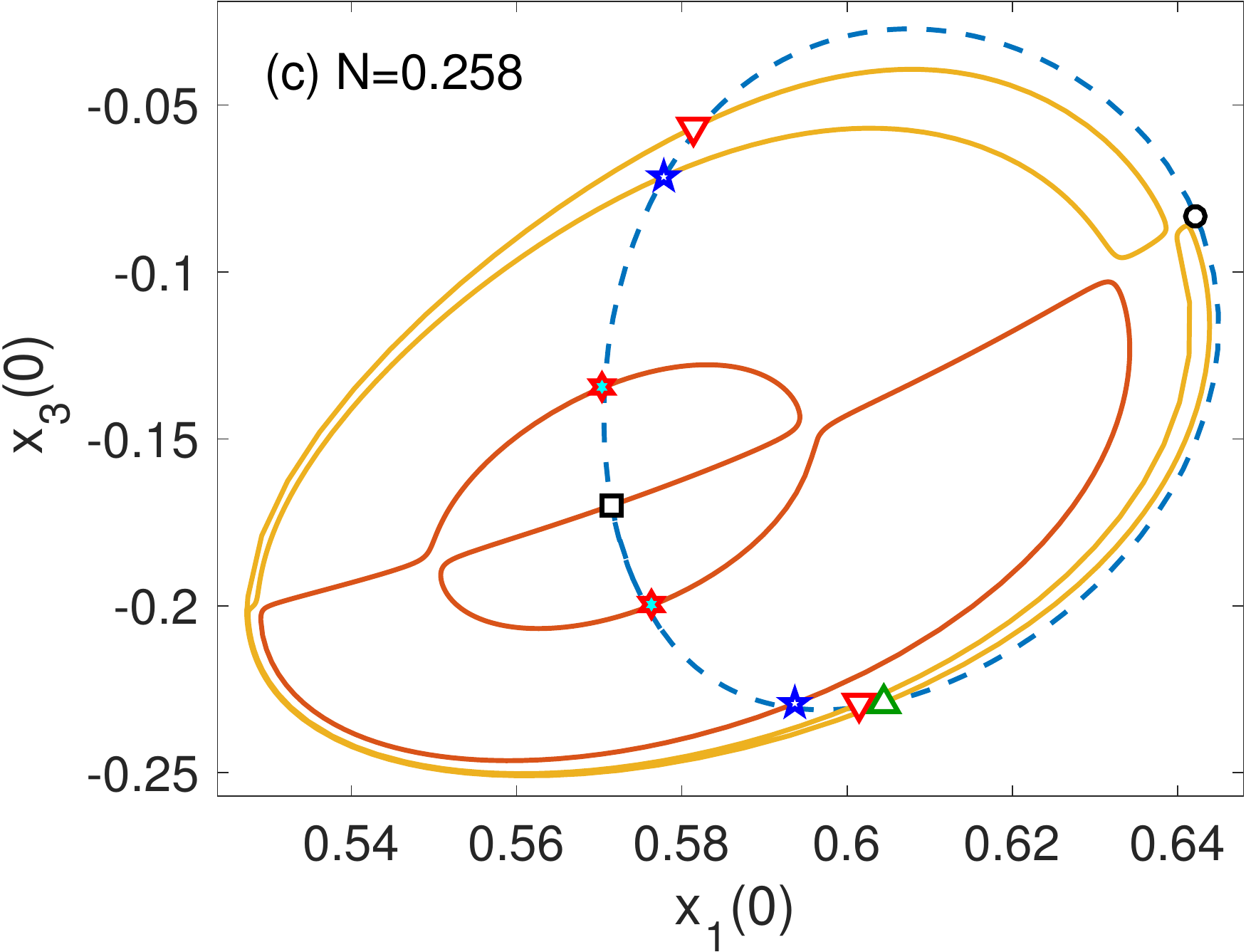}
   \includegraphics[width=3in]{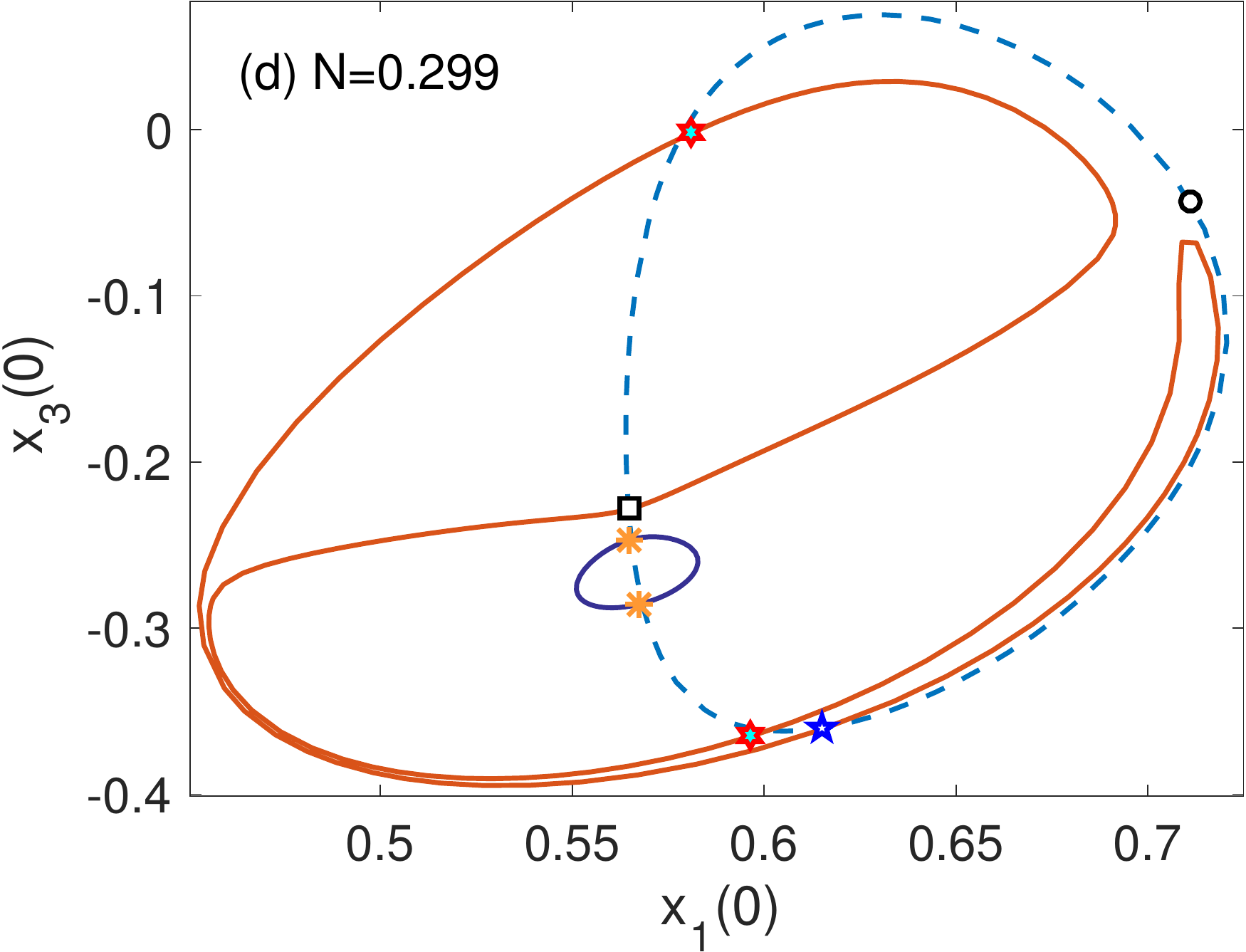}\\ 
   \includegraphics[width=3in]{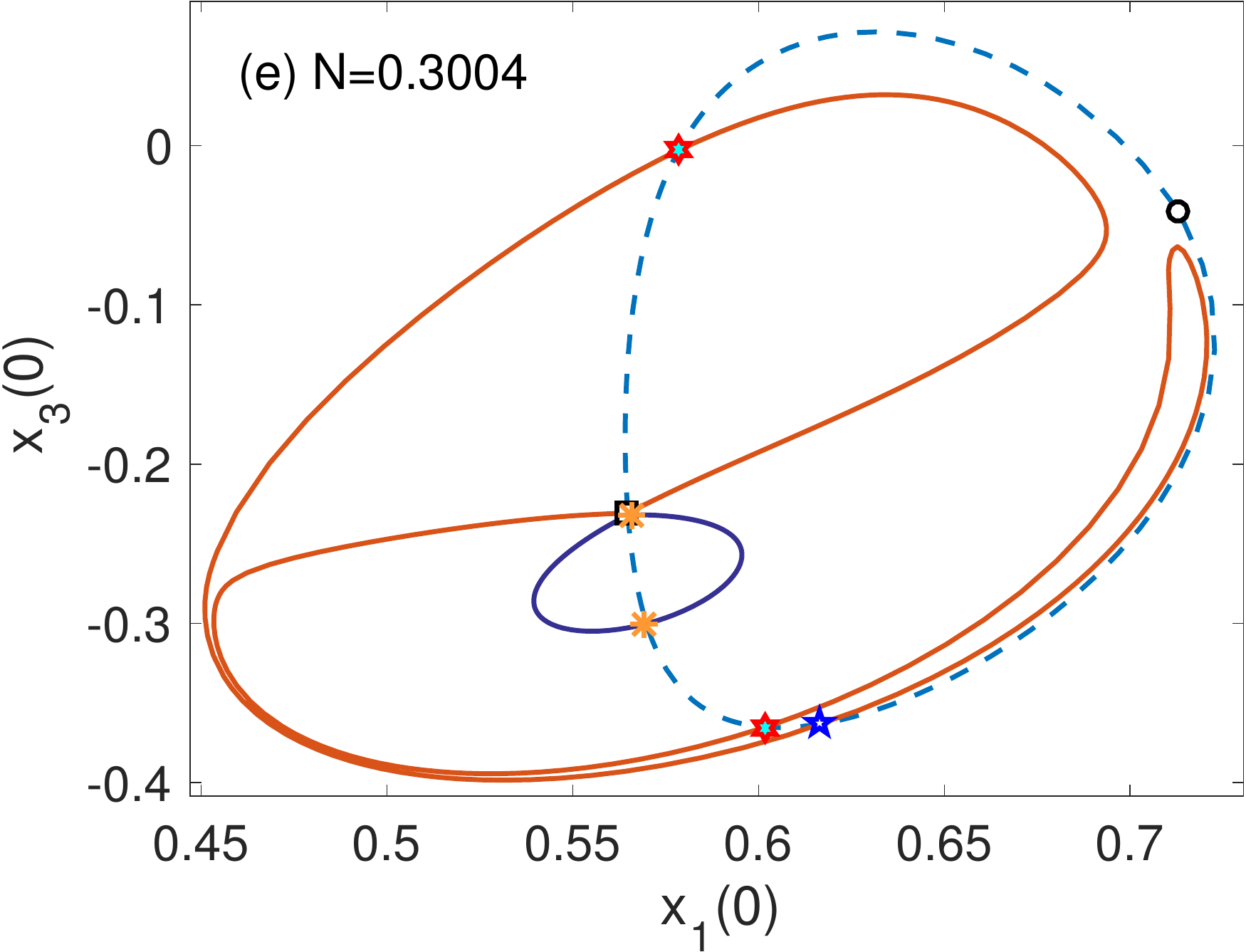} 
   \includegraphics[width=3in]{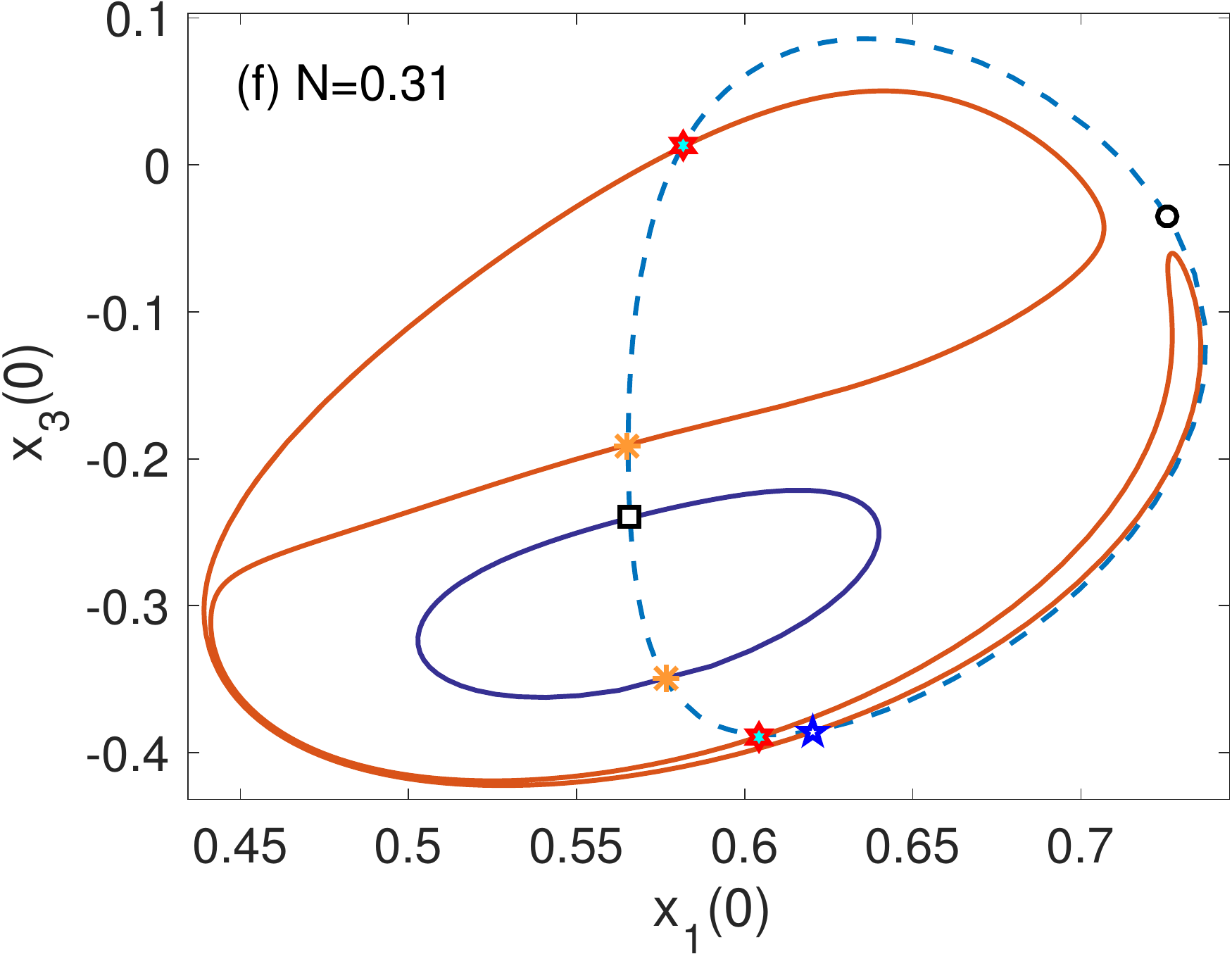} 
   \caption{(Color online) Branches of Lyapunov periodic orbits related to the points $\poo$ at various indicated values of $N$. Coloring of branches is consistent between the subfigures. The dashed curve shows the short-period orbits. (a) A value of $N_4<N<N_3$. (b) Magnification of the region outlined by a black rectangle, showing avoided intersections. (c) $N_3<N<N_2$, a kink and two 3-fold orbits have appeared on the long-period branch. (d) With $N$ slightly less than $N_2$ an isola has appeared containing two 2-fold orbits. (e) At $N$ even closer to $N_2$ the isola approaches the long-period branch at the point $\poo$. (f) For $N>N_2$, the point $\pop$ has swapped branches with a 2-fold orbit. Key: $\square$: $\poo$, $\circ$: $\ppo$, $\ast$: 2-fold orbits, $\medstarofdavid$: 3-fold, $\medstar$: 4-fold, $\triangledown$: 5-fold, $\triangle$: 6-fold, $+$: 7-fold, and $\times$: 8-fold. }
\label{fig:poo}
\end{figure}

We first consider the Lyapunov branches of periodic orbits through $\poo$ and $\ppo$, which appear in a saddle-node bifurcation at $N=N_{a1} \approx 0.246$. The graph in Figure~\ref{fig:poo}(a) has much in common with the branches of periodic orbits for the normal form depicted in Figure~\ref{fig:SN}(a). A closed loop of short-period orbits crosses through both fixed points. A nearly straight curve of long-period orbits extends between them, and there appear to be several concentric ellipses of mixed periodic orbits surrounding the fixed point $\poo$. Figure~\ref{fig:poo}(b) shows a magnified view of the small black rectangle at the lower left. It shows that the long-period and mixed periodic branches do not simply intersect, and cannot be considered as separate entities. Each apparent intersection is in fact a pseudo-intersection: the numerically generated branch follows the path of the long-period branch before taking a sharp turn and following the mixed-periodic branch. 

Figure~\ref{fig:poo}(a) shows three such long/mixed periodic branches, each drawn in a different color and each a simple closed curve. These branches \emph{do} intersect the short-periodic branches in a series of marked points. Since each point on these curve denotes the initial condition for a periodic orbit of an ODE with unique solutions, the two branches must pass through the same periodic orbit (or fixed point). Since the orbit on the long/mixed periodic branch has a longer period, this orbit must consist of an integer number $m$ of copies of the short-period orbit. We call such orbits $m$-fold orbits.

The mixed-periodic branches were shown in Section~\ref{sec:mixed} to arise in bifurcations when the long period is an integer multiple of the short period. The resonant values of $N$ at which this occurs for this system are provided in Table~\ref{table:poo}. How the branches of fast/periodic branches change at these bifurcations is explored in the computations summarized in the rest of Figure~\ref{fig:poo}.

\begin{table}[h]
\centering
\begin{tabular}{|c|c|c|c|c|c|c|}
\hline
Bifurcation & SN & $\ldots$ & {5:1} & {4:1} & {3:1} & {2:1} \\ \hline
$N_k$ & 0.2462 & $\ldots$ & 0.2475 & 0.2494 & 0.2565 & 0.3007 \\ \hline
\end{tabular}
\caption{Values of $N_k$ for which the long and short periods of the mode $\poo$ are in $k:1$ resonance.}
\label{table:poo}
\end{table}

Figure~\ref{fig:poo}(a) depicts the periodic orbits for a value of $N$ between $N_4$ and $N_3$. The long-period branch through this point, marked in red, intersects the short-period branch both at $\poo$ and, at the marked point, a 4-fold orbit. Similarly the next (yellow) curve contains a 4-fold orbit, two 5-fold orbits, and a 6-fold orbit, and the next (green) curve has a 6-fold, two 7-folds, and one 8-fold orbit. There are (presumably) an infinite number of such branches not depicted

As $N$ is further increased, these closed branches deform, and additional branches appear at the resonant values $N_k$. The behavior is different depending on the parity of $k$. No branches disappear as $N$ is increased, but we drop some from the figures for clarity of exposition. The behavior at $N_3$ is typical of resonances when $k$ is odd. The straight portion of long/mixed period branch develops a kink, and, with it, two additional intersections with the short-period branch; see Figure~\ref{fig:poo}(c) which correspond to 3-fold orbits. 

Typical behavior for even values of $k$ is shown next for $k=2$. At some critical value of $N$ just below $N_2$, an isola of periodic orbits bifurcates into existence. This branch intersects the short-period branch in two 2-fold orbits; Figure~\ref{fig:poo}(d). At $N=N_2$, this isola touches the branch $\poo$; Figure~\ref{fig:poo}(e). Finally for $N>N_2$, the point $\poo$ and the 2-fold orbit exchange branches. This is seen in the computation of the bifurcations at the $k:1$ resonances for other even values of $k$.

The bifurcations that occur in a neighborhood of $N\approx N_k$ where the two frequencies are resonant, is long studied, e.g.\ by Schmidt or Duistermaat~\cite{Schmidt:1974ha,Duistermaat:1984gd}. However, we do not know of a study of a sequence of such bifurcations as seen here.

\subsubsection{Branches through $\mpo$ and $\mpp$}
\label{sec:hyperbolicSN}
\begin{figure}[htbp] 
   \centering
   \includegraphics[width=3in]{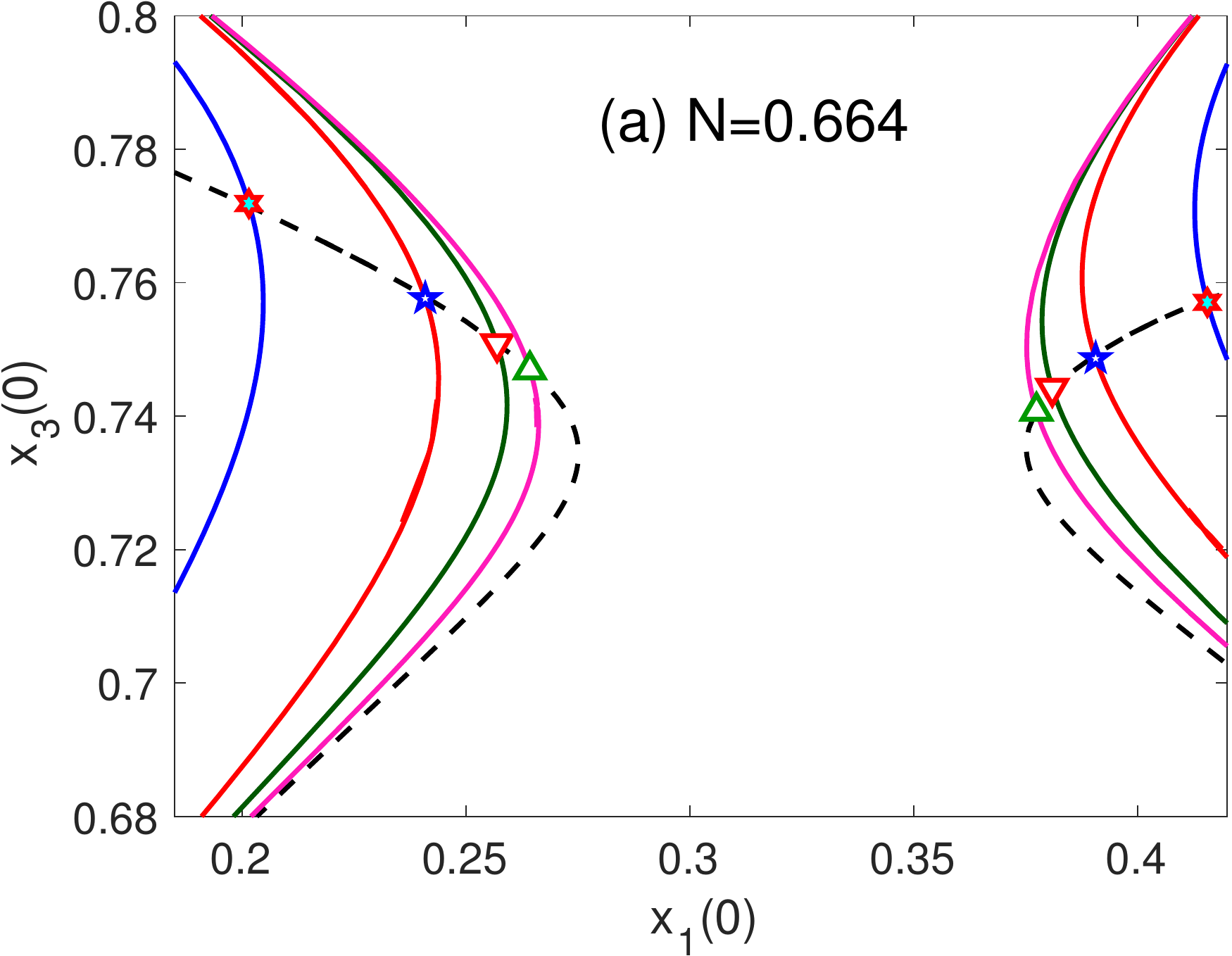} 
   \includegraphics[width=3in]{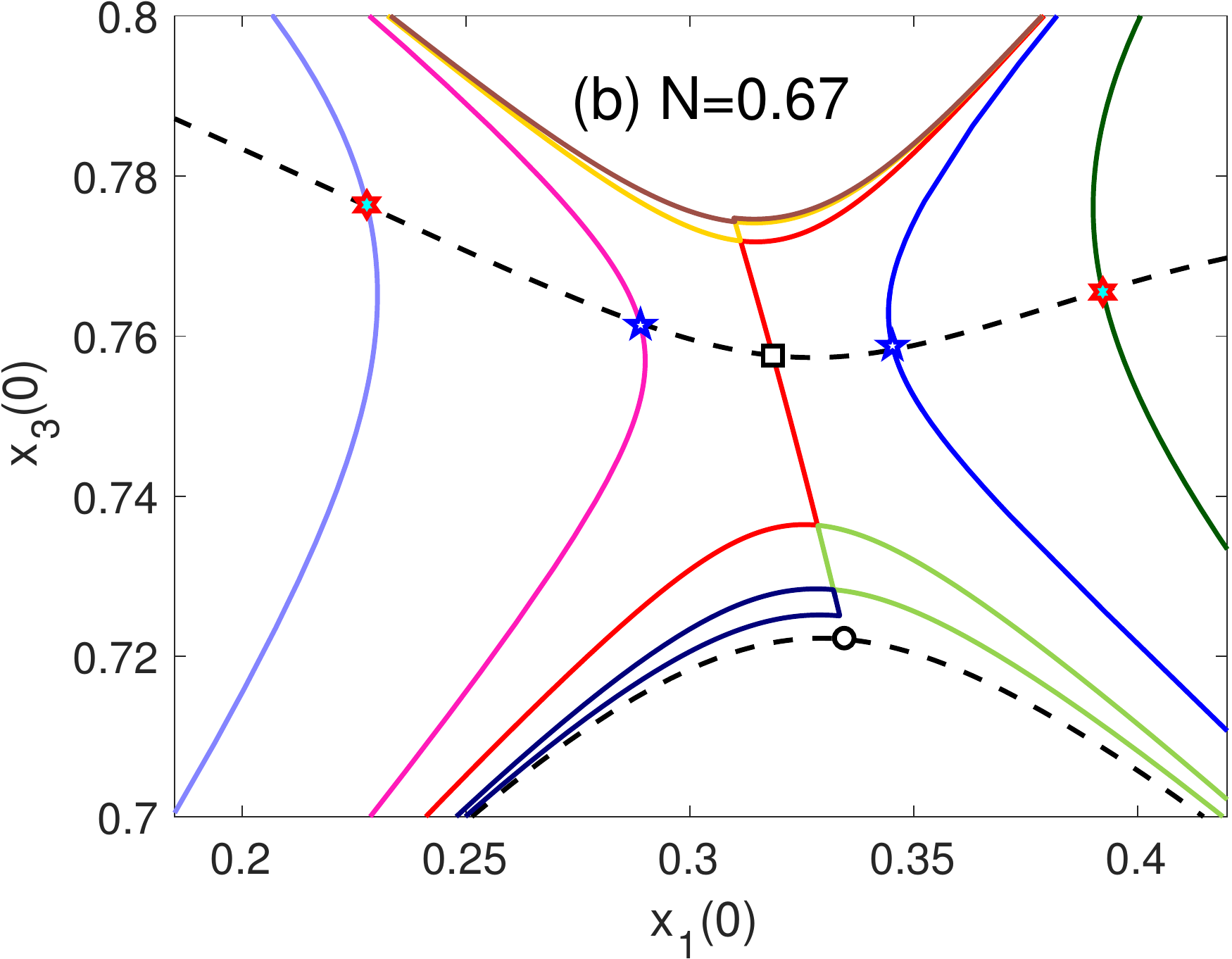} \\
   \includegraphics[width=3in]{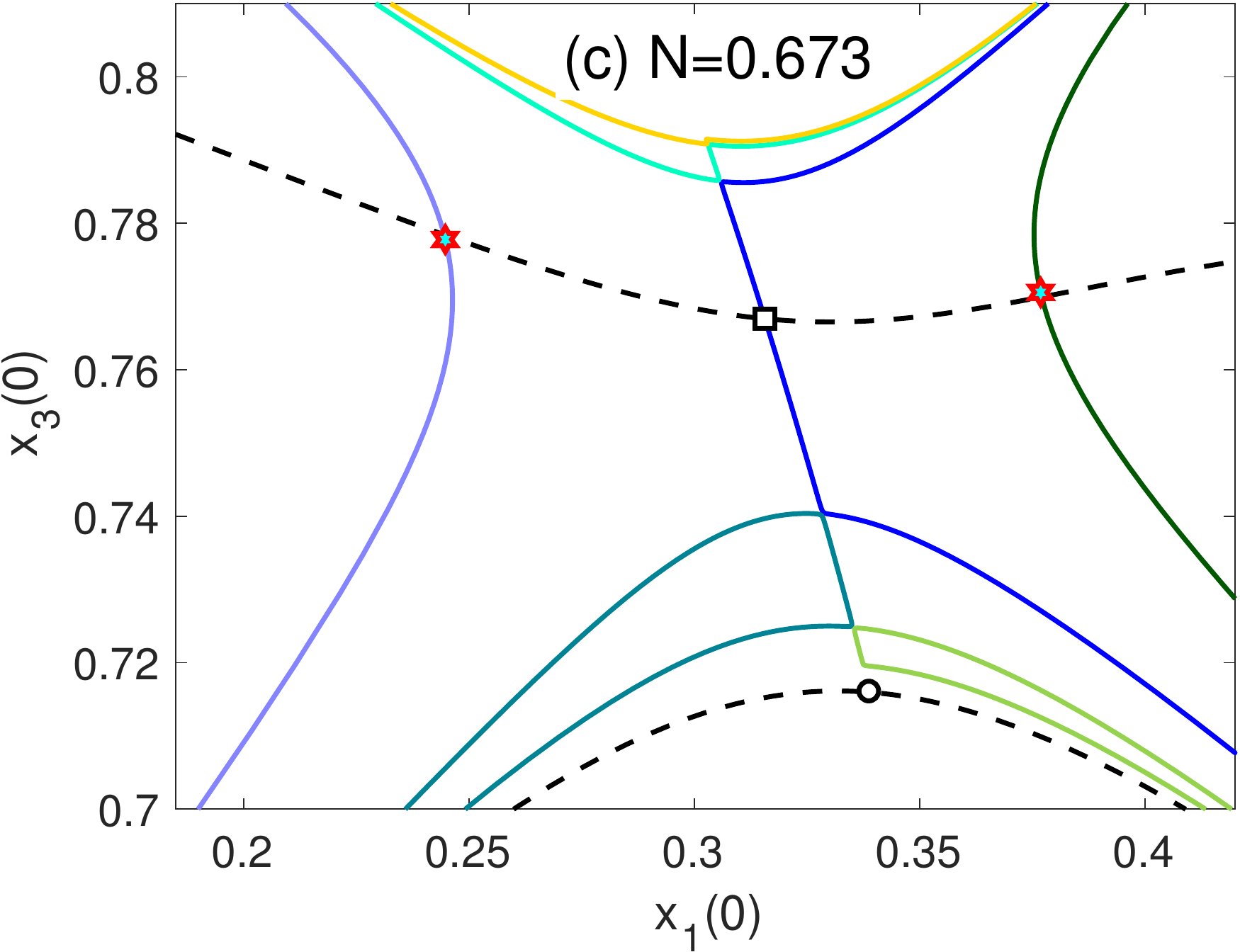} 
   \caption{(Color online) Branches of Lyapunov periodic orbits through the points $\mpo$ and $\mpp$. As $N$ increases through the resonant values, the vertical branches, each of which includes a pair of $k$:1 periodic orbits, collide at the point $\mpo$ and reconnect into a pair of horizontally-oriented hyperbolas, with no $k$:1 periodic orbits. (a) $N<N_{a_2}$. (b) $N_{a2}<N_5< N< N_4$. (c) $N_4<N<N_3$. Markers as in Figure~\ref{fig:poo}. Since this bifurcation involves re-connection, not all curves are identified by color between panels. }
\label{fig:mpo}
\end{figure}

At $N =N_{a2} \approx 0.6672$, the branches $\mpo$ and $\mpp$ arise in a saddle-node bifurcation. The branch $\mpo$ has both short- and long-period Lyapunov families of  periodic orbits while the branch $\mpp$ has only the short-period family. Figure~\ref{fig:mpo} shows some of the branches of periodic orbits that exist for $N=0.664$, $N=0.67$, and $N=0.673$, demonstrating clearly it is a hyperbolic $0^2 i \omega$ bifurcation. At resonant values of $N$, displayed in Table~\ref{table:mpo}, these branches collide with and reconnect with the Lyapunov family of long-period orbits that emerges from $\mpo$.

\begin{table}[h]
\centering
\begin{tabular}{|c|c|c|c|c|c|c|c|c|}
\hline
Bifurcation & SN & $\ldots$ & {5:1} & {4:1} & {3:1} & {2:1} & {1:1}& {1:1} \\ \hline
$N_k$ & 0.6672 & $\ldots$ & 0.6688 & 0.6710 & 0.6786 & 0.7169 & 0.9824 &1.2916\\ \hline
\end{tabular}
\caption{Values of $N_k$ for which the long and short periods of the mode $\poo$ are in $k$:1 resonance. The two 1:1 resonances are the HH bifurcations.}
\label{table:mpo}
\end{table}

Figure~\ref{fig:mpo}(a), for which $N<N_{a2}$, features only the short period and mixed-periodic families, each of which, locally, is a hyperbola. Figure~\ref{fig:mpo}(b) shows a computation with $N_5<N<N_4$. The long-period branch is, as in the elliptic case, broken up by a sequence of pseudo-intersections with mixed-period branches. Figure~\ref{fig:mpo}(c) shows a computation with $N_4<N<N_3$, demonstrating how the branches re-connect following a resonant bifurcation. 

While these branches of periodic orbits look locally like hyperbolas, they extend far outside the region of the points $\mpo$ and $\mpp$, and exist on both sides of the saddle-node bifurcation. These branches extend far outside a region where the local normal form expansion of Section~\ref{sec:saddlenode} applies.

The branch $\mpo$ also undergoes a pair of HH bifurcations at larger values of $N$. These follow the pattern seen in Sections~\ref{sec:HH},~\ref{sec:HH1}, and~\ref{sec:HH2}. First, the periods of the short- and long-period branches approach a common value, and the angle between them at $\mpo$ decreases to zero at the HH bifurcation value $N \approx 0.9824$, with the two branches merging and detaching from the fixed point. Then at the second HH bifurcation at $N \approx 1.2916$, an elliptical HH bifurcation occurs, with behavior similar to Figure~\ref{fig:HH2ODE}.

\subsubsection{What is known about such bifurcations?}
\label{sec:what}
The phenomena seen in this section has been largely described in two separate streams of research, but the pieces have not been assembled in this manner before, as far as we can tell. Schmidt~\cite{Schmidt:1974ha} considers the bifurcations of the short and long periodic branches that occur in a neighborhood of $k$:$l$ bifurcations where $k>l\ge1$. While in Section~\ref{sec:mixed}, we considered only mixed periodic orbits where the long and short period orbits were in $n$:1 resonance, but the argument there works with $n$ replaced by any rational $\tfrac{k}{l}>1$. Schmidt shows that, in fact, those periodic orbits exist in the full system for $k/l>1$ and come in hyperbolic and elliptic flavors. For $l=1$ and $k\ge4$, he proves a similar result, but with the appearance of a gap in the branch of long period orbits as seen in our numerics. For $l=1$ and $k=2$ or $k=3$, the results are a bit more delicate but can be put in correspondence with our numerical results. In particular, he explains the development of the kink in the long period branch in the 3:1 resonance bifurcation shown in Figure~\ref{fig:poo}(c). Duistermaat later recast and extended Schmidt's results in a more geometric setting~\cite{Duistermaat:1984gd}.

The above results take a localized view in a neighborhood of the individual $k$:1 resonances. Gelfreich and Lerman~\cite{Gelfreich:2003bn} take a more global view of the Hamiltonian $0^2i\omega$ bifurcation of system~\eqref{H02iw} for fixed parameter. In a re-scaled system, the  plane $(q_1,p_1)=(0,0)$ becomes a slow manifold that persists for small nonzero $\mu$ and is foliated by long periodic orbits, except for a sequence of annular gaps. These gaps occur precisely at the pseudo-intersections between the long periodic and mixed periodic branches shown in Figures~\ref{fig:poo} and~\ref{fig:mpo}. They illustrate their result with an example that is equivalent to Figure~\ref{fig:poo}. The results of this paper are confined to a small neighborhood of the slow manifold and do not address the global topology of the branches nor the difference between the elliptic and hyperbolic cases.

\section{Conclusion}
\label{sec:conclusion}
In the context of of a finite-dimensional reduced system, we have used normal forms to show how branches of relative periodic orbits behave as the total amplitude in the system is raised. We have supplemented this with numerical continuation studies tracing out families of relative periodic orbits that lie beyond the reach of the local normal forms. Our goal has been to illustrate some of the behaviors found in this system beyond simple standing waves or numerically-generated Poincar\'e sections. An unexpected find has been the complex arrangement of periodic orbits near the two saddle-node bifurcations. Further work will be necessary to more fully specify and to combine previous results into a rigorous global description of the bifurcations of these branches.

The NLS equation with a multi-well potential is an excellent model system for understanding in detail the nonlinear dynamics that occur in nonlinear dispersive equations. Previous studies of the double-well potential showed rigorously that periodic orbits of a reduced system are shadowed for long time by nearly-periodic solutions to NLS~\cite{Pelinovsky:2012,Marzuola:2010,Goodman:2015by}. A motivation for this paper has been to enumerate some of the analogous orbits in the triple-well problem, setting the stage for similar rigorous work in this more complicated case. Sigal has shown that true quasiperiodic orbits to NLS/GP cannot exist~\cite{Sigal:1993wl}, so the long-time breakdown of these nearly quasiperiodic solutions is also an important question.

While this paper has avoided talk of chaotic dynamics, they are of course ubiquitous, and were explored numerically in our previous work~\cite{Goodman:2011}.  One place where we should be able to say something rigorous is in a neighborhood of the HH bifurcations. The normal form~\eqref{HHnormal} is integrable to all orders, so standard Melnikov methods will not show any splitting of separatrices leading to homoclinic chaos; however Gaivao and Gelfreich have showed that an exponentially small splitting does indeed occur~\cite{Gaivao:2011wb} in systems with a nonsemisimple -1:1 resonance.

\section*{Acknowledgments} The author received support from NSF DMS--0807284. He is grateful for conversations with 
Casayndra Basarab,
Panayotis Kevrekidis, 
Richard Kollar,  
Jeremy Marzuola,
Dmitry Pelinovsky,
Arnd Scheel,
and
Eli Shlizerman, and to Stephen Shipman for a reading of the final manuscript. The author gratefully acknowledges the support and hospitality provided by the IMA during his visit which took place from September to December 2016.

\bibliographystyle{ieeetr}
\bibliography{sumof3wells}

\end{document}

%% file: abbrevs.tex
\renewcommand{\k}{\kappa}
\newcommand{\w}{\omega}
\newcommand{\W}{\Omega}
\newcommand{\Th}{\Theta}
\renewcommand{\a}{\alpha}
\renewcommand{\b}{\beta}

\renewcommand{\d}{\delta}
\newcommand{\D}{\Delta}
\newcommand{\ep}{\epsilon}

\newcommand{\cbar}{\bar{c}}
\newcommand{\zbar}{\bar{z}}

\newcommand{\cH}{{\mathcal H}}

\newcommand{\cN}{{\mathcal N}}
\newcommand{\cP}{{\mathcal P}}

\newcommand{\cJ}{{\mathcal J}}

\newcommand{\cNHHa}{{\mathcal N_{\rm HH1}}}
\newcommand{\cNHHb}{{\mathcal N_{\rm HH2}}}
\newcommand{\NSNe}{N_{\rm SN,even}}

\newcommand{\abs}[1] {\left\lvert #1 \right\rvert}

\newcommand{\thalf}{\tfrac{1}{2}}

\renewcommand{\ss}{\mathfrak s}
\newcommand{\cS}{{\mathcal S}}
\newcommand{\cU}{{\mathcal U}}
\newcommand{\cBo}{\cB_{\rm odd}}
\newcommand{\cBe}{\cB_{\rm even}}
\DeclareMathOperator{\sech}{sech}
\DeclareMathOperator{\ran}{ran}
 
\DeclareMathOperator{\sn}{sn} 
\newcommand{\ad}[1]{{\rm ad}_{#1}}

\newcommand{\NHHa}{N_{\rm HH1}}
\newcommand{\NHHb}{N_{\rm HH2}}

\newcommand{\Neven}{N_{\rm even}}

\newtheorem{theorem}{Theorem}

\newcommand{\CC}{{\mathbb C}}
\newcommand{\RR}{{\mathbb R}}
\renewcommand{\SS}{{\mathbb S}}
\newcommand{\ZZ}{{\mathbb Z}}

\newcommand{\innerProd}[2]{\langle #1, #2 \rangle}

\newcommand{\PB}[2]{\left\{ #1, #2 \right\} }
\newtheorem{rem}{Remark}[section]
\newcommand{\Htrunc}{H^{\rm N,trunc}}
\newcommand{\cA}{{\mathcal A}}
\newcommand{\cB}{{\mathcal B}}
\newcommand{\ee}{{\mathbf e}}
\newcommand{\uu}{{\mathbf u}}
\newcommand{\vv}{{\mathbf v}}
\newcommand{\qq}{{\mathbf q}}
\newcommand{\pp}{{\mathbf p}}
\newcommand{\zz}{{\mathbf z}}
\newcommand{\zzbar}{{\mathbf \zbar}}

\newcommand{\xx}{{\mathbf x}}
\newcommand{\cc}{{\mathbf c}}
\newcommand{\ccbar}{\bar{\mathbf c}}

\newcommand{\kk}{{\mathbf k}}
\newcommand{\yy}{{\mathbf y}}
\newcommand{\ww}{{\boldsymbol \omega}}

\newcommand{\aalpha}{{\boldsymbol \alpha}}
\newcommand{\bbeta}{{\boldsymbol \beta}}

\newcommand{\xxi}{{\boldsymbol \xi}}

\DeclareMathOperator{\vecspan}{span}
\newcommand{\mop}{\cU_{\text{\tiny{$-$0+}}}}
\newcommand{\mpm}{\cU_{\text{\tiny{$-$+$-$}}}}
\newcommand{\mpo}{\cU_{\text{\tiny{$-$+0}}}}
\newcommand{\mpp}{\cU_{\text{\tiny{$-$++}}}}
\newcommand{\poo}{\cU_{\text{\tiny{+00}}}}
\newcommand{\pop}{\cU_{\text{\tiny{+0+}}}}
\newcommand{\opo}{\cU_{\text{\tiny{0+0}}}}
\newcommand{\ppo}{\cU_{\text{\tiny{++0}}}}
\newcommand{\ppp}{\cU_{\text{\tiny{+++}}}}
\newcommand{\PPsi}{\mathbf{\Psi}}

\theoremstyle{plain}

\newcommand{\Hhigh}{H_{\rm higher}}

\newcommand{\qmup}{\qq_{\mu+}}
\newcommand{\qmum}{\qq_{\mu-}}
\newcommand{\qmupm}{\qq_{\mu\pm}}

%% file: sumof3wellsELS.bbl
\begin{thebibliography}{10}

\bibitem{Newell:2003}
A.~Newell and J.~Moloney, {\em Nonlinear Optics}.
\newblock Advanced Book Program, Westview Press, 2003.

\bibitem{Chen:2016}
X.~Chen and J.~Holmer, ``Focusing quantum many-body dynamics: The rigorous
  derivation of the 1{D} focusing cubic nonlinear {Schr\"odinger} equation,''
  {\em Arch. Ration. Mech. Anal.}, vol.~221, pp.~631--676, 2016.

\bibitem{ESY}
L.~Erd{\H{o}}s, B.~Schlein, and H.-T. Yau, ``Derivation of the cubic non-linear
  {Schr\"odinger} equation from quantum dynamics of many-body systems,'' {\em
  Invent. Math.}, vol.~167, no.~3, pp.~515--614, 2007.

\bibitem{PS:03}
L.~Pitaevskii and S.~Stringari, {\em Bose-{E}instein condensation}.
\newblock No.~116 in Int. Ser. Monogr. on Phys., Oxford University Press, 2003.

\bibitem{KirKevShl:08}
E.~W. Kirr, P.~G. Kevrekidis, E.~Shlizerman, and M.~I. Weinstein,
  ``Symmetry-breaking bifurcation in nonlinear
  {S}chr{\"o}dinger/{G}ross-{P}itaevskii equations,'' {\em SIAM J. Math.
  Anal.}, vol.~40, pp.~566--604, 2008.

\bibitem{KapKevChe:06}
T.~Kapitula, P.~G. Kevrekidis, and Z.~Chen, ``Three is a crowd: {S}olitary
  waves in photorefractive media with three potential wells,'' {\em SIAM J.
  Appl. Dyn. Syst.}, vol.~5, pp.~598--633, 2006.

\bibitem{Kirr:2011eu}
E.~W. Kirr, P.~G. Kevrekidis, and D.~E. Pelinovsky, ``{Symmetry-Breaking
  Bifurcation in the Nonlinear Schr{\"o}dinger Equation with Symmetric
  Potentials},'' {\em Commun. Math. Phys.}, vol.~308, pp.~795--844, Oct. 2011.

\bibitem{Kirr:2016uy}
E.~W. Kirr, ``{Long time dynamics and coherent states in nonlinear wave
  equations},'' {\em arXiv math.AP}, May 2016.

\bibitem{Goodman:2011}
R.~H. Goodman, ``{Hamiltonian Hopf bifurcations and dynamics of NLS/GP
  standing-wave modes},'' {\em J. Phys. A: Math. Theor.}, vol.~44, p.~425101,
  2011.

\bibitem{Sacchetti:2012tx}
A.~Sacchetti, ``{Nonlinear Schr{\"o}dinger equations with multiple-well
  potential},'' {\em Phys. D}, vol.~241, pp.~1815--1824, 2012.

\bibitem{Goodman:2015by}
R.~H. Goodman, J.~L. Marzuola, and M.~I. Weinstein, ``{Self-trapping and
  Josephson tunneling solutions to the nonlinear Schr{\"o}dinger /
  Gross-Pitaevskii equation},'' {\em Discrete Contin. Dyn. Syst.}, vol.~35,
  pp.~225--246, 2015.

\bibitem{Marzuola:2010}
J.~L. Marzuola and M.~I. Weinstein, ``Long time dynamics near the symmetry
  breaking bifurcation for nonlinear {S}chr{\"o}dinger/{G}ross-{P}itaevskii
  equations,'' {\em DCDS-A}, vol.~28, pp.~1505--1554, 2010.

\bibitem{Pelinovsky:2012}
D.~Pelinovsky and T.~Phan, ``Normal form for the symmetry-breaking bifurcation
  in the nonlinear {S}chr{\"o}dinger equation,'' {\em J. Diff. Eq.}, vol.~253,
  pp.~2796--2824, 2012.

\bibitem{Fukuizumi:2011ku}
R.~Fukuizumi and A.~Sacchetti, ``Bifurcation and stability for nonlinear
  {S}chr{\"o}dinger equations with double well potential in the semiclassical
  limit,'' {\em J. Stat. Phys.}, vol.~145, pp.~1546--1594, 2011.

\bibitem{Albiez:2005}
M.~Albiez, R.~Gati, J.~F{\"o}lling, S.~Hunsmann, M.~Cristiani, and M.~K.
  Oberthaler, ``Direct observation of tunneling and nonlinear self-trapping in
  a single {B}osonic {J}osephson junction,'' {\em Phys. Rev. Lett.}, vol.~95,
  p.~010402, 2005.

\bibitem{Yang:2016hj}
J.~Yang, ``A normal form for {H}amiltonian--{H}opf bifurcations in nonlinear
  {S}chr{\"o}dinger equations with general external potentials,'' {\em SIAM J.
  Appl. Math.}, vol.~76, pp.~598--617, 2016.

\bibitem{Kevrekidis:2015bq}
P.~G. Kevrekidis, D.~E. Pelinovsky, and A.~Saxena, ``When linear stability does
  not exclude nonlinear instability,'' {\em Phys. Rev. Lett.}, vol.~114,
  p.~214101, 2015.

\bibitem{Eilbeck:1985tu}
J.~C. Eilbeck, P.~S. Lomdahl, and A.~C. Scott, ``{The discrete self-trapping
  equation},'' {\em Phys. D}, vol.~16, pp.~318--338, 1985.

\bibitem{Carr:1985}
J.~Carr and J.~C. Eilbeck, ``{Stability of stationary solutions of the discrete
  self-trapping equation},'' {\em Phys. Lett. A}, vol.~109, pp.~201--204, 1985.

\bibitem{Susanto:2009km}
H.~Susanto, ``Few-lattice-site systems of discrete self-trapping equations,''
  in {\em The Discrete Nonlinear Schr{\"o}dinger Equation} (P.~G. Kevrekidis,
  ed.), Berlin, Heidelberg: Springer, 2009.

\bibitem{Banacky:1988ft}
P.~Ba{\v{n}}ack{\'{y}} and A.~Zajac, ``{Theory of particle transfer dynamics in
  solvated molecular complexes: analytic solutions of the discrete
  time-dependent nonlinear Schr{\"o}dinger equation. I. Conservative system},''
  {\em Chem. Phys.}, vol.~123, pp.~267--276, 1988.

\bibitem{Kenkre:1986fe}
V.~M. Kenkre and D.~K. Campbell, ``{Self-trapping on a dimer: Time-dependent
  solutions of a discrete nonlinear Schr{\"o}dinger equation},'' {\em Phys.
  Rev., B Condens. Matter}, vol.~34, pp.~4959--4961, 1986.

\bibitem{Liu:2007cl}
B.~Liu, L.-B. Fu, S.-P. Yang, and J.~Liu, ``{Josephson oscillation and
  transition to self-trapping for Bose-Einstein condensates in a triple-well
  trap},'' {\em Phys. Rev. A}, vol.~75, p.~033601, 2007.

\bibitem{Zhang:2001jg}
S.~Zhang and F.~Wang, ``{Interference effects between three coupled
  Bose{\textendash}Einstein condensates},'' {\em Phys. Lett. A}, vol.~279,
  pp.~231--238, 2001.

\bibitem{Johansson:2004uj}
M.~Johansson, ``{Hamiltonian Hopf bifurcations in the discrete nonlinear
  Schr{\"o}dinger trimer},'' {\em J. Phys. A}, vol.~37, pp.~2201--2222, 2004.

\bibitem{Panayotaros:2012cs}
P.~Panayotaros, ``{Instabilities of breathers in a finite NLS lattice},'' {\em
  Phys. D}, vol.~241, pp.~847--856, 2012.

\bibitem{Basarab:2016}
C.~H. Basarab, {\em Hamiltonian Bifurcations in {S}chr{\"o}dinger Trimers}.
\newblock PhD thesis, New Jersey Institute of Technology, 2016.

\bibitem{Chow:1988}
S.-N. Chow and Y.-I. Kim, ``Bifurcation of periodic orbits for non-positive
  definite {H}amiltonian systems,'' {\em Appl. Anal.}, vol.~31, pp.~163--199,
  1988.

\bibitem{Meyer:2010}
K.~Meyer, G.~Hall, and D.~Offin, {\em Introduction to {H}amiltonian Dynamical
  Systems and the N-Body Problem}, vol.~90 of {\em Applied Mathematical
  Sciences}.
\newblock Springer, 2nd~ed., 2010.

\bibitem{Moser:1976}
J.~Moser, ``Periodic orbits near an equilibrium and a theorem by {A}lan
  {W}einstein,'' {\em Commun. Pure Appl. Math.}, vol.~29, pp.~727--747, 1976.

\bibitem{Burgoyne:1974}
N.~Burgoyne and R.~Cushman, ``{Normal forms for real linear Hamiltonian systems
  with purely imaginary eigenvalues},'' {\em Celest. Mech. Dyn. Astr.}, vol.~8,
  pp.~435--443, 1974.

\bibitem{Broer:1993ch}
H.~W. Broer, S.-N. Chow, Y.-I. Kim, and G.~Vegter, ``{A normally elliptic
  Hamiltonian bifurcation},'' {\em Z. Angew. Math. Phys.}, vol.~44,
  pp.~389--432, 1993.

\bibitem{Gelfreich:2014fn}
V.~Gelfreich and L.~M. Lerman, ``{Separatrix splitting at a Hamiltonian $0^2 i
  \omega$ bifurcation},'' {\em Regul. Chaotic Dyn.}, vol.~19, pp.~635--655,
  2014.

\bibitem{Vis:01}
D.~Viswanath, ``The {L}indstedt-{P}oincar\'e technique as an algorithm for
  computing periodic orbits,'' {\em SIAM Rev.}, vol.~43, pp.~478--495, 2001.

\bibitem{Sigal:1993wl}
I.~M. Sigal, ``{Non-linear wave and Schr{\"o}dinger equations I. Instability of
  Periodic and Quasiperiodic Solutions },'' {\em Commun. Math. Phys.}, 1993.

\bibitem{Dohnal:2007}
T.~Dohnal and T.~Hagstrom, ``{Perfectly matched layers in photonics
  computations: 1D and 2D nonlinear coupled mode equations},'' {\em J. Comput.
  Phys.}, vol.~223, pp.~690--710, 2007.

\bibitem{Kennedy:2003}
C.~A. Kennedy and M.~H. Carpenter, ``{Additive Runge--Kutta schemes for
  convection--diffusion--reaction equations},'' {\em Appl. Numer. Math.},
  vol.~44, pp.~139--181, 2003.

\bibitem{Schmidt:1974ha}
D.~S. Schmidt, ``{Periodic solutions near a resonant equilibrium of a
  Hamiltonian system},'' {\em Celestial Mech.}, vol.~9, pp.~81--103, 1974.

\bibitem{Duistermaat:1984gd}
J.~J. Duistermaat, ``{Bifurcations of periodic solutions near equilibrium
  points of Hamiltonian systems},'' in {\em Bifurcation Theory and
  Applications}, pp.~57--105, Springer Berlin Heidelberg, 1984.

\bibitem{Gelfreich:2003bn}
V.~Gelfreich and L.~M. Lerman, ``{Long-periodic orbits and invariant tori in a
  singularly perturbed Hamiltonian system},'' {\em Phys. D}, vol.~176,
  pp.~125--146, 2003.

\bibitem{Gaivao:2011wb}
J.~P. Gaivao and V.~Gelfreich, ``{Splitting of separatrices for the
  Hamiltonian-Hopf bifurcation with the Swift-Hohenberg equation as an
  example},'' {\em Nonlinearity}, vol.~24, pp.~677--698, 2011.

\end{thebibliography}
